\documentstyle[epsfig,a4p,12pt,rotating,cite]{article}

\begin{document}         

\newcommand{\Rparity}{$R$-parity}
\newcommand{\Rp}  {$R_{p}$}
\newcommand{\lb}  {$\lambda$}
\newcommand{\lbp} {$\lambda^{'}$}
\newcommand{\lbpp} {$\lambda^{''}$}

\newcommand{\ee}{{\mathrm e}^+ {\mathrm e}^-}
\newcommand{\sq}{\tilde{\mathrm q}}
\newcommand{\seff}{\tilde{\mathrm f}}
\newcommand{\sele}{\tilde{\mathrm e}}
\newcommand{\supq}{\tilde{\mathrm u}}
\newcommand{\sdown}{\tilde{\mathrm d}}
\newcommand{\sfer}{\tilde{\mathrm f}}
\newcommand{\sstrange}{\tilde{\mathrm s}}
\newcommand{\scharm}{\tilde{\mathrm c}}
\newcommand{\sbottom}{\tilde{\mathrm b}}
\newcommand{\ellp}{\ell^+}
\newcommand{\ellm}{\ell^-}
\newcommand{\sell}{\tilde{\ell}}
\newcommand{\snu}{\tilde{\nu}}
\newcommand{\smu}{\tilde{\mu}}
\newcommand{\stau}{\tilde{\tau}}
\newcommand{\chp}{\tilde{\chi}^+_1}
\newcommand{\chip}{\tilde{\chi}^+_1}
\newcommand{\chim}{\tilde{\chi}^-_1}
\newcommand{\chpm}{\tilde{\chi}^\pm_1}
\newcommand{\chipm}{\tilde{\chi}^\pm_1}
\newcommand{\nt}{\tilde{\chi}^0}
\newcommand{\qq}{q\bar{q}}
\newcommand{\sleppair}{\sell^+ \sell^-}
\newcommand{\nunu}{\nu \bar{\nu}}
\newcommand{\mumu}{\mu^+ \mu^-}
\newcommand{\tautau}{\tau^+ \tau^-}
\newcommand{\ellell}{\ell^+ \ell^-}
\newcommand{\nulqq}{\nu \ell {\mathrm q} \bar{\mathrm q}'}
\newcommand{\MZ}{M_{\mathrm Z}}

\newcommand {\stopm}         {\tilde{\mathrm{t}}_{1}}
\newcommand {\stopn}         {\tilde{\mathrm{t}}}
\newcommand {\stops}         {\tilde{\mathrm{t}}_{2}}
\newcommand {\stopbar}       {\bar{\tilde{\mathrm{t}}}_{1}}
\newcommand {\stopx}         {\tilde{\mathrm{t}}}
\newcommand {\sneutrino}     {\tilde{\nu}}
\newcommand {\slepton}       {\tilde{\ell}}
\newcommand {\stopl}         {\tilde{\mathrm{t}}_{\mathrm L}}
\newcommand {\stopr}         {\tilde{\mathrm{t}}_{\mathrm R}}
\newcommand {\stoppair}      {\tilde{\mathrm{t}}_{1}
\bar{\tilde{\mathrm{t}}}_{1}}
\newcommand {\gluino}        {\tilde{\mathrm g}}

\newcommand {\chin}          {\tilde{\chi }^{0}_{1}}
\newcommand {\neutralino}    {\tilde{\chi }^{0}_{1}}
\newcommand {\neutrala}      {\tilde{\chi }^{0}_{2}}
\newcommand {\neutralb}      {\tilde{\chi }^{0}_{3}}
\newcommand {\neutralc}      {\tilde{\chi }^{0}_{4}}
\newcommand {\bino}          {\tilde{\mathrm B}^{0}}
\newcommand {\wino}          {\tilde{\mathrm W}^{0}}
\newcommand {\higginoa}      {\tilde{\rm H_{1}}^{0}}
\newcommand {\higginob}      {\tilde{\mathrm H_{1}}^{0}}
\newcommand {\chargino}      {\tilde{\chi }^{\pm}_{1}}
\newcommand {\charginop}     {\tilde{\chi }^{+}_{1}}
\newcommand {\KK}            {{\mathrm K}^{0}-\bar{\mathrm K}^{0}}
\newcommand {\ff}            {{\mathrm f} \bar{\mathrm f}}
\newcommand {\bstopm} {\mbox{$\boldmath {\tilde{\mathrm{t}}_{1}} $}}
\newcommand {\Mt}            {M_{\mathrm t}}
\newcommand {\mscalar}       {m_{0}}
\newcommand {\Mgaugino}      {M_{1/2}}
\newcommand {\rs}            {\sqrt{s}}
\newcommand {\WW}            {{\mathrm W}^+{\mathrm W}^-}
\newcommand {\MGUT}          {M_{\mathrm {GUT}}}
\newcommand {\Zboson}        {{\mathrm Z}^{0}}
\newcommand {\Wpm}           {{\mathrm W}^{\pm}}
\newcommand {\allqq}         {\sum_{q \neq t} q \bar{q}}
\newcommand {\mixang}        {\theta _{\mathrm {mix}}}
\newcommand {\thacop}        {\theta _{\mathrm {Acop}}}
\newcommand {\cosjet}        {\cos\thejet}
\newcommand {\costhr}        {\cos\thethr}
\newcommand {\djoin}         {d_{\mathrm{join}}}
\newcommand {\mstop}         {m_{\stopm}}
\newcommand {\msell}         {m_{\sell}}
\newcommand {\mchi}          {m_{\neutralino}}
\newcommand {\pp}{p \bar{p}}

\newcommand{\epair}{\mbox{${\mathrm e}^+{\mathrm e}^-$}}
\newcommand{\mupair}{\mbox{$\mu^+\mu^-$}}
\newcommand{\taupair}{\mbox{$\tau^+\tau^-$}}
\newcommand{\qpair}{\mbox{${\mathrm q}\overline{\mathrm q}$}}
\newcommand{\eeee}{\mbox{\epair\epair}}
\newcommand{\eemumu}{\mbox{\epair\mupair}}
\newcommand{\eetautau}{\mbox{\epair\taupair}}
\newcommand{\eeqq}{\mbox{\epair\qpair}}
\newcommand{\fs}{ final states}
\newcommand{\epairf}{\mbox{\epair\fs}}
\newcommand{\mupairf}{\mbox{\mupair\fs}}
\newcommand{\taupairf}{\mbox{\taupair\fs}}
\newcommand{\qpairf}{\mbox{\qpair\fs}}
\newcommand{\eeeef}{\mbox{\eeee\fs}}
\newcommand{\eemumuf}{\mbox{\eemumu\fs}}
\newcommand{\eetautauf}{\mbox{\eetautau\fs}}
\newcommand{\eeqqf}{\mbox{\eeqq\fs}}
\newcommand{\ffff}{four fermion final states}
\newcommand{\llnunu}{\mbox{\lpair\nul\nubar}}
\newcommand{\lnuqq}{\mbox{\lept\nubar\qpair}}
\newcommand{\zee}{\mbox{Zee}}
\newcommand{\zzg}{\mbox{ZZ/Z$\gamma$}}
\newcommand{\wenu}{\mbox{We$\nu$}}

\newcommand{\el}{\mbox{${\mathrm e}^-$}}
\newcommand{\selem}{\mbox{$\tilde{\mathrm e}^-$}}
\newcommand{\smum}{\mbox{$\tilde\mu^-$}}
\newcommand{\staum}{\mbox{$\tilde\tau^-$}}
\newcommand{\slept}{\mbox{$\tilde{\ell}^\pm$}}
\newcommand{\sleptm}{\mbox{$\tilde{\ell}^-$}}
\newcommand{\lept}{\mbox{$\ell^-$}}
\newcommand{\Hl}{\mbox{$\mathrm{L}^\pm$}}
\newcommand{\Hm}{\mbox{$\mathrm{L}^-$}}
\newcommand{\Hnu}{\mbox{$\nu_{\mathrm{L}}$}}
\newcommand{\nul}{\mbox{$\nu_\ell$}}
\newcommand{\nubar}{\mbox{$\overline{\nu}_\ell$}}
\newcommand{\spair}{\mbox{$\tilde{\ell}^+\tilde{\ell}^-$}}
\newcommand{\lpair}{\mbox{$\ell^+\ell^-$}}
\newcommand{\staupair}{\mbox{$\tilde{\tau}^+\tilde{\tau}^-$}}
\newcommand{\smupair}{\mbox{$\tilde{\mu}^+\tilde{\mu}^-$}}
\newcommand{\selepair}{\mbox{$\tilde{\mathrm e}^+\tilde{\mathrm e}^-$}}
\newcommand{\ch}{\mbox{$\tilde{\chi}^\pm_1$}}
\newcommand{\chpair}{\mbox{$\tilde{\chi}^+_1\tilde{\chi}^-_1$}}
\newcommand{\chm}{\mbox{$\tilde{\chi}^-_1$}}
\newcommand{\chmp}{\mbox{$\tilde{\chi}^\pm_1$}}
\newcommand{\chz}{\mbox{$\tilde{\chi}^0_1$}}
\newcommand{\dch}{\mbox{\chm$\rightarrow$\chz\lept\nubar}}
\newcommand{\dslept}{\mbox{\sleptm$\rightarrow$\chz\lept}}
\newcommand{\dH}{\mbox{\Hm$\rightarrow$\lept\nubar\Hnu}}
\newcommand{\mch}{\mbox{$m_{\tilde{\chi}^\pm_1}$}}
\newcommand{\mslept}{\mbox{$m_{\tilde{\ell}}$}}
\newcommand{\mstau}{\mbox{$m_{\staum}$}}
\newcommand{\msmu}{\mbox{$m_{\smum}$}}
\newcommand{\msele}{\mbox{$m_{\selem}$}}
\newcommand{\mchz}{\mbox{$m_{\tilde{\chi}^0_1}$}}
\newcommand{\dm}{\mbox{$\Delta m$}}
\newcommand{\dmch}{\mbox{$\Delta m_{\ch-\chz}$}}
\newcommand{\dmslept}{\mbox{$\Delta m_{\slept-\chz}$}}
\newcommand{\dmhl}{\mbox{$\Delta m_{\Hl-\Hnu}$}}
\newcommand{\w}{\mbox{W$^\pm$}}

\newcommand{\acopc}{\mbox{$\phi^{\mathrm{acop}}$}}
\newcommand{\acolc}{\mbox{$\theta^{\mathrm{acol}}$}}
\newcommand{\acop}{\mbox{$\phi_{\mathrm{acop}}$}}
\newcommand{\acol}{\mbox{$\theta_{\mathrm{acol}}$}}
\newcommand{\pt}{\mbox{$p_{t}$}}
\newcommand{\pz}{\mbox{$p_{\mathrm{z}}^{\mathrm{miss}}$}}
\newcommand{\ptevt}{\mbox{$p_{t}^{\mathrm{miss}}$}}
\newcommand{\ptaxic}{\mbox{$a_{t}^{\mathrm{miss}}$}}
\newcommand{\stevt}{\mbox{$p_{t}^{\mathrm{miss}}$/\Ebeam}}
\newcommand{\staxic}{\mbox{$a_{t}^{\mathrm{miss}}$/\Ebeam}}
\newcommand{\dptaxic}{\mbox{missing $p_{t}$ wrt. event axis \ptaxic}}
\newcommand{\cosevt}{\mbox{$\mid\cos\theta_{\mathrm{p}}^{\mathrm{miss}}\mid$}}
\newcommand{\axicos}{\mbox{$\mid\cos\theta_{\mathrm{a}}^{\mathrm{miss}}\mid$}}
\newcommand{\pthet}{\mbox{$\theta_{\mathrm{p}}^{\mathrm{miss}}$}}
\newcommand{\athet}{\mbox{$\theta_{\mathrm{a}}^{\mathrm{miss}}$}}
\newcommand{\dcosevt}{\mbox{$\mid\cos\theta\mid$ of missing p$_{t}$}}
\newcommand{\daxicos}{\mbox{$\mid\cos\theta\mid$ of missing p$_{t}$ wrt. event
axis}}
\newcommand{\efdsw}{\mbox{$x_{\mathrm{FDSW}}$}}
\newcommand{\acopf}{\mbox{$\Delta\phi_{\mathrm{FDSW}}$}}
\newcommand{\acopm}{\mbox{$\Delta\phi_{\mathrm{MUON}}$}}
\newcommand{\acopt}{\mbox{$\Delta\phi_{\mathrm{trk}}$}}
\newcommand{\po}{\mbox{$E_{\mathrm{isol}}^\gamma$}}
\newcommand{\qprod}{\mbox{$q1$$*$$q2$}}
\newcommand{\lcode}{lepton identification code}
\newcommand{\nctro}{\mbox{$N_{\mathrm{trk}}^{\mathrm{out}}$}}
\newcommand{\necao}{\mbox{$N_{\mathrm{ecal}}^{\mathrm{out}}$}}
\newcommand{\mout}{\mbox{$m^{\mathrm{out}}$}}
\newcommand{\nctec}{\mbox{\nctro$+$\necao}}
\newcommand{\gfract}{\mbox{$F_{\mathrm{good}}$}}
\newcommand{\zz}       {\mbox{$|z_0|$}}
\newcommand{\dz}       {\mbox{$|d_0|$}}
\newcommand{\sint}      {\mbox{$\sin\theta$}}
\newcommand{\cost}      {\mbox{$\cos\theta$}}
\newcommand{\mcost}     {\mbox{$|\cos\theta|$}}
\newcommand{\dedx}     {\mbox{$dE/dx$}}
\newcommand{\wdedx}     {\mbox{$W_{dE/dx}$}}
\newcommand{\xe}     {\mbox{$x_E$}}

\newcommand{\ssix}     {\mbox{$\sqrt{s}$~=~161~GeV}}
\newcommand{\sthree}     {\mbox{$\sqrt{s}$~=~130--136~GeV}}
\newcommand{\mrecoil}     {\mbox{$m_{\mathrm{recoil}}$}}
\newcommand{\llmass}     {\mbox{$m_{ll}$}}
\newcommand{\sml}{\mbox{Standard Model \lpair$\nu\nu$ events}}
\newcommand{\sme}{\mbox{Standard Model events}}
\newcommand{\sig}{events containing a lepton pair plus missing transverse momentum}
\newcommand{\wpair}{\mbox{$W^+W^-$}}
\newcommand{\dW}{\mbox{W$^-\rightarrow$\lept\nubar}}
\newcommand{\dsele}{\mbox{\selem$\rightarrow$\chz e$^-$}}
\newcommand{\eeeell}{\mbox{\epair$\rightarrow$\epair\lpair}}
\newcommand{\eell}{\mbox{\epair\lpair}}
\newcommand{\llgam}{\mbox{$\ell\ell(\gamma)$}}
\newcommand{\nunugam}{\mbox{$\nu\bar{\nu}\gamma\gamma$}}
\newcommand{\acope}{\mbox{$\Delta\phi_{\mathrm{EE}}$}}
\newcommand{\nee}{\mbox{N$_{\mathrm{EE}}$}}
\newcommand{\eesum}{\mbox{$\Sigma_{\mathrm{EE}}$}}
\newcommand{\at}{\mbox{$a_{t}$}}
\newcommand{\spp}{\mbox{$p$/\Ebeam}}
\newcommand{\acoph}{\mbox{$\Delta\phi_{\mathrm{HCAL}}$}}

\newcommand{\roots}     {\sqrt{s}}
%
%
\newcommand{\thrust}    {T}
\newcommand{\nthrust}   {\hat{n}_{\mathrm{thrust}}}
\newcommand{\thethr}    {\theta_{\,\mathrm{thrust}}}
\newcommand{\phithr}    {\phi_{\mathrm{thrust}}}
\newcommand{\acosthr}   {|\cos\thethr|}
\newcommand{\thejet}    {\theta_{\,\mathrm{jet}}}
\newcommand{\acosjet}   {|\cos\thejet|}
\newcommand{\thmiss}    { \theta_{\mathrm{miss}} }
\newcommand{\cosmiss}   {| \cos \thmiss |}

\newcommand{\Evis}      {E_{\mathrm{vis}}}
\newcommand{\Emiss}     {E_{\mathrm{miss}}}
\newcommand{\Rvis}      {E_{\mathrm{vis}}\,/\roots}
\newcommand{\Mvis}      {m_{\mathrm{vis}}}
\newcommand{\Rbal}      {R_{\mathrm{bal}}}

\newcommand{\Ecm}{\mbox{$E_{\mathrm{cm}}$}}
\newcommand{\Ebeam}{\mbox{$E_{\mathrm{beam}}$}}
\newcommand{\ipb}{\mbox{pb$^{-1}$}}
\newcommand{\wrt}{with respect to}
\newcommand{\sm}{Standard Model}
\newcommand{\smb}{Standard Model background}
\newcommand{\smp}{Standard Model processes}
\newcommand{\smc}{Standard Model Monte Carlo}
\newcommand{\mc}{Monte Carlo}
\newcommand{\btb}{back-to-back}
\newcommand{\tp}{two-photon}
\newcommand{\tpb}{two-photon background}
\newcommand{\tpp}{two-photon processes}
\newcommand{\lp}{lepton pairs}
\newcommand{\vto}{\mbox{$\tau$ veto}}
\newcommand{\gsim}{\;\raisebox{-0.9ex}
           {$\textstyle\stackrel{\textstyle >}{\sim}$}\;}
\newcommand{\lsim}{\;\raisebox{-0.9ex}{$\textstyle\stackrel{\textstyle<}
           {\sim}$}\;}
\newcommand{\degree}    {^\circ}

\newcommand{\phiacop}   {\phi_{\mathrm{acop}}}


%
%
\newcommand{\ZP}[3]    {Z. Phys. {\bf C#1} (#2) #3.}
\newcommand{\PL}[3]    {Phys. Lett. {\bf B#1} (#2) #3.}
\newcommand{\etal}     {{\it et al}.,\,\ }
\newcommand{\PhysLett}  {Phys.~Lett.}
\newcommand{\PRL} {Phys.~Rev.\ Lett.}
\newcommand{\PhysRep}   {Phys.~Rep.}
\newcommand{\EuroPhys}  {Eur.~Phys. \ J.}
\newcommand{\PhysRev}   {Phys.~Rev.}
\newcommand{\NPhys}  {Nucl.~Phys.}
\newcommand{\NIM} {Nucl.~Instr.\ Meth.}
\newcommand{\CPC} {Comp.~Phys.\ Comm.}
\newcommand{\ZPhys}  {Z.~Phys.}
\newcommand{\IEEENS} {IEEE Trans.\ Nucl.~Sci.}
%
%
\newcommand{\OPALColl}   {OPAL Collab.}
\newcommand{\ALEPHColl}  {ALEPH Collab.}
\newcommand{\DELPHIColl} {DELPHI Collab.}
\newcommand{\XLColl}     {L3 Collab.}
\newcommand{\JADEColl}   {JADE Collab.}
\newcommand{\CDFColl}    {CDF Collab.}
\newcommand{\DXColl}     {D0 Collab.}
\newcommand{\HXColl}     {H1 Collab.}
\newcommand{\ZEUSColl}   {ZEUS Collab.}
%
\newcommand{\onecol}[2] {\multicolumn{1}{#1}{#2}}
\newcommand{\ra}        {\rightarrow}   


 


%
%
\newcommand{\PPEnum}    {CERN-EP-2003-xxx}
\newcommand{\PNnum}     {OPAL Physics Note PN-499}
\newcommand{\TNnum}     {OPAL Technical Note TN-xxx}
\newcommand{\Date}      {\today}

\newcommand{\Author}    
{Authors: S.~Braibant, M.~Fanti, I.~Fleck, \\
P.~Giacomelli, M.~Harin-Dirac, G.~Masetti, A.~Mutter, G.~P\'{a}sztor }
\newcommand{\MailAddr}  {Sylvie.Braibant@cern.ch, Paolo.Giacomelli@cern.ch}
\newcommand{\EdBoard}   {Editorial Board: C.~Rembser, I.~Trigger,
S.~Yamashita, B.~Vachon}
\newcommand{\DraftVer}  {Version 2.0}
\newcommand{\DraftDate} {\today}
\newcommand{\TimeLimit} {Friday, 12$^{th}$ July 2002}


\begin{titlepage}
%
%
\begin{center}
    \large
    EUROPEAN ORGANISATION FOR NUCLEAR RESEARCH
\end{center}
\begin{flushright}
    \large
     CERN-EP/2003-036\\
    23rd June 2003
\end{flushright}

%
%
\bigskip
\bigskip
\bigskip
\bigskip
\begin{center}
    \huge\bf\boldmath
 Search for R-Parity Violating Decays \\
      of Scalar Fermions at LEP

\normalsize

\vspace{0.5cm}

\LARGE

The OPAL Collaboration \\

\large
\end{center}

\vspace{1.0cm}

%
%
\begin{abstract}

A search for pair-produced scalar fermions 
under the assumption 
that \Rparity\ is not conserved has been performed 
using data  collected with the OPAL detector at LEP.
The data samples analysed correspond to an integrated luminosity of
about 610~pb$^{-1}$ collected 
at centre-of-mass energies of $\sqrt{s}=$ 189$-$209~GeV.
An important consequence of \Rparity\ violation is that 
the lightest supersymmetric particle is expected to be unstable. 
Searches for \Rparity\ violating decays of charged sleptons, 
sneutrinos and squarks have been 
performed under the assumptions that the 
lightest supersymmetric particle decays promptly and that only one of 
the \Rparity\ violating couplings is dominant for each of the decay 
modes considered. Such processes would yield final states consisting of
leptons, jets, or both, with 
or without missing energy.
No significant signal-like excess of events has been observed with 
respect to the \sm\ expectations.
Limits on the production cross-sections of 
scalar fermions in \Rparity\ violating scenarios are obtained. 
Constraints on the supersymmetric particle masses 
are also presented in an \Rparity\ violating framework analogous to 
the Constrained Minimal Supersymmetric Standard Model.
\end{abstract}

\vspace{1.5cm}

\begin{center}
{\large  (To be submitted to European Physical Journal C)}
\end{center}

\smallskip
\begin{center}
\smallskip
\end{center}
 
\bigskip
\bigskip
\smallskip

\begin{center}

\end{center}

\end{titlepage}

\newpage

\begin{center}{\Large        The OPAL Collaboration
}\end{center}\bigskip
\begin{center}{
G.\thinspace Abbiendi$^{  2}$,
C.\thinspace Ainsley$^{  5}$,
P.F.\thinspace {\AA}kesson$^{  3}$,
G.\thinspace Alexander$^{ 22}$,
J.\thinspace Allison$^{ 16}$,
P.\thinspace Amaral$^{  9}$, 
G.\thinspace Anagnostou$^{  1}$,
K.J.\thinspace Anderson$^{  9}$,
S.\thinspace Arcelli$^{  2}$,
S.\thinspace Asai$^{ 23}$
% ...
}\end{center}\bigskip
\bigskip
$^{  1}$School of Physics and Astronomy, University of Birmingham,
Birmingham B15 2TT, UK
\newline
$^{  2}$Dipartimento di Fisica dell' Universit\`a di Bologna and INFN,
I-40126 Bologna, Italy
\newline
$^{  3}$Physikalisches Institut, Universit\"at Bonn,
D-53115 Bonn, Germany
\newline
% ...
$^{  5}$Cavendish Laboratory, Cambridge CB3 0HE, UK
\newline
% ...
$^{  9}$Enrico Fermi Institute and Department of Physics,
University of Chicago, Chicago IL 60637, USA
\newline
% ...
$^{ 16}$Department of Physics, Schuster Laboratory, The University,
Manchester M13 9PL, UK
\newline
% ...
$^{ 22}$Department of Physics and Astronomy, Tel Aviv University,
Tel Aviv 69978, Israel
\newline
$^{ 23}$International Centre for Elementary Particle Physics and
Department of Physics, University of Tokyo, Tokyo 113-0033, and
Kobe University, Kobe 657-8501, Japan
\newline
% ...
$^{  s}$ now at Universitaire Instelling Antwerpen, Physics Department, 
B-2610 Antwerpen, Belgium
\newline
% ...

\newpage
\section{Introduction}
\label{sec:intro}

In Supersymmetric (SUSY)~\cite{ref:SUSY} models, each elementary
particle is accompanied by a
supersymmetric partner whose spin differs by half a unit.
Many of the searches for these supersymmetric particles (``sparticles'') 
are performed within the Minimal Supersymmetric extension of the 
\sm\ (MSSM)~\cite{ref:MSSM}, assuming \Rparity\ conservation. 
\Rparity~\cite{ref:rparity} is a multiplicative quantum number defined as 
$R_{p}=(-1)^{2S+3B+L}$ where $S$, $B$ and $L$ are the spin, baryon and
lepton numbers of the particle, respectively. 
\Rparity\ discriminates between ordinary
and supersymmetric particles: \Rp\ = +1 for \sm\ particles 
and \Rp\ = --1 for their supersymmetric partners.
\Rparity\ conservation implies that supersymmetric particles are 
always pair-produced and always decay through
cascade decays to ordinary particles and the lightest supersymmetric 
particle (LSP). In this context, the LSP is often assumed to be
the lightest neutralino, $\neutralino$, which is then expected to be 
stable and to escape detection due to its weakly interacting nature. 
The  characteristic signature of the supersymmetric \Rparity\ 
conserving decays is therefore missing energy. 

If \Rparity\ is violated, 
sparticles can be pair-produced 
but may decay directly to \sm\ particles.
Scalar neutrinos could also be singly-produced 
via an s- or t-channel sneutrino exchange~\cite{ref:singleprod, ref:barger1}. 
In the present paper, the possible direct manifestations of \Rparity\
breaking couplings via 
pair-production of sparticles and their subsequent \Rparity\ violating 
decays to \sm\ particles are studied.  
The signatures sought in the analyses of 
this paper therefore differ from the 
missing energy signatures of \Rparity\ conserving processes.

\Rparity\ violating interactions of the particles of the MSSM are 
parametrised with a gauge-invariant superpotential that includes the 
following Yukawa coupling terms~\cite{ref:dreiner1}: 

\begin{eqnarray}
{\sl W}_{RPV}  = 
    \lambda_{ijk}      L_i L_j {\overline E}_k
 +  \lambda^{'}_{ijk}  L_i Q_j {\overline D}_k
 +  \lambda^{''}_{ijk} {\overline U}_i {\overline D}_j {\overline D}_k
 +  \epsilon_i L_i H_2, 
\label{lagrangian}
\end{eqnarray}
where $i,j,k$ are the generation indices of the superfields 
$L, Q,E,D$ and $U$. $L$ and $Q$ are lepton and quark left-handed doublets,  
respectively. 
$\overline E$, $\overline D$ and $\overline U$ are right-handed 
singlet charge-conjugate superfields for the charged 
leptons, down- and up-type quarks, respectively. $H_2$ is the Higgs 
doublet superfield with a weak hypercharge $Y$ = 1. The last term 
in (\ref{lagrangian}) which mixes the lepton and Higgs superfields is 
not considered in this paper. 
The interactions corresponding to these superpotential terms are
assumed to respect the gauge symmetry SU(3)$_{\rm C}$ $\times$ 
SU(2)$_{\rm L}$ $\times$ U(1)$_{\rm Y}$ of the \sm.
The \lb$_{ijk}$ are 
non-vanishing only if $i < j$, so that at least two different 
generations are coupled in the purely leptonic vertices. 
The \lbpp$_{ijk}$ 
are non-vanishing only for $ j < k $.
The \lb\ and \lbp\ couplings both violate lepton number conservation
and the \lbpp\ couplings violate baryon number conservation. 
There are nine \lb\ couplings for the triple lepton vertices, 27 \lbp\ 
couplings for the lepton-quark-quark vertices and nine \lbpp\ couplings 
for the 
triple quark vertices. There are therefore a total of 45
\Rparity\ violating couplings.  

In the Constrained Minimal Supersymmetric Standard Model (CMSSM), 
which assumes a 
common gaugino mass, $m_{1/2}$, and a common sfermion mass, $m_0$, at the 
Grand Unification (GUT) scale,  
all sparticle masses 
and \Rparity\ conserving couplings are completely determined 
by $m_0$ and a set of three parameters: 
$M_2$, the SU(2) gaugino mass parameter at 
electroweak scales\footnote{
$M_1$, the U(1) gaugino mass at electroweak scales,
is related to $M_2$ by the usual gauge unification
condition:  $M_1 =  \frac{5}{3} \tan^2 \theta_W M_2$.}, 
$\mu$, the mixing parameter of the 
two Higgs doublets
and $\tan \beta= v_2/v_1$, the ratio of the 
vacuum expectation values for the two Higgs doublets.

From low energy experiments, there exist several 
upper bounds
on the \Rparity\ violating Yukawa couplings, \lb, \lbp\ and \lbpp. 
A list of upper limits on individual couplings
can be found 
in~\cite{
ref:mohapatra,
ref:barger1,
ref:godbole,
ref:ellis,
ref:goity, 
ref:hirsh,
ref:agashe,
ref:smir,
ref:bhatta,
ref:allanach}.  
The experimental limits on processes such as proton decay imply that it is
reasonable to assume that no more than one of the 45 couplings is
significantly different from zero~\cite{ref:dreiner1}.

This paper describes searches for \Rparity\ violating decays of 
pair-produced scalar 
fermions  (``sfermions"), such as the charged and neutral scalar leptons 
($\slepton$ and $\sneutrino$ respectively), scalar top quark ($\stopn$) 
and scalar light quarks ($\sq$).
The results of similar searches performed with the OPAL data at centre-of-mass
energies up to 183~GeV have already been published~\cite{ref:opal_sfer_183}.

Two different scenarios are probed. In the first scenario, the decays 
of sfermions via the lightest neutralino, $\nt_1$, are considered, 
where $\nt_1$ is
treated as the LSP and assumed to decay via an \Rparity\ violating
interaction.
These are denoted ``indirect decays''. 
SUSY cascade decays via particles other than the LSP are not 
considered; however, when calculating limits in the 
\Rparity\ violating framework analogous to 
the CMSSM, these cascade decays are taken into account by assuming 
a zero efficiency for their detection. 
In the second scenario,
``direct'' decays
of sparticles to \sm\ particles are investigated.  
In this case, the sparticle is assumed
to be the LSP, such that \Rparity\ conserving decay modes do not
contribute. 

In both the direct and indirect decay scenarios, 
it is assumed that only one of the 
45 Yukawa couplings is significantly non-zero at a time. 
Moreover, only values of the \lb-couplings (\lb, \lbp, \lbpp) larger than 
$ {\cal{O}} (10^{-5})$ are relevant to this analysis. 
For smaller couplings,  the lifetime of sparticles would be
sufficiently long 
to produce a secondary decay vertex clearly detached 
from the primary vertex, or even outside the detector. These topologies
have not been considered in this paper. Decays outside the 
detector have been treated in~\cite{ref:stable-part}.

Results published by the other LEP Collaborations can be found 
in~\cite{ref:otherLEP}. Results have also been obtained by the CDF and D0
Collaborations~\cite{ref:CDF, ref:D0} at the Tevatron
and by the H1 and ZEUS Collaborations~\cite{ref:H1, ref:ZEUS} at HERA.

The production and \Rparity\ violating decays
of $\sell$, $\snu$ via \lb\ and \lbp, $\stopx$ via \lbp\ and 
$\sq$ via \lbpp\ are described 
in Section~\ref{sec:production}, 
together with the possible signal topologies 
resulting from these processes. 
Short descriptions of the OPAL detector and of the data samples used 
are presented in Section~\ref{sec:opaldet}. 
The signal and background Monte Carlo simulations used in the 
different analyses are described in Section~\ref{sec:MC}. 
Sections~\ref{sec:multileptons}
to \ref{sec:multijetsnoemiss}
describe the specific analyses optimised to search for \Rparity\ violating
processes. The physics interpretation is given in 
Section~\ref{sec:results} which presents 
cross-section limits and interpretations in the CMSSM. A summary is 
given in Section~\ref{sec:conclusions}.

\section{Sparticle production and decays}
\label{sec:production}

This section briefly describes the production and 
decay modes of different sfermion species.
A detailed description of the various 
decay channels, decay widths and assumptions can be found 
in~\cite{ref:opal_sfer_183}.  
The decay modes studied, resulting from non-zero \lb, \lbp\ and \lbpp\
couplings, are presented 
in Table~\ref{tab:relation}. 
Table~\ref{tab:relation} also summarises 
the production and decay mechanisms as well as the coupling 
involved in each decay and
the final state topologies.
In the indirect decays, the particles
resulting from the $\nt_1$ decay are shown in parentheses. 
The sneutrino indirect \lbp\ decays followed by
$\nt_1 \ra \ell \qq$, 
leading to topologies with jets and one or more leptons, are not investigated 
because the branching ratio into these final states is not significant. 
In the available phase space and for the CMSSM parameter space relevant 
to these analyses, the decay $\nt_1 \ra \nu \qq$ is 
preferred with respect
to the decay  $\nt_1 \ra \ell \qq$ because of the light sneutrino.  

\begin{table}[htbp]
\begin{center}
\renewcommand{\arraystretch}{1.3}
\begin{tabular}{|ll|l|l|c|}
\hline
Production and Decay  & & Coupling & Topology &  Section \\
\hline

$ \sell^- \sell^+ \rightarrow $ &  
$ \nu  \ell^-   $  $ \bar{\nu}  \ell^+   $ & 
\lb\ direct & 2 $\ell$ + $E_{miss}$ 
&  \ref{sec:2leptons}\\

\hline
$ \snu \bar{\snu} \rightarrow  $ 
$\nu \chin $ $ \bar{\nu} \chin \ra $ & 
$\nu (\stackrel{(-)}{\nu} \ellp \ellm)$  $ \bar{\nu} (\stackrel{(-)}{\nu} \ellp \ellm)$  &  
\lb\ indirect & 
4 $\ell$ + $E_{miss}$ & 
 \ref{sec:4leptons} \\

$ \snu \bar{\snu} \rightarrow $ & 
$ \ellm \ellp  $  $  \ellp \ellm  $  &  \lb\ direct  & 
4 $\ell$  & 
 \ref{sec:4leptons} \\ 

\hline
$ \sell^- \sell^+ \rightarrow $   
$ \ell^- \chin  \ell^+ \chin$  $\ra$  &
$\ellm (\stackrel{(-)}{\nu} \ellp \ellm)$  $\ellp (\stackrel{(-)}{\nu} \ellp \ellm)$ &  
\lb\ indirect & 6 $\ell$ + $E_{miss}$ & 
 \ref{sec:6leptons} \\

\hline
$ \stoppair \rightarrow $   &
$ e^+ q  $  $ e^- \bar{q}  $ &  \lbp\ direct & 
2 e + 2 jets  &
 \ref{sec:2jets2leptons} \\

$ \stoppair \rightarrow $   &
$ \mu^+ q  $  $ \mu^- \bar{q}  $  & \lbp\ direct & 
2 $\mu$ + 2 jets  &
 \ref{sec:2jets2leptons} \\

$ \stoppair \rightarrow $   &
$ \tau^+ q  $  $ \tau^- \bar{q}  $ & \lbp\ direct  & 
2 $\tau$ + 2 jets  &
 \ref{sec:2jets2leptons} \\

\hline

$ \sell^- \sell^+ \rightarrow $   
$ \ell^-  \chin    \ell^+ \chin$   $\ra$  &
$ \ell^-  (\ell^\pm q \bar{q'})$ $ \ell^+ (\ell^\pm q \bar{q'})$ &  \lbp\ indirect & 
$\geq 2 \ell$ + jets &
 \ref{sec:jetsleptons} \\
&
$ \ell^- (\stackrel{(-)}{\nu} q \bar{q})$ $\ell^+ (\ell^\pm q \bar{q'})$  & \lbp\ indirect & 
$\geq 2 \ell$ + jets &
 \ref{sec:jetsleptons} \\
&
$ \ell^- (\stackrel{(-)}{\nu} q \bar{q}) $ $\ell^+ (\stackrel{(-)}{\nu} q \bar{q})$ &  \lbp\ indirect & 
$\geq 2 \ell$ + jets &
 \ref{sec:jetsleptons} \\

\hline 
$ \snu \bar{\snu} \rightarrow  $ 
$\nu \chin $ $ \bar{\nu} \chin \ra $ & 
$ \nu (\stackrel{(-)}{\nu} q \bar{q})$  $\bar{\nu} (\stackrel{(-)}{\nu} q \bar{q} )$ & 
\lbp\  indirect & 
4 jets + $E_{miss}$  &
 \ref{sec:multijetsemiss} \\

\hline 
$ \snu \bar{\snu} \rightarrow$   & 
 $ q \bar{q} \; \bar{q} q$  & \lbp\  direct & 
4 jets   &
 \ref{sec:multijetsnoemiss} \\
$ \sell^- \sell^+ \rightarrow $ &  
  $\bar{q} q' \; q \bar{q'}$ &  \lbp\ direct & 
4 jets   &
 \ref{sec:multijetsnoemiss} \\
$ \tilde{q}\bar{\tilde{q}} \rightarrow $ &  
  $\bar{q} \bar{q'} \; q q'$ &  \lbpp\ direct & 
4 jets   &
 \ref{sec:multijetsnoemiss} \\

\hline
\end{tabular}
\end{center}
\caption{\it
List of production and decay mechanisms 
of the channels that are covered by the various analyses
described in this paper. The couplings involved and decay type 
together with the 
resulting topologies are given in the
second and third columns, respectively.
The corresponding section number, where details of the 
relevant analysis are presented,  is indicated
in the last column. 
} 
\label{tab:relation}
\end{table}

\section{Data samples and OPAL detector}
\label{sec:opaldet}

The analyses presented in this paper use the data collected at LEP
from centre-of-mass energies ($\sqrt{s}$) between 189~GeV and
209~GeV, including the data obtained at the highest energies
attained by LEP in the year 2000. 
The total data sample corresponds 
to an integrated luminosity of
about 610~pb$^{-1}$. 
The data set collected at energies 
$\sqrt{s} \geq $~206 GeV has a luminosity weighted  
average centre-of-mass energy of 206.6~GeV.
The results presented here are then combined with 
results obtained using data at lower 
energies~\cite{ref:opal_sfer_183}.
The various integrated luminosities of the data taken 
between 1998 and 2000 are summarised in Table~\ref{tab:lumi}. 
The integrated luminosity is measured 
with a precision of 0.3$-$0.4\%~\cite{ref:lumi2}.
These uncertainties are taken into 
account in the limit calculation in all channels considered.

\begin{table}[htbp]
\begin{center}
\begin{tabular}{|l||c|c|c|c|c|c|c|c|c|}
\hline
Year                  & 1998         
                      & \multicolumn{4}{c|}{1999}    
                      & \multicolumn{2}{c|}{2000}            
                                       \\
\cline{2-2}
\cline{3-6}
\cline{7-8}
$\roots$ (GeV)        &  189              
                      &  192     & 196       & 200 & 202 
                      &  $<$ 206 & $\ge 206$ 
                                        \\
${\cal L}$ (pb$^{-1}$)& 179    
                      &  29      &  73       &  74 &  37
                      &  87      & 133        
                                        \\
\hline
\end{tabular}
\end{center}
\caption{\it
Integrated luminosities recorded with the OPAL detector between 1998 and
2000. Due to different requirements on the operation of the OPAL subdetectors
the precise integrated luminosity differs from one analysis to another. 
}
\label{tab:lumi}
\end{table}

A complete description of the  OPAL detector can be found 
in Ref.~\cite{ref:OPAL-detector}, and only a brief overview is given here.

The central detector consisted of
a system of tracking chambers
providing charged particle tracking
over the polar 
angle\footnote
   {The OPAL coordinate system is defined so that the $z$ axis is in the
    direction of the electron beam, the $x$ axis points towards the centre 
    of the LEP ring, and  
    $\theta$ and $\phi$
    are the polar and azimuthal angles, defined relative to the
    $+z$ and $+x$ axes, respectively. The radial coordinate is denoted
    as $r$.}
range of $|\cos \theta |<0.96$    
inside a 0.435~T uniform magnetic field parallel to the beam axis. 
The central tracking system was composed of 
a large volume jet chamber containing a high precision drift chamber and 
a two-layer
silicon microstrip vertex detector~\cite{ref:silicon}, and was 
surrounded by a set of $z$ chambers measuring 
track coordinates along the beam direction. 
A lead-glass electromagnetic
calorimeter located outside the magnet coil
covered the full azimuthal range with excellent hermeticity
in the polar angle range of $|\cos \theta |<0.82$ in the barrel
region  and $0.81<|\cos \theta |<0.984$ in the endcap regions.
The magnet return yoke was instrumented 
to comprise the barrel and the endcap sections of the
hadron calorimeter.
With the
hadron pole tip detectors included, the hadron calorimeter
covered the region with $|\cos \theta |<0.99$.
Four layers of muon chambers 
covered the outside of the hadron calorimeter. 
A system of forward detectors close to the beam axis 
completed the geometrical acceptance down to 24 mrad, except
for the regions where a tungsten shield was present to protect
the detectors from synchrotron radiation.
The forward detectors included 
a lead-scintillator sandwich calorimeter, a
ring of lead-scintillator modules, scintillating tile
counters~\cite{ref:mip-plug} and the silicon-tungsten 
luminometer~\cite{ref:SW} 
located on both sides, a few meters away from the interaction point.

The tracks and calorimeter clusters used in the analyses were required to 
satisfy minimum quality criteria. These were slightly different for the 
analysis of purely leptonic final states and for the analyses with jets, 
either with or without leptons. For leptonic final states, a track was 
required to have at least
20 hits in the wire tracking chambers, 
the first of which was required to be at a radius of less than 75~cm, 
and to have a momentum transverse to the beam
direction larger than 0.1~GeV. Its distance of closest approach to the
interaction point in the transverse plane ($d_{0}$)
had to be  less than 1~cm and the distance along the
beam direction  less than 40~cm.
Clusters of energy deposited in the barrel region of the 
electromagnetic calorimeter were taken into account in the analyses if 
their energy was larger than 0.1~GeV. Similarly, 
endcap electromagnetic energy clusters were taken into account 
in the analyses if their energy was larger
than 0.2~GeV and if 
the cluster was composed of at least two adjacent lead blocks.
Hadron calorimeter energy clusters 
had to contain an energy in excess of 0.6~MeV in the barrel and
endcap regions, and 3 GeV in the pole tips.
For leptons with jets and for purely hadronic channels, 
tracks in the central detector and energy clusters
in the calorimeters were required to satisfy the quality
criteria employed in~\cite{ref:newsmhiggs}.

In calculating the total visible energies and
momenta of events and individual jets, corrections were applied to prevent
double-counting of energy in the case of tracks and associated
energy clusters. For the leptonic channels, 
an algorithm based on a global approach described in~\cite{ref:OPAL-Higgs}
was used. 
For jets with leptons and for purely hadronic 
channels, an algorithm matching tracks and calorimeter energy clusters, 
described in~\cite{ref:newsmhiggs}, was
employed.

\section{Monte Carlo simulation}
\label{sec:MC}

Monte Carlo samples were generated corresponding to 
charged slepton, sneutrino, stop and squark 
pair-production processes as well as  
to \smp.
All generated events were processed
through the full simulation of the OPAL detector~\cite{ref:GOPAL},
and the same analysis chain was applied to simulated events
as to the data.

All sparticle production and decays considered in this paper,
except for stop production described below, 
were simulated at various 
centre-of-mass energies between 189~GeV and 209~GeV and for various
sfermion masses between 45 and 103~GeV
using the Monte Carlo program SUSYGEN~\cite{ref:SUSYGEN}. 
All MC samples of signal events contain 1000 events per mass point.

Charged and neutral sleptons decaying via a \lb\ 
coupling 
were simulated at $\sqrt{s} = $ 189 and 205~GeV. 
Charged sleptons decaying indirectly via a \lbp\ coupling 
were simulated at $\sqrt{s} = $ 183 and 205~GeV. 
The selection efficiencies were interpolated (or extrapolated)
for the centre-of-mass energies where no samples were generated.
A few test samples were produced 
to test the validity of the interpolation.
  
For charged and neutral sleptons decaying directly via a \lbp\ coupling, 
production was done at 
$\sqrt{s} = $ 189, 196, 200 and 206 GeV. 
Squark pair-production samples were
generated at $\sqrt{s} = $ 189, 192, 196, 200, 202 and 206~GeV. 
In both cases, the signal samples with the 
closest $\sqrt{s}$ were used for the centre-of-mass energies 
where no samples were simulated.


For the indirect decays, events were produced with 
three sparticle masses ($m_{\sfer}$) and the $\nt_1$ mass was fixed such that
$\Delta m = m_{\sfer} - m_{\nt_1} = m_{\sfer}/2$.
Samples were also simulated 
at the kinematic limit 
with $\Delta m = m_{\sfer} - m_{\nt_1} = 5$~GeV to account for changes in
the event topologies. The values of  
$\Delta m$ were chosen to cover a wide range 
of event topologies 
for a limited number of 
Monte Carlo events. For charged sleptons decaying indirectly  
with a \lb\ and a \lbp\ coupling, selection efficiencies
were also obtained using 
samples generated with $\Delta m$ equal to the mass of the 
lepton plus 0.1~GeV
in order to cover the small $\Delta m$ region.
To study the selection efficiency dependency on large $\Delta m$, 
samples of sneutrinos decaying indirectly with a \lb\ and a \lbp\ coupling
were also generated with $m_{\nt_1}$ = 10, 25 and 40 GeV,
respectively for $m_{\snu} <$ 50 GeV, 70 $< m_{\snu} <$ 80 GeV 
and 90 GeV $< m_{\snu}$.    
For charged slepton indirect decays via \lbp, 
events were simulated for each lepton flavour 
corresponding to the first index of
\lbp. The quark flavours corresponding to the second and third indices
of \lbp\ were fixed to the first and second generation, with 
the exception of a few additional 
samples containing bottom quarks for systematic checks. 
All possible sparticle pair-production processes were simulated
only at 183 GeV~\cite{ref:opal_sfer_183} in order to determine  
the \lb-like couplings 
which give the lowest and the highest selection efficiencies. 
At higher centre-of-mass energies, only the 
decays governed by 
these two couplings were simulated and considered
when computing results.

Charged and neutral sleptons decaying directly via a \lbp\ coupling 
were generated using 
the Lund string fragmentation scheme of PYTHIA~\cite{ref:JETSET1, ref:JETSET2}.
In this scheme, 
the final state quarks are evolved according to the ``Lund Parton Shower''
model. A colour string is formed between colour singlet $q\bar{q}$ 
states. The two quarks are subsequently evolved in time and space, 
emitting gluon radiation between the two quarks. The jets are 
relatively soft owing to the gluon radiation between the quark states. 
In the case of squark production,
only the independent fragmentation model was implemented at the 
generator level. In this scheme, 
the final state quarks
are treated as independent particles with respect to each other. 
There is
no QCD radiation between the final state quarks, and the hadronization
process produces relatively hard jets.
For 
systematic studies additional smuon samples were generated by the
independent fragmentation model.

For a stop decaying via the \lbp\ coupling into a quark and a lepton, 
all nine combinations of quark and lepton flavours
in the final state were generated at $\sqrt{s} = $ 189, 192, 196, 
200, 202 and 207~GeV.
Since stop masses below 73~GeV were excluded 
in~\cite{ref:opal_sfer_183}, only stop masses 
between 65 and 103~GeV were considered here.
The production and decay of the stop were simulated as described 
in~\cite{ref:stoppaper}.
The stops were hadronised to form colourless hadrons
using the Peterson fragmentation function~\cite{ref:peterson}, 
and associated fragmentation particles
according to the Lund string fragmentation scheme. 
For the decay, a colour string was stretched between the spectator 
quark and the quark from the stop decay. Further hadronisation 
was also done using the Lund scheme.
The left-right stop mixing angle, $\theta_{\stopx\ }$, was
set to zero.
The dependences of the detection efficiencies on the fragmentation function, 
on the mixing angle and on the Fermi motion were evaluated by generating 
additional Monte Carlo samples with the appropriate parameters varied. 

The main sources of background arise from 
\sm\ two-photon processes ($\ee \ra \ee \gamma \gamma  \ra \ee X$), 
and from four-fermion ($\ee \ra \ f \bar{f} f \bar{f}$) 
and two-fermion 
(lepton-pair $\ee \ra (Z/\gamma)^* \ra l \bar{l}$
and multi-hadronic $\ee \ra (Z/\gamma)^* \ra q \bar{q}$) 
final states. 
For two-photon processes, the PHOJET~\cite{ref:PHOJET} and 
HERWIG~\cite{ref:herwig} generators were used to simulate
hadronic final states.
The Vermaseren~\cite{ref:VERMASEREN} generator was
used to estimate the background contribution from all 
two-photon $\ee \ell^+ \ell^-$
final states.
Four-fermion final states, other than two-photon events,  
were simulated 
with grc4f~\cite{ref:grace4f}, which takes into 
account all interfering four-fermion diagrams 
with the exception of the multiperipheral two-photon processes. 
An $\ee \ell^+ \ell^-$ four-fermion sample 
generated with grc4f v2.2 with 
all interfering four-fermion diagrams  
including the multiperipheral two-photon processes
was compared with the standard combination of grc4f v2.1 and 
Vermaseren events for a systematic cross-check. 
For final states with hadronic jets, 
systematic studies were performed using 
KORALW~\cite{ref:KORALWW}, which internally uses 
grc4f matrix elements. For purely hadronic final states, 
additional special samples generated with KandY~\cite{ref:KANDY}, a
special version of KORALW running concurrently 
with YFSWW3~\cite{ref:YFSWW3}, were also used. 
For the two-fermion final states, BHWIDE~\cite{ref:BHWIDE} 
was used for the ee$(\gamma)$ 
final state and KORALZ~\cite{ref:KORALZ} and KK2F~\cite{ref:KK2F} for the 
$\mu \mu$ and the $\tau \tau$ states. The multi-hadronic events, 
${\rm qq}(\gamma)$, 
were simulated using PYTHIA~\cite{ref:JETSET1} and KK2F.

Final states with six or more
primary fermions 
were not included in the background Monte Carlo
samples but they are expected to make only a negligible contribution to 
the background. 

\section{Multi-lepton final states}
\label{sec:multileptons}

This section describes the searches for purely leptonic final states
resulting from pair-production of neutral or charged sleptons,
with subsequent direct or indirect \lb\ decays 
(see Table~\ref{tab:relation}). The inefficiencies 
and systematic uncertainties 
associated with each analysis are estimated in a similar way 
for all multi-lepton
final state searches, and are
presented in Section~\ref{sec:syserr}.
The selections of purely leptonic final states use the data 
collected at centre-of-mass energies between 189~GeV and 209~GeV
corresponding to an integrated luminosity of 613.6~pb$^{-1}$.

\subsection{Event and lepton selection}
\label{sec:preselections}

Multi-hadronic, cosmic and Bhabha 
scattering events were vetoed~\cite{ref:slept161}.
A loose high multiplicity veto was also applied:
events were rejected if the total multiplicity of tracks in the 
central detector and  energy clusters in the electromagnetic calorimeters 
satisfying the quality criteria described in Section~\ref{sec:opaldet}
was greater than 26. 
At the preselection level, it was also required that
the ratio of the number of such ``good'' tracks as defined in 
Section~\ref{sec:opaldet}
to the total number of reconstructed tracks be greater than 0.2
in order to reduce backgrounds from beam-gas and beam-wall events.
At least two good tracks
were required.

Only tracks with $|\cos \theta| <$ 0.95 were considered  
for lepton identification. 
Tracks resulting from photon conversion
were rejected using the algorithm described in
\cite{ref:conversion}.
A track was considered ``isolated'' 
if the total energy of other charged particle tracks within 
$10\degree$ of
the lepton candidate was less than 2~GeV.
In these searches, a very loose lepton identification 
is sufficient. 
A track was selected as an electron candidate if one of the following three
criteria was satisfied: {\it (i)} the output probability of the 
neural net algorithm described in
\cite{ref:NN} was larger than 0.8;
{\it (ii)} the electron selection 
algorithm described in \cite{ref:elecbarrel}
for the barrel region or in \cite{ref:elecendcap} for
the endcap region was satisfied;
{\it (iii)} $0.5 < E/p < 2.0$,  where $p$ is 
the momentum of the electron candidate
and $E$ is the energy of the electromagnetic calorimeter energy cluster
associated with the track. 
The simple criterion {\it (iii)} identifies most electrons. 
The algorithms {\it (i)} and {\it (ii)} are based predominantly 
on $E/p$, dE/dx and shower shape information, and are used to increase 
the electron selection efficiency. These algorithms are also
used in the lepton plus jets analyses.  
A track was selected as a muon candidate according to the 
criteria employed in OPAL's
analysis of \sm\ muon pairs~\cite{ref:leptpairs}, 
that is, the track had associated activity in the muon chambers or hadron
calorimeters or it had a high momentum but was associated with only a
small amount of energy deposited in the electromagnetic calorimeter.
Identified electrons or muons  
were also considered as tau candidates. 
Taus
were selected by requiring that there be at most three 
tracks within a 35$^\circ$ cone.
The invariant mass computed using all good tracks and electromagnetic 
energy clusters
within the above cone had to be less than 3~GeV.
For muon and electron candidates, the momentum was estimated 
from the charged particle track momentum measured in the central 
detector, while
for tau candidates, 
the momentum was estimated from the vector sum of
the measured momenta of the charged particle tracks within the tau cone.

For the two- and six-lepton final states, 
the large background from two-photon
processes was reduced by requiring that  
the measured amount of energy deposited 
in each silicon tungsten calorimeter,
in each forward calorimeter, and in each side of the gamma-catcher
be less than 5~GeV (``forward detector vetoes'').

\subsection{Final states with two leptons and missing energy }
\label{sec:2leptons}

Final states with two charged leptons and missing energy may result from 
direct charged slepton decays via a $\lambda$ coupling. 
The analysis was optimised to retain good signal efficiency while reducing 
the background, mainly due to
$\mathrm{W^+W^- \longrightarrow \ell^+\nu \ell^-\nu}$ events and 
two-photon processes. 

The following criteria were applied in addition to those described in 
Section~\ref{sec:preselections}.

\begin{description}

\item[(1)]
Events had to contain 
exactly two identified 
and oppositely-charged leptons, each with a transverse 
momentum with respect to the beam axis greater than 2~GeV.

\item[(2)]
The background from two-photon processes and ``radiative return" events
($\ee \ra {\mathrm Z} \gamma$, where the $\gamma$ escapes
down the beam pipe) was reduced  by requiring
that the polar angle of the missing momentum, $\thmiss$,
satisfy \mbox{$\cosmiss < 0.9$}. 

\item[(3)]
To reduce further the residual background from 
Standard Model lepton pair events,
it was required that $\Mvis /\sqrt{s} < 0.80$, 
where $\Mvis$ is the invariant mass derived from the 
measured energy and momentum
of all the tracks and electromagnetic energy 
clusters observed in the event, hereafter
referred to as event visible mass.

\item[(4)]
The acoplanarity angle\footnote
   {The acoplanarity angle, $\acop$,
    is defined as 180$\degree$ minus the angle
    between the two lepton momentum vectors  
    projected 
    into the $x-y$ plane.}
($\acop$) between the two leptons was required to be
greater than 10$\degree$ in order to reject Standard Model leptonic events,
and smaller than 175$\degree$ in order to reduce the
background due to photon conversions.
The acoplanarity angle distribution is shown
in Figure~\ref{fig:multilepton}
after cuts (1) to (3).
Figure~\ref{fig:multilepton} demonstrates the good discriminating 
power of this quantity.
There is a disagreement 
between data and \sm\ expectation 
which is characteristic of 
the presence in 
the data of un-modelled two-photon events with low acoplanarity.
The acollinearity angle\footnote
   {The acollinearity angle, $\acol$,
    is defined as 180$\degree$ minus the 3-dimensional space-angle
    between the two lepton momentum vectors.}
($\acol$) was also required to be greater than 10$\degree$
and smaller than 175$\degree$.

\begin{figure}[htbp]
\centering
\includegraphics*[scale=0.5]{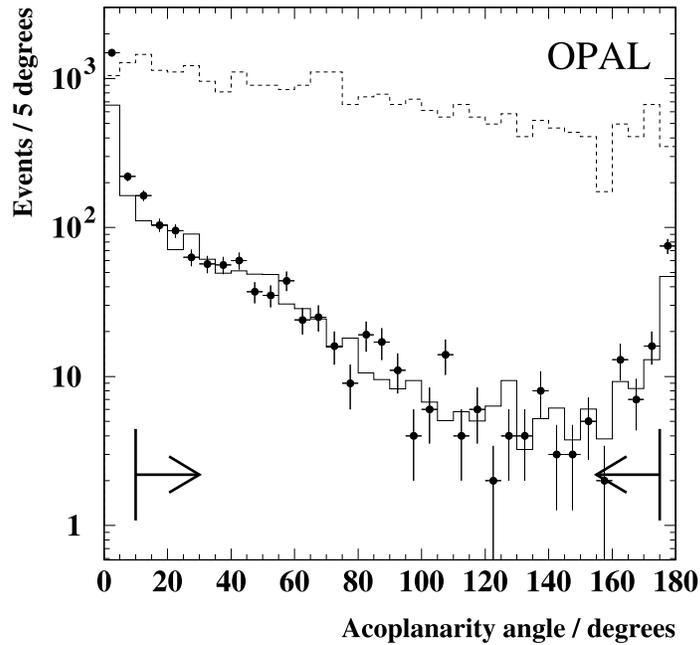}
\caption[]{\it
  Final states with two leptons and missing energy: 
  distribution of the
  acoplanarity angle after applying cuts (1) to (3).
  The dashed histogram 
  shows signal Monte Carlo events for direct decays of 
  $\sele$ via \lb$_{122}$
  with $m_{\sele} = $ 78~GeV.
  Data are shown as points and the sum of all Monte Carlo background 
  processes is shown as a solid line. 
  Data used for this plot were collected in the year 2000 at 
  a luminosity weighted average centre-of-mass energy of 206.6~GeV.
  The simulated signal events have arbitrary normalisation. 
  The arrows point into the region accepted by the cuts.
  The disagreement 
  between data and \sm\ expectation, visible in a few bins, 
  is due to the presence of un-modelled two-photon events
  in the data.

 } 
\label{fig:multilepton}
\end{figure}

\item[(5)]
It was required that there be no 
electromagnetic energy cluster with an energy larger 
than 25~GeV and no associated track and no hadronic 
energy cluster with an energy larger than 10~GeV and no
associated track.

\item[(6)]
In order to reduce the background from two-photon events 
with high transverse momentum, 
complementary cuts were applied for different regions of acoplanarity.
For events with small acoplanarity ($\acop <$ 1.2~rad), a cut 
was made on 
$a_t^{\mathrm {miss}}$, 
the component of the missing momentum vector perpendicular to the 
event thrust axis in the
plane transverse to the beam axis.
For events with large acoplanarity ($\acop >$ 1.2~rad), a cut was made 
on $p_t^{\mathrm {miss}}$, the missing transverse momentum.
In the low acoplanarity region, $a_t^{\mathrm {miss}}$ provides a good
discriminating power since it is less sensitive than
$p_t^{\mathrm {miss}}$ to the presence of neutrinos from tau decays 
or poorly measured tracks. 
However, in the large acoplanarity region, $a_t^{\mathrm {miss}}$ 
does not discriminate sufficiently between signal and background and 
$p_t^{\mathrm {miss}}$ is used.  
A further subdivision was applied according to the event 
visible energy, $E_{{\mathrm{vis}}}$,
which is defined as 
the scalar sum of the energy of all tracks and electromagnetic 
energy clusters
of the event.
This selection provides a good discrimination between events with genuine
missing energy and events which have missing energy due to secondary decays
such as $\tau$ decays. It is optimized to retain good signal efficiency also
in the case of very small visible energy:

\begin{enumerate}

\item
Cuts in the small acoplanarity region:
\begin{enumerate}
\item
If $\Evis / \sqrt{s} < 0.06$, then $a_t^{\mathrm {miss}}/ \sqrt{s} >$  0.015,
\item
If $\Evis / \sqrt{s} > 0.06$, then $a_t^{\mathrm {miss}}/ \sqrt{s} >$  0.0175.
\end{enumerate}

\item
Cuts in the large acoplanarity region:
\begin{enumerate}
\item
If $\Evis / \sqrt{s} < 0.06$, then $p_t^{\mathrm {miss}}/ \sqrt{s} >$  0.025,
\item
If $\Evis / \sqrt{s} > 0.06$, then $p_t^{\mathrm {miss}}/ \sqrt{s} >$  0.035.
\item
The event transverse momentum 
computed without the hadron calorimeter
was required to be larger than 0.02 $\times \rs$. This cut rejects events with
a large measured transverse momentum due to fluctuations in the 
hadron calorimeter.
\end{enumerate}

\end{enumerate}

\item[(7)]
The background from two-photon processes and 
$\WW$ production was further reduced by categorising the events
in different classes 
according to the flavour of the leptons in the final state.
The flavours of the two identified leptons were required to match
the flavours of the leptons expected for each signal class. 
Events were selected by applying cuts on the momentum
of the two leptons and on the visible energy:

\begin{list}{$\bullet$}{\itemsep=0pt \parsep=0pt \topsep=-5pt \leftmargin=30pt}

\item $e \nu e \nu$ selection: 
$E_{\mathrm{vis}} / \sqrt{s} < 0.8$ and 
$p_{\mathrm{max}} / E_{\mathrm{beam}} > 0.1$,  
where $p_{\mathrm{max}}$ is the momentum of the most 
energetic electron and $E_{\mathrm{beam}}$ is the beam energy.

\item $\mu \nu \mu \nu$ selection: 
$E_{\mathrm{vis}} / \sqrt{s} < 0.7$, 
$0.05 < p_{\mathrm{min}} / E_{\mathrm{beam}} < 0.60$ and 
$0.05 < p_{\mathrm{max}} / E_{\mathrm{beam}} < 0.95$,
where $p_{\mathrm{min}}$ is the momentum of the less 
energetic muon and $p_{\mathrm{max}}$ is the momentum of the most 
energetic muon.

\item $\tau \nu \tau \nu$ selection: 
$E_{\mathrm{vis}} / \sqrt{s} < 0.4$, 
$p_{\mathrm{min}} / E_{\mathrm{beam}} < 0.30$ and 
$0.04 < p_{\mathrm{max}} / E_{\mathrm{beam}} < 0.5$.

\item $e \nu \mu \nu$ selection: 
$0.12 < E_{\mathrm{vis}} / \sqrt{s} < 0.8$ and 
$p_{\mathrm{max}} / E_{\mathrm{beam}} > 0.15$.

\item $e \nu \tau \nu$ selection: 
$p_{\mathrm{min}} / E_{\mathrm{beam}} < 0.5$ and 
$p_{\mathrm{max}} / E_{\mathrm{beam}} > 0.1$.

\item $\mu \nu \tau \nu$ selection: 
$p_{\mathrm{min}} / E_{\mathrm{beam}} < 0.5$ and 
$p_{\mathrm{max}} / E_{\mathrm{beam}} > 0.1$.

\end{list}

\end{description}

Table \ref{table:2l} shows the results for the different 
selections 
at centre-of-mass energies from 189~GeV to 209~GeV. Since 
the detection efficiencies are independent of the charged 
slepton species within
the statistical uncertainties, 
results are quoted irrespective of the charged slepton 
species but only for the different possible final states. 
The lowest efficiencies arise from final states with two taus 
and missing energy
and the highest ones from final states with two muons and missing energy.
They range from 18\% ($\tau \nu \tau \nu $) to 
66\% ($\mu \nu \mu \nu $)
for charged slepton masses between 45 and 103~GeV. 
The main contribution to the background comes from four-fermion 
processes which,
depending on the final state lepton flavours,
account for 90$-$100\% of the \sm\ expectation.

\begin{table}[htbp]
  \centering
\renewcommand{\arraystretch}{1.3}
  \begin{tabular}[c]{|c|c|c|c|c|c|}
\hline  
  Physics process   & Eff. (\%) &  Data & Tot. bkg MC & Data $>$ 206~GeV & 
MC $>$ 206~GeV   \\\hline
$\tilde{\ell}^+ \tilde{\ell}^- \rightarrow $ & & & & & \\
$ e \nu e \nu $ 
                    & 54 $-$ 64& 135 & 145.3 $\pm$ 0.9 $\pm$ 10.7 & 31 & 28.6 $\pm$ 0.1 $\pm$ 3.0 \\
$ \mu \nu \mu \nu $ 
                  & 48 $-$ 66& 99 & 101.8 $\pm$ 0.6 $\pm$ 11.1 & 22 & 21.9 $\pm$ 0.1 $\pm$ 3.3 \\
$ \tau \nu \tau \nu $ 
                    & 18 $-$ 27& 166 & 161.5 $\pm$ 2.2 $\pm$ 22.0 & 25 & 35.6 $\pm$ 0.6 $\pm$ 6.7 \\

$ e \nu \mu \nu $ 
                    & 55 $-$ 64& 463 & 502.4 $\pm$ 2.0 $\pm$ 37.1 & 94 & 106.2 $\pm$ 0.3 $\pm$ 10.5 \\

$ e \nu \tau \nu $ 
                    & 37 $-$ 44& 429 & 459.9 $\pm$ 2.7 $\pm$ 39.7 & 85 & 98.0 $\pm$ 0.7 $\pm$ 11.4 \\
$ \mu \nu \tau \nu $ 
                    & 38 $-$ 47& 434 & 451.0 $\pm$ 2.7 $\pm$ 41.0 & 84 & 97.4 $\pm$ 0.7 $\pm$ 11.9 \\

\hline 
\end{tabular}
\caption{\it
Final states with two leptons and missing energy
for $\sqrt{s} = 189 - 209$~GeV and for 
$\sqrt{s} > 206$~GeV:
signal selection efficiencies, observed numbers of events in the data and 
background estimates from \sm\ processes. 
The first uncertainty on the background estimate is statistical 
and the second is systematic.
}
\label{table:2l}
\end{table}

\subsection{Final states with four leptons, with or without missing energy}
\label{sec:4leptons}

Final states with four charged leptons and no missing energy 
may result from direct sneutrino decays via a \lb\ coupling 
while final states with
missing energy may result from indirect sneutrino  decays via a \lb\ coupling. 
Two analyses have been developed and optimised separately for these two final
states. 

The following criteria were applied to select a possible signal with
four leptons and missing energy:

\begin{description}

\item[(1)]
The scaled event total momentum was required to be  
$p_{\mathrm{vis}} /\sqrt{s} > 0.035$
and the direction of the event missing momentum was required to satisfy
$|\cos \theta_{\mathrm miss}| < 0.9$.

\item[(2)]
Events were required to have between three and eight tracks, each 
with a transverse 
momentum relative to the beam axis greater than 1.5~GeV.

\item[(3)]
Events had to contain at least three identified 
leptons, each with a transverse 
momentum relative to the beam axis greater than 1.5~GeV.


\item[(4)]
 It was also required that $\Evis /\sqrt{s} < 0.9$.

\item[(5)]
The total leptonic energy, defined as the sum of the 
energies of all identified leptons,
was required to be greater than $0.65 \times \Evis$.

\item[(6)]
It was required that the scaled longitudinal component of the
event momentum satisfy $p_{\mathrm{vis,z}} /\sqrt{s} < 0.25$.

\end{description}

\begin{table}[htbp]
  \centering
  \begin{tabular}[c]{|c|c|c|c|c|c|}
\hline  
  Physics process   & Eff. (\%) &  Data & Tot. bkg MC & Data $>$ 206~GeV & 
MC $>$ 206~GeV   \\\hline
$\tilde{\nu} \tilde{\nu} \rightarrow$ &&&&& \\
$\mu \mu \nu \nu \mu \mu \nu \nu $ & 72 $-$ 83& 
 &   &    &    \\

& & 16 & 15.6 $\pm$ 0.3 $\pm$ 2.5  &  4  &   4.8  $\pm$ 0.1 $\pm$ 0.7 \\

$ \tau \tau \nu \nu \tau \tau \nu \nu $ & 13 $-$ 28& 
 &   &    &    \\

\hline 
\end{tabular}
\caption{\it
Final states with four leptons and missing energy 
for $\sqrt{s} =$ 189 - 209~GeV and separately for 
$\sqrt{s} >$ 206~GeV:
signal selection efficiencies 
for $\Delta m \leq m_{\snu}/2$, observed number of events in the data and 
background estimates from \sm\ processes. The uncertainties on the 
background estimates are the quadratic sum of the statistical and 
systematic uncertainties.
Many more final states have been studied; only the final states with 
the highest and lowest efficiencies are reported here.
}
\label{table:4lemiss}
\end{table}

Table \ref{table:4lemiss}  
shows the results for the selections 
with the highest and the lowest efficiencies 
at centre-of-mass energies from 189~GeV to 209~GeV.
Efficiencies 
are quoted for $\Delta m \leq m_{\snu}/2$.
Since 
the detection efficiencies are independent of the sneutrino species within
the statistical uncertainties, results are quoted irrespective of the sneutrino
species. 
These analyses do not differentiate the different lepton 
flavours present in the final states and the number of observed events 
in the data and the background estimates are therefore the same 
for the various signal final states. 
However, the signal detection efficiencies depend strongly on the
charged lepton flavours present in the final states.  
The lowest efficiencies arise from final states with four taus 
and missing energy
and the highest ones from final states with four muons and missing energy.
They range from 13\% ($\tau \tau \nu \nu \tau \tau \nu \nu $) 
to 83\% ($\mu \mu \nu \nu \mu \mu \nu \nu $)
for a sneutrino mass between 45 and 103~GeV and $\Delta m \leq m_{\snu}/2$.
As the value of  $\Delta m$ is increased above 
$m_{\snu}/2$, the selection efficiencies drop by a factor 2 
to 3, depending on $m_{\snu}$ and $\Delta m$, due to the larger
fraction of missing energy and lower energy of the charged leptons
in the final state.   
The main contribution to the background comes from four-fermion processes, 
which
account for about 75\% of the \sm\ expectation. The remaining 25\% arises from
processes yielding two-lepton final states.

To select final states without missing energy, the following 
requirements were imposed:

\begin{description}

\item[(1)]
Events were required to have at least three tracks with a transverse 
momentum relative to the beam axis greater than 1.0~GeV.

\item[(2)]
Events had to have between three and ten identified 
leptons, each with a transverse 
momentum relative to the beam axis greater than 1.5~GeV.

\item[(3)]
It was required that $\Evis /\sqrt{s} > 0.9$.

\item[(4)]
To eliminate events containing photons, events were required not to have
electromagnetic energy 
clusters with energies greater than 10 GeV  not associated to a track.

\item[(5)]
The total leptonic energy
was required to be greater than $0.7 \times \Evis$.

\item[(6)]
To reduce the residual four-fermion background,
pairs were formed with the four most energetic particle tracks, 
and the invariant
mass was computed for each pair. Events were selected if 
one of the three possible pairings satisfied
$ |m_{i,j} - m_{k,l}|/(m_{i,j} + m_{k,l}) < 0.4 $, where 
$m_{i,j}$ is the invariant mass of the pair $(i,j)$. Only pairs 
with invariant mass $m_{i,j}$ greater than 20~GeV were used in the computation.

\end{description}

\begin{table}[htbp]
  \centering
  \begin{tabular}[c]{|c|c|c|c|c|c|}
\hline  
  Physics process   & Eff. (\%) &  Data & Tot. bkg MC & Data $>$ 206~GeV & 
MC $>$ 206~GeV   \\\hline
$\tilde{\nu} \tilde{\nu} \rightarrow $ &&&&& \\
$\mu \mu  \mu \mu $ 
                    & 75 $-$ 81 & & & & \\ 

 & & 42 & 32.0  $\pm$ 0.3 $\pm$ 3.2 & 8 & 6.7  $\pm$ 0.1 $\pm$ 0.9 \\

$ \tau \tau \tau \tau $ 
                    & 32 $-$ 36 & & & & \\


\hline 
\end{tabular}
\caption{\it
Final states with four leptons and no missing energy 
(including candidates from the analysis for four leptons with
missing energy)
for $\sqrt{s} = $ 189 - 209~GeV and for 
$\sqrt{s} > $ 206~GeV:
signal selection efficiencies, observed number of events in the data and 
background estimates from \sm\ processes. 
The uncertainties on the background estimates are the quadratic sum of the 
statistical and systematic uncertainties.
Many more final states have been studied; only the final states with the
highest and lowest efficiencies are given in the table.
}
\label{table:4l}
\end{table}


To maximise the detection efficiencies for 
final states without missing energy, 
especially for the four tau final states which contain missing energy 
from the neutrinos present in the tau decays, 
this analysis was combined with the analysis for
final states with missing energy, previously described. 
Events passing either set of criteria 
were accepted as candidates for the pair-production of sneutrinos 
followed by a direct decay. 
Table \ref{table:4l} 
shows the results for the selections 
with the highest and the lowest efficiencies
at centre-of-mass energies from 189~GeV to 209~GeV.
The lowest efficiencies arise for final states with four taus 
and the highest ones from final states with four muons.
They range from 32\% ($\tau \tau \tau \tau$) 
to 81\% ($\mu \mu \mu \mu$)
for a sneutrino mass between 45 and 103~GeV. 
The main contribution to the background comes 
from four-fermion processes, which
account for about 80\% of the \sm\ expectation. The remaining 20\% arise from
processes yielding two-lepton final states.

\subsection{Final states with six leptons and missing energy}
\label{sec:6leptons}

Events with
six charged leptons and missing energy in the final state
may  result for example   
from indirect charged slepton decays with a \lb\ coupling.
Here, the strategy adopted was to design two complementary
searches for events with six leptons either with or without missing
energy. Events satisfying either selection were accepted, which improved
the detection efficiencies compared to only allowing events with
significant missing energy.

The following criteria were applied for the selection of final states 
with six leptons and missing energy:

\begin{description}

\item[(1)]
The direction of the event missing momentum was required to satisfy
$|\cos \theta_{\mathrm miss}| < 0.9$.

\item[(2)]
The scaled event total momentum was required to be  
$p_{\mathrm{vis}} /\sqrt{s} > 0.035$.

\item[(3)]
To reduce the background from 
two-photon processes and di-lepton final states,
it was required that $0.2 < \Evis /\sqrt{s} < 1.0$.


\item[(4)]
Events were required to have at least three tracks with a transverse 
momentum relative to the beam axis greater than 0.3~GeV.
Tracks originating from photon conversions were 
excluded from this analysis.  

\item[(5)]
Events had to contain between four and ten identified 
leptons, at least two of them  with a transverse 
momentum relative to the beam axis greater than 1.5~GeV, and the
third one with a transverse 
momentum relative to the beam axis greater than 0.3~GeV.

\item[(6)]
The total leptonic energy
was required to be greater than $0.55 \times \Evis$. 

\end{description}

These criteria were used 
for the search for final states with six leptons without missing energy:

\begin{description}

\item[(1)]
To reduce the background from 
two-photon processes and di-lepton final states,
it was required that $0.2 < \Evis /\sqrt{s} < 1.2$.

\item[(2)]
Events were required to have at least five tracks with a transverse 
momentum with respect to the beam axis greater than 1.0~GeV.
Tracks from photon conversions were excluded.  

\item[(3)]
Events had to contain between four and twelve identified 
leptons, each with a transverse 
momentum with respect to the beam axis greater than 1.5~GeV.

\item[(4)]
The total leptonic energy
was required to be greater than $0.40 \times \Evis$.

\end{description}

\begin{table}[htbp]
  \centering
\renewcommand{\arraystretch}{1.2}
  \begin{tabular}[c]{|c|c|c|c|c|c|}
\hline  
  Physics process   & Eff. (\%) &  Data & Tot. bkg MC & Data $>$ 206~GeV & 
MC $>$ 206~GeV   \\\hline
$\tilde{\ell}^+ \tilde{\ell}^-  \rightarrow$&&&&&\\ 
$ e \tau \tau \nu e \tau \tau \nu $
                    & 38 $-$ 63  &&&& \\
$ e \mu \mu \nu e \mu \mu \nu $ 
                    & 78 $-$ 92  &&&& \\
$ \mu \tau \tau \nu \mu \tau \tau \nu $ 
                    & 59 $-$ 69  
& 9 & 8.5  $\pm$ 0.2 $\pm$ 1.3 & 1 & 1.5  $\pm$ 0.04 $\pm$ 0.3 \\
$ \mu \mu \mu \nu \mu \mu \mu \nu $ 
                    & 90 $-$ 96  &&&& \\
$ \tau \tau \tau \nu \tau \tau \tau \nu $ 
                    & 24 $-$ 41  &&&& \\
$ \tau \mu \mu \nu \tau \mu \mu \nu $ 
                    & 79 $-$ 89  &&&& \\
\hline 
\end{tabular}
\caption{\it
Final states with six leptons and missing energy 
for $\sqrt{s} = $ 189 - 209~GeV and for 
$\sqrt{s} > $ 206~GeV:
signal selection efficiencies, observed number of events in the data and 
background estimates from \sm\ processes. The uncertainties on the background 
estimates are the quadratic sum of the statistical and systematic 
uncertainties.
}
\label{table:6lemiss}
\end{table}

Table \ref{table:6lemiss}  shows the results for 
the events passing either selection
at centre-of-mass energies from 189~GeV to 209~GeV. Since 
the detection efficiencies are independent of the charged slepton 
species within
the statistical uncertainties, 
results are quoted irrespective of the charged slepton
species.
These analyses do not differentiate the different lepton 
flavours present in the final states and the number of observed events 
in the data and the background estimates are therefore the same 
for the various signal final states. 
However, the signal detection efficiencies depend strongly on the
charged lepton flavours present in the final states.  
The lowest efficiencies arise from final states with six taus 
and missing energy
and the highest ones from final states with six muons and missing energy.
They range from 24\% ($\tau \tau \tau \nu \tau \tau \tau \nu $) 
to 96\% ($\mu \mu \mu \nu \mu \mu \mu \nu $)
for a charged slepton mass between 45 and 103~GeV.
The background is entirely due to four-fermion processes.

In the small $\Delta m$ region, the final state charged leptons 
resulting from the charged slepton decay into a charged lepton and a
neutralino may not be detected due to the 
small phase space available. In this case, the analysis 
searching for four leptons and missing energy is
applied to maximise the detection efficiencies. The lowest efficiencies in
the $\Delta m <$~5~GeV range were used.  
Signal selection efficiencies are of the order of 12-14\%, depending 
on the charged slepton flavour. 
A total of 16 events was observed in the data and a 
background of $15.6  \pm 0.3 \pm 2.5$ was estimated from \sm\ processes
for $\sqrt{s} = $ 189 - 209~GeV. For $\sqrt{s} > $ 206~GeV, the numbers
are respectively 4 (data) and $4.8  \pm 0.1 \pm 0.7$ (background).
The uncertainty quoted on the background estimate is the quadratic sum of the 
statistical and systematic uncertainties.  

\subsection{Inefficiencies and systematic uncertainties for 
leptonic final states}
\label{sec:syserr}

The inefficiency due to false forward detector vetoes, described 
in Section~\ref{sec:preselections},  caused by
beam-related backgrounds or detector noise was estimated 
to range from 1.4 to 2.8\%, depending on the data samples,
from a study of randomly triggered beam crossings.
The quoted efficiencies and background expectations
take this effect into account. 

The following systematic uncertainties on the signal 
detection efficiencies were considered:

\begin{enumerate}
\item
The statistical uncertainty on the determination of the efficiency from the 
Monte Carlo simulation.

\item
The systematic uncertainty on the integrated luminosity (0.3-0.4\%).

\item
The uncertainty due to the interpolation of the efficiencies (4.0\%).

\item 
The lepton identification uncertainty 
(2.4\% for the muons, 3.9\% for the electrons and 4.7\% for the taus). 

\item
The systematic uncertainty arising from the modelling of
the variables used in the selections. 
This was estimated by recalculating the 
detection efficiencies using a new cut value shifted from its original value 
by a ratio of the means of the cut variable distributions of the data
and of the \sm\ samples. The difference between the original efficiency and 
the new efficiency, usually less than 5\%,  
is taken as systematic uncertainty due to the modelling 
of a cut variable. 

\end{enumerate}

The systematic uncertainty due to the trigger efficiency is
negligible because of the high lepton transverse momentum requirement.

All cut variables are 
treated as independent; hence, 
systematic uncertainties originating from each variable are added in 
quadrature. 
The total systematic uncertainty was calculated by summing in quadrature 
the individual uncertainties.


\begin{table}[htbp]
\begin{center}
\begin{tabular}{|l|r|r|r|}
\hline

Channel   &  $\sigma^{\rm stat}_{\epsilon}$ (\%) & $\sigma^{\rm sys}_{\epsilon}$ (\%) &  
             $\sigma^{\rm tot}_{\rm N_{bkg}}$ (\%)  \\
\hline
2 $\ell$ + $\Emiss$   & 2.2 -- 6.7 & 4.3 -- 31.2  &  7.4 -- 18.9 \\
4 $\ell$ + $\Emiss$   & 1.5 -- 8.3 & 7.6 -- 52.7  & 13.7 -- 17.6 \\   

4 $\ell$ + $\Emiss$ OR 4 $\ell$
                        & 1.5 -- 4.6 & 7.7 -- 20.1  &  7.0 -- 12.7 \\   
6 $\ell$ + $\Emiss$ OR 6 $\ell$
                        & 0.7 -- 5.6 & 6.5 -- 27.8  & 12.2 -- 31.8 \\   
\hline
\end{tabular}
\end{center}
\caption{\it
Relative statistical and systematic uncertainties on the selection efficiency, 
$\sigma^{\rm stat}_{\epsilon}$ and   
$\sigma^{\rm sys}_{\epsilon}$, and relative total uncertainty on the number 
of expected \sm\ background events,  $\sigma^{\rm tot}_{\rm N_{bkg}}$,  
for the various multi-lepton final state searches.} 
\label{tab:errors}
\end{table}

The systematic uncertainty on the number of 
expected \sm\ background events is estimated 
in a similar way. The statistical uncertainty on this number is small, 
typically less than 1\%, due to the large size of the \sm\ event samples.
For final states with more than two leptons,
an additional systematic uncertainty on the number of expected background 
events, arising from the imperfect simulation of four-fermion processes 
is taken into account. It is determined by comparing predictions of 
grc4f version 2.2 including all interfering four-fermion diagrams to the 
default Monte Carlo predictions, which use Vermaseren for the two-photon
processes and grc4f version 2.1 for the other four-fermion processes, 
and neglect
the interference between the two. This 
systematic uncertainty ranges from 10 to 17\%, 
depending on the analysis. 

The relative statistical and systematic uncertainties 
on the selection efficiency, 
$\sigma^{\rm stat}_{\epsilon}$ and   
$\sigma^{\rm sys}_{\epsilon}$, 
and the relative total uncertainty on the number 
of expected \sm\ background events, $\sigma^{\rm tot}_{\rm N_{bkg}}$, 
are summarised in Table~\ref{tab:errors}
for the multi-lepton final state searches presented in the previous 
sections.





\section{Final states with two jets and two leptons}
\label{sec:2jets2leptons}

This section describes the selection for final states 
with two charged leptons, two jets and no missing energy. 
These final states may result
from the direct decay of pair-produced stops via a \lbp\ coupling.
In contrast to the purely leptonic final states described in the previous
section, the topologies searched for in this analysis involve hadronic jets; 
more stringent  cuts are needed to obtain a purer lepton sample.  
Particles are considered as electrons or muons if they are either identified by
the  selection algorithms described in~\cite{ref:elecbarrel} 
and~\cite{ref:elecendcap}, or by an algorithm used for selecting 
semileptonic W decays, as
described in~\cite{ref:wwpaper}.
A Neural Net (NN) based on track properties~\cite{ref:tauid} 
is used to
identify taus.
Events were preselected by requiring the following criteria to be 
satisfied (the same criteria were also used for the analysis presented 
in Section~\ref{sec:jetsleptons}):

The ratio of good tracks satisfying the criteria 
of Section~\ref{sec:opaldet}
to the total number of reconstructed tracks
had to be greater than 0.2 
to reduce beam-gas
and beam-wall background events.
Events with fewer than 
eight good tracks were not considered, in order to reduce the
background from Bhabha scattering.
Events had to contain at least one identified 
electron or muon with a momentum $p       $
greater than 3~GeV, to
reduce the background  from final states with 
low energy leptons (electron or muon) arising for instance from
semi-leptonic quark decays or misidentified particles.
To reduce background from two-photon processes, it was required that 
the visible energy
normalised to the centre-of-mass energy
$\Evis/\sqrt{s}$ be greater than 0.3.

The following cuts were then applied:


\begin{description}

\item [(1)]
The visible energy had to be close to the centre-of-mass energy, i.e.\ 
$0.75 < \Evis / \sqrt{s}  < 1.25$ for electron and muon and
$0.5 < \Evis / \sqrt{s} < 1.0$ for tau final states.

\item [(2)]
It was required that four jets be reconstructed 
using the Durham~\cite{ref:durham} algorithm,
with $y_{34} > 0.001$, where $y_{34}$ is the jet resolution 
at which the number of jets changes from 3 to 4. 
Both hadronic and leptonic objects are used in the jet reconstruction.

\item [(3)]
Events had to contain at least one pair of identified oppositely-charged 
lepton candidates of the same flavour.

\item [(4)]
To make use of the signal topology of two leptons and 
two jets,
where a lepton and a jet stem from the same object, 
a five-constraint (5C) kinematic fit was
performed for the two possible 
combinations of each lepton with each jet.
The kinematic constraints are: the vector sum of all 
momenta has to be
equal to zero, the total energy of all objects has to 
be equal to the 
centre-of-mass energy and the masses of the two 
reconstructed particles
have to be equal.
From the three most energetic leptons of the same 
flavour, 
the two most isolated\footnote{
The most isolated lepton is the one with the largest 
angle to the closest other track.}
were selected and the rest of the event was 
reconstructed as two jets.  
The combination with the highest fit 
probability was
selected. The probability for the fit, based on the 
$\chi^2$, was
required to be larger than 0.01.

\item [(5)]
The scaled momenta $p/\roots$ of the most and second most energetic leptons had
 to be 
greater than 0.082 and 0.055, respectively, for final states with electrons 
and muons and greater than 0.055 and 0.0275, respectively, for taus.

\item [(6)]
It was required that there be no  track 
within a cone of half opening angle of 
$15^{\circ}$ around the most energetic 
lepton candidate.
\end{description}

\begin{figure}[htbp]
\centering
\includegraphics*[scale=0.5]{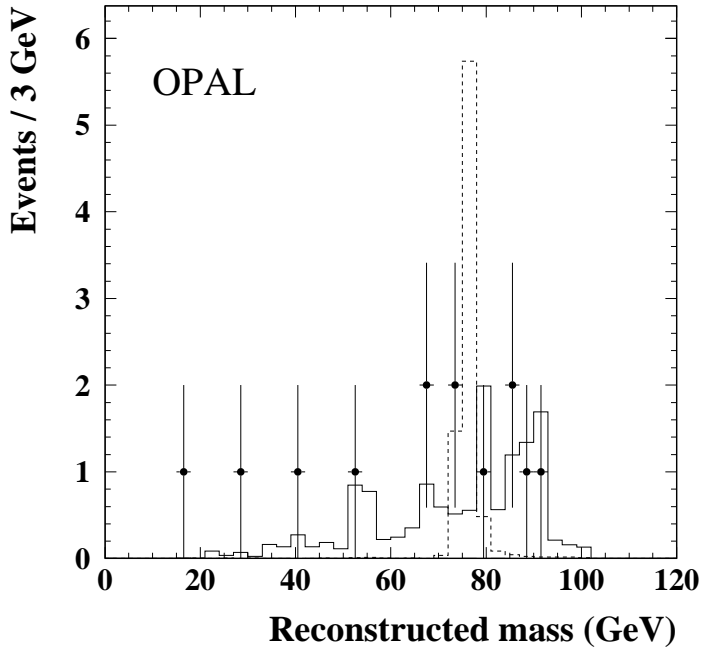}
\caption[]{\it
Final states with two jets and two leptons: distribution of the
reconstructed stop mass after cut (6) for the direct decay of stops into
final states with two jets and two electrons. The signal is shown  
as the dashed line for a stop mass of 75~GeV. Data
are shown as points and the sum of all Monte Carlo background processes is
shown as a solid line. 
Data used for this plot were collected at
centre-of-mass energies of 189--209~GeV. 
The simulated signal events have arbitrary normalisation. 
Uncertainties are statistical only.
}
\label{figure:stop}
\end{figure}

Figure~\ref{figure:stop} shows the 
distribution of the
reconstructed stop mass after cut (6). 
The kinematic fit achieves a very good mass resolution.
For stop masses above 75~GeV, the mass resolution is better than 0.8~GeV 
for final states containing either electrons or muons and better than 
1.3~GeV for final states with taus.
Events are counted only in a 
two sigma mass window around the mean reconstructed stop mass. For masses
near the kinematic limit the mass resolution approaches zero. 
If the mass resolution is below 0.5~GeV, it is set to 0.5~GeV.

These cuts, including the cut of the two sigma mass window,
 yield an efficiency of more than 46\% for a stop mass 
above 75~GeV 
for final states with either electrons or muons.
For final states with taus, efficiencies above  15~\% are reached for 
stop masses above 75~GeV.
Table~\ref{tab:ivor:stop} shows the efficiencies, 
the number of candidate events observed in the data and
the number of background events for the selection at centre-of-mass 
energies from 189~GeV to 209~GeV, corresponding to an integrated 
luminosity of 599.6 pb$^{-1}$. 
The main
contribution to the background comes from four-fermion processes which
amount to 100\% of the Standard Model expectation for final states with
muons and to 75$-$95\% for final states with taus.   
In the electron channel, 
four-fermion events account for 45$-$75\% of the \sm\ background, 
two-fermion processes contribute less than 5\% 
and the remaining events are expected from two-photon interactions.

\begin{table}[htbp]
\renewcommand{\arraystretch}{1.3}
\centering
{\small
\begin{tabular}{|c|c|c|c|c|c|}\hline
Physics process & Eff. (\%)     & Data  & Tot.\ bkg MC  & Data $>$ 206 GeV
& MC $>$ 206 GeV \\\hline
$\tilde{t}\bar{\tilde{t}} \to$ & & & & & \\
  $2\;\mbox{jets} + e^+ + e^-$          &  46 $-$ 55         
 & 13   & $ 13.8\pm1.4\pm1.2 $    & 3    & $ 3.3\pm0.5\pm0.3 $\\
  $2\;\mbox{jets} + \mu^+ + \mu^-$      &  47 $-$ 55 
 & 7    & $ 7.1\pm0.4 \pm1.5 $    & 0    & $  1.5\pm0.2\pm0.1$\\
  $2\;\mbox{jets} + \tau^+ + \tau^-$    &  15 $-$ 25         
 & 30   & $ 34.4\pm1.0\pm4.1 $     & 7    & $ 7.9\pm0.6\pm0.9$\\\hline
\end{tabular}
}
\caption{\it
Final states with two jets and two leptons
for $\sqrt{s} = $ 189 - 209~GeV and for 
$\sqrt{s} > $ 206~GeV:
signal selection efficiencies, observed number of events in the data and 
background estimates from \sm\ processes. 
The first uncertainty on the background estimate is statistical 
and the second is systematic.
The efficiencies are given for stop masses greater than
$75$~{\rm GeV}
for the signal at $\sqrt{s} = $ 207~{\rm GeV}.
}
\label{tab:ivor:stop}
\end{table}

\subsubsection*{Systematic uncertainties}

The following systematic uncertainties on the signal efficiencies 
were considered:

\begin{enumerate}

\item 
The statistical uncertainties from the limited size of the Monte Carlo samples.

\item 
The systematic uncertainties on the integrated luminosity (0.3 - 0.4\%).

\item 
The uncertainty due to the interpolation of efficiencies for mass values 
between the generated stop masses (less than 3\%). 

\item 
For the lepton identification, the highest uncertainty (4\% for electrons) 
was taken for all flavours.

\item
The systematic uncertainty arising from the modelling of the 
variables used in the selection, as described below. 

\item 
The fragmentation of the stop was simulated using the fragmentation 
function from Peterson {\it et al.} with the $\epsilon$ parameter 
extrapolated from measurements of charm and bottom~\cite{ref:opalstop}.
To check the model dependence of the fragmentation,
it was also performed using the function from 
Bowler~\cite{ref:bowler}.
No significant change in the efficiency due to the difference in the 
fragmentation function was found. 
The difference was at most 0.5\%, where a variation of the 
$\epsilon$ parameter of the $\stopx\ $ in the Peterson {\it et al.} scheme 
was included. This uncertainty on $\epsilon_{\stopx}$
was propagated from the uncertainty of 
$\epsilon_{\rm b}$ and the uncertainty on the b-quark
mass as described in detail in Ref.~\cite{ref:opalstop}.

\item 
The signal events were produced for a zero mixing angle between the two
stop eigenstates. The mixing angle describes the coupling between the stop 
and the Z$^0$, and therefore the energy distribution of the initial state 
radiation depends on this mixing angle. 
To check the dependence of the detection efficiency on this angle, 
events were generated with $\theta_{\stopx\ } = 0.98$~rad, 
where the stop decouples from the Z$^0$.
The change in efficiency was less than 0.5\% for the two extreme cases.

\item 
The Fermi motion of the spectator quark in the stop-hadron influences 
its  measured mass. The Fermi motion was increased from 220~MeV to
520~MeV and the efficiency changes by no more than 1\%, which was taken as a 
systematic uncertainty.

\end{enumerate}

The systematic uncertainty 
due to the uncertainty in the trigger efficiency was 
estimated to be negligible, because of the requirement of at least eight good
tracks.

All of the above contributions to the systematic uncertainties result in 
a total systematic uncertainty of
less than 7\%.

The uncertainty in the signal efficiency and in the number of background
events due to the modelling of the selection variables was estimated in
the following way.
First the mean and the r.m.s.\ of the distribution of the variable ($y$) was
determined for the data and the \sm\ background simulation (SM MC). 
Then for each
MC event (both signal and background) the variable was transformed as
$$y' = (y - \mathrm{ mean(SM \ MC)) \cdot r.m.s.(data) / 
r.m.s.(SM \ MC) + mean(data)}$$
The standard analysis was applied to 
signal and \sm\ background events with 
all variables transformed; the difference in 
the MC expectation due to the transformation was considered as the
systematic uncertainty. 
The contributions for the different variables were added
in quadrature.

The systematic uncertainty on the expected number of
background events estimated in this way  
was found to be less than 20\%, the largest contribution
($\sim$ 15\%)
stemming from the variable $y_{34}$.
In addition, the difference in the number of selected events by comparing
different Monte Carlo generators, as described in Section~\ref{sec:MC},
was also taken as a systematic uncertainty.


\section{Final states with four jets and at least two charged leptons}
\label{sec:jetsleptons}

\begin{figure}[htbp]
\centering
\includegraphics*[scale=0.5]{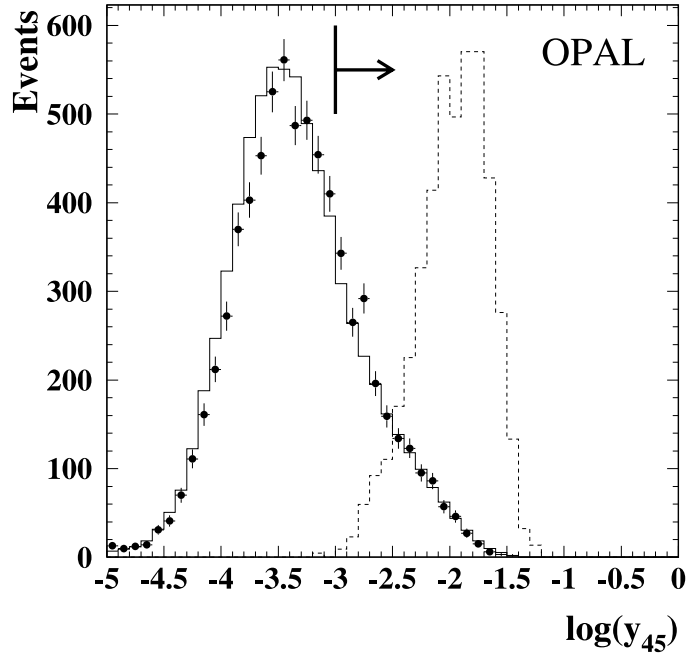}
\caption[]{\it
Final states with four jets and at least two charged leptons:
distribution of the variable $y_{45}$ before the cut on this variable
for the indirect decay of selectrons into final states with four jets
and four electrons.
The signal is shown as the dashed line for
a selectron mass of 102~GeV and a mass difference $\Delta m = m/2$.
Data are shown as points and the sum of all Monte Carlo background 
  processes is shown as a solid line. 
  Data used for this plot were collected in the year 2000 at 
  a luminosity weighted average centre-of-mass energy of 206.6~GeV.
  The visible disagreement in the peak position 
  between data and \sm\ expectation is taken into account as a systematic
  uncertainty. 
  The simulated signal events have arbitrary normalisation. 
  The arrow points into the region accepted by the cut.
Uncertainties are statistical only. 
}
\label{figure:selec}
\end{figure}

This section describes the event selection for final states resulting 
from the indirect 
decay of selectrons, smuons and staus via the coupling \lbp.
The final states consist of two charged leptons of the same flavour as the 
sleptons plus the decay products of the two neutralinos.
These will be two jets plus a neutral or charged lepton for each neutralino.
The identification of leptons and the preselection are the same as described 
in Section~\ref{sec:2jets2leptons}.
The selection cuts were as follows:

\begin{description}
\item [(1)]
Lower and upper cuts in the ranges 0.5--0.7 and 0.75--1.2, respectively,
were applied
to the visible energy scaled by the centre-of-mass energy
$\Evis/\sqrt{s}$,
depending on the expected number of neutrinos.
In addition, if some missing momentum was expected, a cut was 
made on the angle 
of the missing momentum 
with respect to the beam direction
at $|\cos \theta| <$~0.95.

\item [(2)]
The jets in the event were reconstructed using the Durham algorithm.
The jet resolution parameter $y_{45}$  
at which the number of jets changes from four to five
was required to be greater than  0.002. 
The  ${\rm log(y_{45})}$  distribution is shown
in Figure~\ref{figure:selec}.
It demonstrates the good discriminating power of this quantity.
The visible disagreement in the peak position 
between data and \sm\ expectation was taken into account as a systematic
uncertainty. 
This cut takes into account the high multiplicity of the signal events.

\item [(3)]
At least two leptons of the flavour of the slepton  
had to be identified.
To retain sensitivity to small mass differences 
between the slepton and
the neutralino, 
the required scaled energy greater than 0.022 
for the two electrons
in the selectron case,
the required momentum had to be greater than 4~GeV
for both muons in the smuon case and 
the required momentum had to be greater than 3~GeV
for both taus 
in the tau case, respectively.

\item [(4)]
In addition to the leptons required in the previous cut (3), 
the leptons from 
the neutralino decay had to be identified. If two additional charged leptons 
with a different flavour to the slepton
were expected, both had to be identified except in the case of two taus,
where only
one had to be identified.
If a total of four leptons of the same flavour was expected, 
including those in cut (3), only three of them had to be identified.
If only one additional lepton was expected, it had to be identified.

The scaled energy or momentum of the most energetic lepton had to be above
a cut value varying between 0.044 and 0.08, 
depending on the topology.
If a total of four leptons was required, for the second most energetic 
lepton a scaled energy or momentum larger than a cut value varying 
between 0.016 and 0.022, 
depending on the topology, was required.

\item [(5)]
To make use of the isolation of the leptons in the signal, one or two
of the identified leptons, depending on the expected topology,
must be isolated.
The isolation criterion was that  
there be no charged track within a cone of 
half opening angle $\varphi$, such that $|\cos \varphi| =$~0.99,
around the track of the lepton. 
\end{description}

\begin{table}[htbp]
\begin{center}
\begin{tabular}{|c|c|c|c|c|c|}\hline
Physics process  & Eff. (\%)             & Data  & Tot.\ bkg MC     &
 Data $>$ 206 GeV       & MC $>$ 206 GeV \\\hline
$\tilde{e}^+ \tilde{e}^- \to$ & & & & & \\
         $4 \;\mbox{jets} + 2 e + 2 e$         & 39 $-$ 69       & 13    & $ 13.3\pm0.6\pm1.2 $       &  2    & $  2.5\pm0.3\pm0.3 $\\
         $4 \;\mbox{jets} + 2 e + 2\mu$        & 25 $-$ 53       &  2    & $  2.6\pm0.3\pm0.4 $       &  0    & $  0.7\pm0.1\pm0.1 $\\
         $4 \;\mbox{jets} + 2 e + 2\tau$       &  6 $-$ 25       & 14    & $ 12.7\pm0.6\pm1.6 $       &  5    & $  2.7\pm0.3\pm0.2 $\\
         $4 \;\mbox{jets} + 2 e + e\nu$        & 17 $-$ 40       & 16    & $ 17.6\pm0.7\pm1.5 $       &  2    & $  3.3\pm0.3\pm0.3 $\\
         $4 \;\mbox{jets} + 2 e +\mu\nu$       & 21 $-$ 52       & 16    & $ 12.3\pm0.6\pm1.4 $       &  4    & $  2.8\pm0.3\pm0.4 $\\
         $4 \;\mbox{jets} + 2 e +\tau\nu$      &  2 $-$ 12       &  7    & $  7.3\pm0.5\pm0.9 $       &  1    & $  1.5\pm0.2\pm0.2 $\\
         $4 \;\mbox{jets} + 2 e + 2\nu$        &  6 $-$ 17       &  6    & $  5.8\pm0.4\pm0.4 $       &  2    & $  1.4\pm0.3\pm0.1 $\\
\hline
\hline
$\tilde{\mu}^+ \tilde{\mu}^- \to$    & & & & & \\
         $4 \;\mbox{jets} + 2\mu + 2 e$        & 34 $-$ 54        &  2    & $  1.6\pm0.2\pm0.2 $       &  0    & $  0.2\pm0.1\pm0.0 $\\
         $4 \;\mbox{jets} + 2\mu + 2\mu$       & 52 $-$ 79        &  4    & $  2.9\pm0.3\pm0.6 $       &  2    & $  0.7\pm0.1\pm0.2 $\\
         $4 \;\mbox{jets} + 2\mu + 2\tau$      &  9 $-$ 32        &  2    & $  3.6\pm0.3\pm0.5 $       &  0    & $  0.7\pm0.1\pm0.1 $\\
         $4 \;\mbox{jets} + 2\mu + e\nu$       & 25 $-$ 52        &  5    & $  4.4\pm0.3\pm0.8 $       &  0    & $  1.0\pm0.2\pm0.3 $\\
         $4 \;\mbox{jets} + 2\mu +\mu\nu$      & 30 $-$ 51       &  4    & $  3.6\pm0.3\pm0.6 $       &  2    & $  0.9\pm0.2\pm0.2 $\\
         $4 \;\mbox{jets} + 2\mu +\tau\nu$     &  6 $-$ 22        &  2    & $  2.3\pm0.2\pm0.3 $       &  0    & $  0.4\pm0.1\pm0.0 $\\
         $4 \;\mbox{jets} + 2\mu + 2\nu$       & 15 $-$ 35        &  6    & $  3.8\pm0.3\pm0.4 $       &  0    & $  0.8\pm0.1\pm0.0 $\\
\hline
\hline
$\tilde{\tau}^+\tilde{\tau}^-\to$  & & & & & \\
         $4 \;\mbox{jets} + 2\tau + 2 e$       & 19 $-$ 52        &  9    & $  9.4\pm0.5\pm0.6 $       &  1    & $  1.9\pm0.3\pm0.1 $\\
         $4 \;\mbox{jets} + 2\tau + 2\mu$      & 19 $-$ 56        &  7    & $  8.8\pm0.4\pm1.4 $       &  0    & $  2.2\pm0.2\pm0.5 $\\
         $4 \;\mbox{jets} + 2\tau + 2\tau$     &  7 $-$ 22        & 53    & $ 46.0\pm1.1\pm3.9 $       &  6    & $  9.3\pm0.6\pm0.6 $\\
         $4 \;\mbox{jets} + 2\tau + e\nu$      &  4 $-$ 16        & 15    & $ 12.1\pm0.6\pm1.1 $       &  1    & $  2.1\pm0.3\pm0.2 $\\
         $4 \;\mbox{jets} + 2\tau +\mu\nu$     &  4 $-$ 14        & 12    & $ 12.8\pm0.5\pm1.4 $       &  1    & $  2.8\pm0.3\pm0.4 $\\
         $4 \;\mbox{jets} + 2\tau +\tau\nu$    &  8 $-$ 21        & 26    & $ 24.9\pm0.8\pm2.3 $       &  6    & $  5.1\pm0.4\pm0.5 $\\
         $4 \;\mbox{jets} + 2\tau + 2\nu$      &  3 $-$  8        & 83    & $ 66.3\pm1.2\pm1.9 $       & 16    & $ 13.6\pm0.6\pm0.2 $\\\hline
\end{tabular}

\end{center}
\caption{\it 
  Final states with four jets and at least two charged leptons
for $\sqrt{s} = $ 189 - 209~GeV and for 
$\sqrt{s} > $ 206~GeV.
Signal selection efficiencies, observed number of events in the data and 
background estimates from \sm\ processes. 
The first uncertainty on the background estimate is statistical 
and the second is systematic.
The efficiencies are given for slepton masses larger than
45~GeV and for a signal at $\sqrt{s} = $ 207~GeV.
}
\label{tab:sleptons_ivor}
\end{table}

Table~\ref{tab:sleptons_ivor} shows the efficiencies, 
the number of candidate events observed in the data and
the number of background events for the selection at centre-of-mass 
energies from 189~GeV to 209~GeV, corresponding to an integrated 
luminosity of 599.6 pb$^{-1}$. 
Good agreement was found between the numbers of events expected and
observed. 
The largest difference is observed
for the final state with two taus and two neutrinos, where there is 
a 1.8 $\sigma_{\rm stat}$ excess in the number of data events.
The main contribution to the \sm\ background comes 
from four-fermion processes; two-fermion multi-hadron events
contribute up to 30\% and other processes were negligible.

\subsubsection*{Systematic uncertainties}

The systematic uncertainties on the signal efficiencies 
and on the expected number of background events 
were estimated in the same manner as in
Section~\ref{sec:2jets2leptons}:

\begin{enumerate}

\item 
The statistical uncertainty from the limited size of the Monte Carlo samples.

\item 
The systematic uncertainty on the integrated luminosity (0.3 - 0.4\%).

\item 
The uncertainty due to the interpolation of efficiencies for mass values 
between the generated mass points (4\%). 

\item 
For the lepton identification, the highest uncertainty 
(4\% for electrons) was taken for all flavours.

\item
The systematic uncertainty arising from the modelling of the 
variables used in the selection. 
This results in a total systematic uncertainty on the efficiency of
5.2 - 13.4\% for the selectron selection, 
of 5.2 - 12.0\% for the smuon selection and
of 5.2 - 19.3\% for the stau selection.

\item
From the studies of the fragmentation in Section~\ref{sec:2jets2leptons}, 
the systematic uncertainty 
for this analysis was estimated to be less than 1\%. 

\end{enumerate}

The systematic uncertainty due to the uncertainty in the trigger efficiency 
was negligible, because of the requirement of at least eight good
tracks.

 The systematic uncertainty on the expected number of background events was
 again estimated to be less than 20\% for all cases.

\section{Final states with four jets and missing energy}
\label{sec:multijetsemiss}

Indirect decays of sneutrinos via a $\lambda'$ coupling can lead to final 
states with four jets and a large amount of missing energy due to the four 
undetected neutrinos. The dominant background contribution comes from 
four-fermion processes, mainly 
$\mathrm{ W^+W^- \rightarrow q\overline{q}\ell\nu}$, 
and radiative or mis-measured two-fermion events. 
The selection procedure closely follows the one described 
in~\cite{ref:opal_sfer_183} with
small changes required at the higher centre-of-mass energies. 
The selection procedure is described below:
\begin{description}

\item[(1)] The event had to be classified as a multi-hadron final-state as 
described in \cite{ref:LEP2MH}.

\item[(2)] The visible energy of the event was required to be less than 
$0.75 \times \sqrt{s}$ to account for the undetectable neutrinos in the 
final state. 

\item[(3)] To reject two-photon and radiative two-fermion events,
the transverse momentum of the event should be larger than 
$0.075 \times \sqrt{s}$, 
the total energy measured in the forward calorimeter, gamma-catcher 
and silicon tungsten calorimeter should be less than 15 GeV, 
and there should be no significant energy deposit in the 
scintillating tile counters.
The missing momentum should not point along the beam 
direction ($| \cos \theta_{\mathrm{miss}}| <$ 0.96). 

\item[(4)] The events were forced into four jets using the Durham 
jet-finding algorithm, and rejected if the jet resolution parameter 
$y_{34}$ was less than 0.0008. All jets must contain at least one charged
particle track.

\item[(5)] An additional cut was applied 
against semi-leptonic four-fermion events, vetoing events containing 
isolated leptons.
The lepton veto is based on a NN routine~\cite{ref:tauid}, which
was designed to identify tau leptons. The algorithm to select 1-prong tau
candidates looks for high momentum isolated tracks, and therefore it is
also suitable for vetoing leptons of other flavours.
If any lepton candidate was found
with a NN output larger than 0.97, the event was rejected. 
\end{description}

Finally, a likelihood selection
was employed to classify the remaining events
as two-fermion, four-fermion or signal-like processes.
The method is described in~\cite{ref:mssmpaper}. 
The signal reference histograms were produced separately for first
generation sneutrinos where $t$-channel production is also expected and
for second or third generation sneutrinos with $s$-channel production
only. For a given centre-of-mass energy, all the generated signal events
with the coupling $\lambda^\prime_{121}$, whatever their masses,
were used with equal weight to
form the reference distributions. 
The information from the following variables was combined:
\begin{list}{$\bullet$}{\itemsep=0pt \parsep=0pt \topsep=-5pt \leftmargin=30pt}
\item the effective centre-of-mass energy~\cite{ref:sprime} of the event;
\item the transverse momentum of the event;
\item the cosine of the polar angle of the missing momentum vector;
\item the D event-shape parameter~\cite{ref:dpar} which provides
information about the planar nature of an event;
\item the aplanarity of the event which measures the transverse 
momentum component out of the event plane and which is 
defined as 3/2 times the smallest eigenvalue of the sphericity tensor;
\item the logarithm of $y_{34}$;
\item the highest track momentum;
\item the highest electromagnetic cluster energy;
\item the number of lepton candidates in the event 
with a NN output larger than 0.5;
\item the mass of the hadronically decaying W after 
a kinematic fit 
to the W$^+$W$^-$ $\rightarrow$ q\=q$\ell\nu$ 
hypothesis;
\item the cosine of the smallest jet opening angle, defined by the 
half-angle of the smallest cone containing 68\% of the jet energy; 
\end{list}

\begin{figure}[htbp]
\centering
\includegraphics*[scale=0.5]{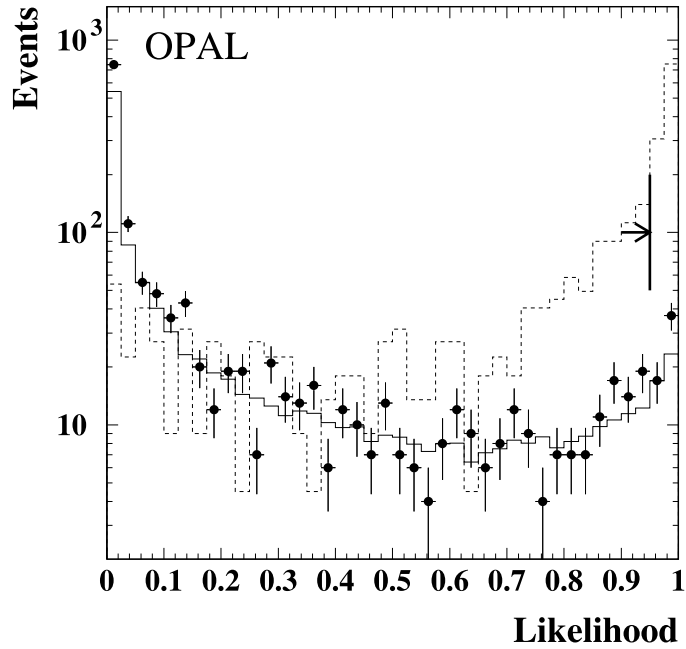}
\caption[]{\it
Final states with four jets and missing energy,
electron sneutrino selection
at $\sqrt{s}= $ 189-209~GeV:
distribution of the likelihood output 
for the selected events in the data (points with error bars), 
the estimated background normalised to the integrated luminosity of the data
(solid line)
and a simulated sample of electron sneutrino pair-production for
$m_{\tilde{\nu}_e} = $ 102.5~GeV with arbitrary normalisation (dashed line). 
Uncertainties are statistical only. The arrow indicates the cut position.
}
\label{figure:4jEmiss}
\end{figure}

The distribution of the likelihood output is shown in 
Figure~\ref{figure:4jEmiss}
for the data, the estimated background and a simulated signal sample 
in the electron sneutrino selection.
The event was rejected if its likelihood output was smaller than 0.95.

Table~\ref{table:4jEmiss} shows the results of the different
selections on the data 
collected at centre-of-mass energies between 189~GeV and 209~GeV
corresponding to an integrated luminosity of 610.6~pb$^{-1}$.
The efficiencies shown in 
Table~\ref{table:4jEmiss} are quoted for 
the coupling $\lambda'_{121}$ and $\Delta m \leq m_{\snu}/2$.
For  $\Delta m > m_{\snu}/2$, the detection efficiency drops 
by a factor 2 to 4, depending on $m_{\snu}$ and $\Delta m$, due to the larger
fraction of missing energy and lower energy final state jets.
For a mass difference of 5~GeV, the primary neutrinos carry away less 
energy and the efficiency increases 
by an absolute value of about 10\%.
The efficiency is small for light sneutrino masses due to the effect of the 
initial-state radiation and the larger boost of the particles, 
which make the event similar to the QCD two-fermion background.
About 94\% of the events expected from the \sm\ originate 
from four-fermion processes.

\begin{table}[htbp]
\renewcommand{\arraystretch}{1.3}
\centering
{\small
\begin{tabular}[c]{|@{\hspace{1.0mm}}c@{\hspace{1.0mm}}|@{\hspace{1.0mm}}c@{\hspace{1.0mm}}|@{\hspace{1.0mm}}c@{\hspace{1.0mm}}|@{\hspace{1.0mm}}c@{\hspace{1.0mm}}|@{\hspace{1.0mm}}c@{\hspace{1.0mm}}|@{\hspace{1.0mm}}c@{\hspace{1.0mm}}|}
\hline  
  Physics process   & Eff. (\%) &  Data & Tot. bkg MC & Data $>$ 206~GeV & 
MC $>$ 206~GeV   \\\hline
$\tilde{\nu}_e \overline{\tilde{\nu}}_e \rightarrow 
4 \ jets + E_{miss}$ 
                    & 1 $-$ 25& 54     & 39.5 $\pm$ 0.5  $\pm$ 4.3 & 8 & 8.2 $\pm$ 0.2 $\pm$ 1.3\\
$\tilde{\nu}_\mu \overline{\tilde{\nu}}_\mu, \tilde{\nu}_\tau \overline{\tilde{\nu}}_\tau \rightarrow
4 \ jets + E_{miss}$                    
                    & 1 $-$ 22& 57     & 39.4 $\pm$ 0.5 $\pm$ 4.3  & 9 & 8.3 $\pm$ 0.2 $\pm$ 2.0\\
\hline 
\end{tabular}
}
\caption{\it
Final states with four jets and missing energy
for $\sqrt{s} = $ 189 - 209~GeV and for 
$\sqrt{s} > $ 206~GeV:
signal selection efficiencies for sparticle masses above 45~GeV
and $\Delta m \leq m_{\snu}/2$, 
observed numbers of events in the data and 
background estimates from \sm\ processes.  
The first uncertainty on the background estimate is statistical 
and the second is systematic.
The numbers of observed events are highly
correlated between the selections.
}
\label{table:4jEmiss}
\end{table}


The observed approximately 2 $\sigma_{\mathrm stat}$ excess in the data 
spreads over the different centre-of-mass energies and it is concentrated
in the $0.075 < p_{\mathrm T}/\sqrt{s} < 0.15$ region, not favoured by events
with pair-produced high mass sleptons. An excess appears from the first
steps of the selection.
After cut (5), 1445 events were 
observed for 1240
expected from \sm\ processes. 
As seen in Figure~\ref{figure:4jEmiss}, 
there is a significant 
excess in the low likelihood region, which is expected to be 
almost equally populated by two- and four-fermion events.  

\subsubsection*{Systematic uncertainties}

The following sources of uncertainties were considered on the selection 
efficiencies (all quoted uncertainties are relative):


\begin{enumerate}
\item The limited MC statistics (5.4$-$28.7\%).
\item The statistical and systematic uncertainties on the luminosity measurement
(0.3$-$0.4\%).
\item The modelling of the preselection variables (2.2$-$2.5\%).
\item The modelling of the variables used in the likelihood selection
(20.9$-$23.4\%).
\item The modelling of the lepton veto (1\%).
\item The interpolation of efficiencies (2.7$-$5.2\%).
\end{enumerate}

Similarly the background estimate is affected by uncertainties due to:

\begin{enumerate}
\item The limited MC statistics (1.2\%).
\item The statistical and systematic uncertainties on the luminosity measurement
(0.3$-$0.4\%).
\item The modelling of the preselection variables (8.6$-$9.6\%).
\item The modelling of the variables used in the likelihood selection
(27.8$-$29.1\%).
\item The modelling of the lepton veto (1\%).
\item The imperfect knowledge of \sm\ cross-sections (2\%).
\item The imperfect simulation of four-fermion processes 
determined by comparing the predictions of grc4f, KORALW and KandY (16\%).
\end{enumerate}
The total systematic uncertainty 
excluding MC statistics ranges from 21.2 to 24.3\%
on the signal detection efficiency and from 33.3 to 34.6\% on the background
rate.

The effect of the modelling of the preselection and likelihood variables
was estimated for each variable using the transformation described 
in Section~\ref{sec:2jets2leptons}.
Applying
the same analysis with unchanged reference histograms to
calculate the likelihood, the difference in 
the MC expectation due to the transformation was considered as the
systematic uncertainty. 
The largest contributions both for the signal and for the background come 
from the modelling of the logarithm of $y_{34}$, the energy
of the most energetic electromagnetic cluster and the cosine of the smallest
jet opening angle in the likelihood selection.

The uncertainty introduced by interpolating the signal efficiencies between 
simulated signal masses was estimated in the following way: 
the signal efficiencies were randomly smeared within
their statistical uncertainty and the resulting numbers were used in the
interpolation.  This process was repeated 100 times, 
and the absolute differences
between the original and the smeared fits were averaged. For each slepton mass,
the mean difference was taken as a measure of the systematic uncertainty.
There is no interpolation between the generated centre-of-mass energies:
the efficiency determined from the signal sample with closest generated
centre-of-mass energy was always taken. The largest difference 
between the actual and
the generated centre-of-mass energies is about 3 GeV, and its effect 
on the result is negligible.

The inefficiency due to the forward energy veto,  by which both the signal
efficiencies and the background  estimates were decreased, was found to vary 
between 1.9 and 3.7\% by studying randomly triggered events.

\section{Final states with four jets and no missing energy}
\label{sec:multijetsnoemiss}

Direct decays of sleptons via $\lambda'$ couplings and squarks via
$\lambda''$  couplings can 
result in final states with four well-separated, high multiplicity 
hadron jets and a large amount of visible energy. 
The main  background comes from four-fermion
processes, dominantly  $\mathrm{ W^+W^- \rightarrow }$ q\=qq\=q, with some
contribution from
$\mathrm{q\overline{q}}$($\gamma$) events with hard  gluon  emission.

The analysis is similar to the one of~\cite{ref:opal_sfer_183}. 
It consists of a cut-based
preselection to reduce the two-photon and two-fermion backgrounds  and a
likelihood selection to suppress the contribution from \sm\ four-fermion
events.
The preselection consists of the following steps:
\begin{description}

\item[(1)] The event has to be classified as a multi-hadron final state as 
described in~\cite{ref:LEP2MH}.

\item[(2)] The effective centre-of-mass energy of the 
event, $\sqrt{s'}$~\cite{ref:sprime}, 
was required to be greater than $0.82 \times \sqrt{s}$ to reject events 
with large initial state radiation.

\item[(3)] To ensure that the events were well contained in the active
region of the detector, the   
visible energy should be larger than $0.7 \times \sqrt{s}$. 

\item[(4)] The events were forced into four jets using the Durham 
jet-finding algorithm, and rejected if the jet resolution parameter 
$y_{34}$ was less than 0.0025. Moreover, all jets must contain at least one
charged particle.

\item[(5)] A four-constraint kinematic fit (4C-fit)
requiring energy and momentum conservation for the jet four-momenta
should yield a $\chi^2$ probability larger than $10^{-5}$.

\item[(6)] To test the compatibility 
with pair-produced equal mass objects and
to obtain the best possible 
resolution for the reconstructed mass of a hypothetical sfermion,
the jet four-momenta were refitted requiring energy and momentum
conservation and equal jet pair invariant masses (5C-fit). 
The event was kept if at least one of the three jet pairing combinations
has a $\chi^2$ probability larger than $10^{-5}$.
The combination with the highest $\chi^2$ probability was considered later
in the mass reconstruction.

\item[(7)] A cut was applied on the $C$ event shape
parameter~\cite{ref:dpar} which provides an effective measure of the multi-jet
structure of the event, $C > 0.45$.
\end{description}

Finally, a likelihood selection was employed to classify the remaining events.
Three event classes were used: signal, two-fermion and four-fermion. 
The signal reference histograms were produced separately for selectrons
where $t$-channel production plays an important role, for
electron-sneutrinos where the $t$-channel process also contributes but to
a lesser extent and for second or third generation neutral and charged
sleptons with $s$-channel production only. Since squarks are coloured
particles they were treated separately. For a given centre-of-mass
energy, all the generated masses were used with equal weight to form the
reference distributions.
The following variables were used as inputs to the likelihood calculation:
\begin{list}{$\bullet$}{\itemsep=0pt \parsep=0pt \topsep=-5pt \leftmargin=30pt}
\item the cosine of the polar angle of the thrust axis;
\item 
the cosine of the smallest angle between the directions of any two of
the four reconstructed jets;
\item the difference between the largest and smallest jet energy
after the 4C-fit;
\item 
the smallest difference between the reconstructed masses of the two
jet pairs from any of the three possible jet pair combinations; 
\item 
the cosine of the direction of the jet pair momentum multiplied by the 
charge of the jet pair\footnote{
The charge of the jet pair was calculated as
$\Sigma q_{(i)} p_{{} \mathrm{L} (i)}^{0.5}$, 
where the sum goes over each track within the two jets, 
$q_{(i)}$ is the charge of the track and 
$p_{{} \mathrm{L} (i)} $  is its momentum parallel to the jet 
direction. 
A charge of +1 was assigned to the jet pair with the larger
charge, and a charge of $-1$ to the other.} 
for the combination with the highest $\chi^2$ probability given by
the 5C-fit.

\end{list}
Events were  accepted if their
likelihood output was larger than  0.5 for selectrons, 0.55  for
electron sneutrinos, 0.6 for second and third generation neutral and charged
sleptons and 0.7 for squarks.

\begin{figure}[htbp]
\centering
\includegraphics*[scale=0.5]{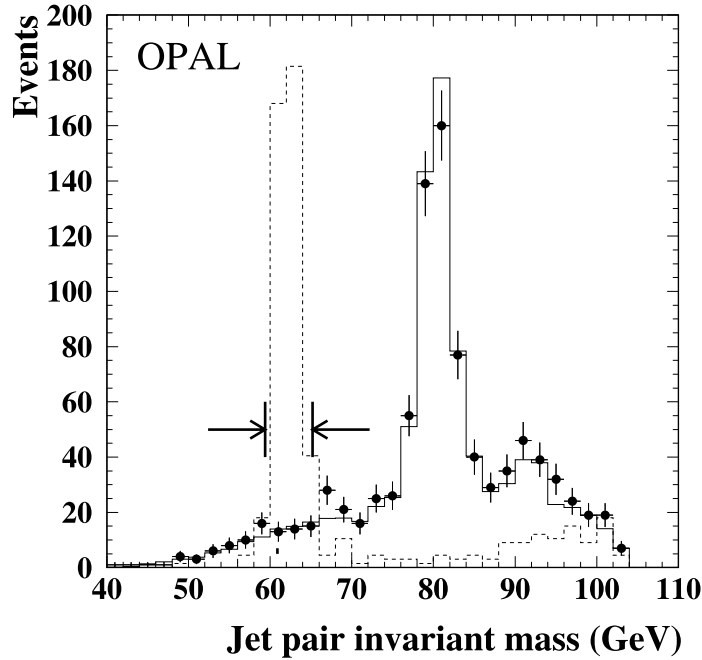}
\caption[]{\it
Final states with four jets and no missing energy, selectron selection
at $\sqrt{s}= $ 189-209~GeV:
distribution of the reconstructed mass
for the selected events in the data (points with error bars), 
the estimated background normalised to the integrated luminosity of the data
(solid line)
and a simulated sample of selectron pair-production for
$m_{\tilde{e}} = $ 62 GeV with arbitrary normalisation (dashed line). 
Uncertainties are statistical only. 
The arrows
indicate the mass window which would be selected for the specific case
of the signal shown.
}
\label{figure:4j}
\end{figure}

Table~\ref{table:4j} shows the numbers of selected data and 
expected background 
events for the different selections at centre-of-mass energies from 189~GeV  
to 209~GeV, corresponding to an integrated luminosity of 610.6~pb$^{-1}$. 
More than 92\% of the events expected from the \sm\ originate from four-fermion
processes. Figure~\ref{figure:4j} shows, as an example, 
the jet pair invariant mass
distribution of the selected events after the 5C-fit in the selectron
selection: the dominance of W$^+$W$^-$ production in the \sm\
background is clearly visible from the strongly peaked distribution around the
W boson mass. The jet pair invariant mass resolution for signal events 
is $\sigma \approx
0.6-1.6$~GeV, depending on the type and the mass of  the produced sparticle. 
To profit from the good mass resolution, events were selected if they  were 
in a
$\pm2\sigma$ mass window around the test mass.  The efficiencies in
Table~\ref{table:4j} are given
after the mass selection and all correspond to
the Yukawa coupling giving the worst  result.

\begin{table}[htbp]
  \centering
{ 
  \begin{tabular}[c]{|@{\hspace{1.0mm}}c@{\hspace{1.0mm}}|@{\hspace{1.0mm}}c@{\hspace{1.0mm}}|@{\hspace{1.0mm}}c@{\hspace{1.0mm}}|@{\hspace{1.0mm}}c@{\hspace{1.0mm}}|@{\hspace{1.0mm}}c@{\hspace{1.0mm}}|@{\hspace{1.0mm}}c@{\hspace{1.0mm}}|}
\hline  
  Physics process   & Eff. (\%) &  Data & Tot. bkg MC & Data $>$ 206~GeV & 
MC $>$ 206~GeV   \\\hline
$\tilde{e}^+ \tilde{e}^- \rightarrow 4 \ jets $ 
                    & 14 $-$ 46& 922  & 917 $\pm$ 10 $\pm$ 69 & 193 & 
210 $\pm$ 5 $\pm$ 38 \\
$\tilde{\nu}_e \overline{\tilde{\nu}}_e \rightarrow 4 \ jets $ 
                    & 15 $-$ 43& 760  & 762 $\pm$ 8 $\pm$ 56 & 158 & 
175 $\pm$ 4 $\pm$ 30 \\
$\tilde{\mu}^+ \tilde{\mu}^-, \tilde{\tau}^+ \tilde{\tau}^-,$ 
& & & & & \\
$\tilde{\nu}_\mu \overline{\tilde{\nu}}_\mu,
\tilde{\nu}_\tau \overline{\tilde{\nu}}_\tau \rightarrow 4 \ jets $ 
                     & 9 $-$ 39& 584  & 589 $\pm$ 6 $\pm$ 43 & 123 & 
138 $\pm$ 3 $\pm$ 24 \\
$\tilde{q} \overline{\tilde{q}} \rightarrow 4 \ jets $ 
                     & 9 $-$ 37& 393  & 393 $\pm$ 5 $\pm$ 29 & 81 & 
89 $\pm$ 2 $\pm$ 15\\
\hline 
\end{tabular}
}
\caption[]{\it
Final states with four-jets and no missing energy 
for $\sqrt{s} = $ 189 - 209~GeV and for 
$\sqrt{s} > $ 206~GeV:
signal selection efficiencies for sparticle masses above 45~GeV, 
observed numbers of events in the data and 
background estimates from \sm\ processes.  
The first uncertainty on the background estimate is statistical 
and the second is systematic.
The efficiencies are given after the mass selection described in the
text
while the data and the background correspond to the whole mass region.
The inefficiency related to the fragmentation model
was taken into account for the squark result.
There is a significant correlation amongst the different
selections at the same centre-of-mass energy.  
}
\label{table:4j}
\end{table}

\subsubsection*{Systematic uncertainties}

The systematic uncertainties 
were estimated in the same manner as in the search for
four jets and missing energy. For the selection efficiency the following 
sources of uncertainties were considered:
\begin{enumerate}
\item The limited MC statistics (3.1$-$11.9\%).
\item The statistical and systematic uncertainties on the luminosity measurement
(0.3$-$0.4\%).
\item The modelling of the preselection variables (15.9$-$18.2\%, 
largest contribution of 10.2$-$15.3\% from the $\chi^2$ probability 
of the 4C-fit).
\item The modelling of the variables used in the likelihood selection
(6.6$-$10.9\%).
\item The interpolation of efficiencies (1.6$-$3.2\%).
\end{enumerate}

The background estimate is subject to uncertainties due to

\begin{enumerate}
\item The limited MC statistics (1.1\%).
\item The statistical and systematic uncertainties on the luminosity measurement
(0.3$-$0.4\%).
\item The modelling of the preselection variables (13.2$-$16.3\%, 
largest contribution of 8.7$-$13.3\% from the 
$\chi^2$ probability of the 4C-fit).
\item The modelling of the variables used in the likelihood selection
(4.5$-$10.7\%).
\item The imperfect knowledge of \sm\ cross-sections (2\%).
\item The imperfect simulation of four-fermion processes (3.3$-$6.4\%).
\end{enumerate}

The total systematic uncertainty excluding MC statistics ranges from 17.3 to 21.5\%
on the signal detection efficiency and from 16.4 to 18.3\% on the background
rate.

As explained in Section~\ref{sec:MC}, the squark pair Monte Carlo samples
were produced using an
independent fragmentation model which gives narrower, better separated jets
and therefore optimistic selection efficiencies. 
To estimate this effect, the efficiencies were compared for smuon Monte Carlo 
samples with independent and string fragmentation models. From this study
an additional relative inefficiency of 7.9\% to 36.9\% was derived, 
by which the squark
detection efficiencies were decreased. 
For the same reason, the squark results at 183~GeV~\cite{ref:opal_sfer_183} 
were updated using a similar
procedure and these new results are used later in the 
combination.

\section{Interpretation}
\label{sec:results}

No significant excess of signal-like events was observed in the 
data with respect to the expected background
for any analysis listed in Table~\ref{tab:relation}.
Production cross-section and mass limits were therefore computed
using the data collected at LEP
from centre-of-mass energies ($\sqrt{s}$) between 189~GeV and
209~GeV and then combined with 
results obtained using data at lower 
energies~\cite{ref:opal_sfer_183}.
These limits take into account 
indirect limits obtained from the study of the Z$^0$ width at LEP1 and 
therefore concern only sparticle masses above 45 GeV.
All limits presented here are quoted at the 95\% confidence level.

Two approaches were used to present sfermion production limits.
In the first, upper limits
on the production cross-sections were calculated as functions of the sfermion 
masses with minimal model assumptions.
These upper limits in general do not depend
on the details of the SUSY models, except for the assumptions 
that the sparticles are pair-produced and that only one \lb-like coupling
at a time is non-zero~\cite{ref:opal_sfer_183}.
In the second approach, limits on the sfermion masses were calculated 
in an \Rparity\ violating framework analogous to the CMSSM, 
where mass limits were derived 
for $\tan \beta=1.5$
and $\mu = -200$~GeV.
This choice of parameters, 
which generally results in small sfermion
production cross-sections,  
is a convenient benchmark for limit setting and is used 
by other collaborations, so results may be compared and combined. 
For the indirect sfermion decays, we used the branching ratios
for the decay $\tilde{f} \rightarrow f \nt_1$ predicted 
by the CMSSM, and we conservatively assumed no 
experimental sensitivity to any other decay mode.
The branching ratio for direct decay was always assumed to be unity,
as 
only one \lb\ coupling at a time was allowed to be different from zero.

As in \cite{ref:opal_sfer_183}, the relative branching ratios
of the neutralino into a final state with a charged or a neutral
lepton were varied between 0 and 1
to avoid a dependence of the results on the CMSSM parameters. 
A likelihood ratio method~\cite{likelihood} was used 
to determine an upper limit for the 
cross-section. 
This results in a cross-section limit as a function of the 
branching ratio and the sfermion mass. A limit independent of the
branching ratio was determined by taking the 
lowest limit at each sfermion mass. 
For the direct decays, the final states are fully
determined by the indices of the coupling considered. 

In the following sections, cross-section limits are shown for the
various direct and indirect decays studied in this paper and are listed in 
Table~\ref{tab:relation}. 
In each cross-section plot for the decays via \lb\ couplings,
the curves corresponding to the highest and the lowest 
cross-section limits are 
shown. 
Generally, the highest excluded cross-section comes from final states
with a maximum number of muons and no taus, while the worst results come from
final states with many taus, due to their lower detection efficiency.
In the other cases, only the curve corresponding to the 
worst cross-section limit is
shown amongst all possible cross-section limits resulting from the
couplings considered. 

In the CMSSM framework, the exclusion regions for the indirect
decays are valid 
for $m_{\nt_1} \ge 10$~GeV, to ensure prompt decays.
The region in the plane ($m_{\sell}$, $m_{\nt_1}$) 
corresponding to  $m_{\nt_1} < 10$~GeV, 
where the lifetime of 
sparticles would be sufficiently long 
to produce a secondary decay vertex, clearly detached 
from the primary vertex, or even outside the detector, is labelled 
``Lifetime signature'' and is not excluded by these searches.  

For indirect sparticle decays, mass lower limits are  
quoted for $\nt_1= 10$~GeV, referred to as a low-mass 
$\nt_1$.
In the small $\Delta m$ region
($\Delta m = m_{\sell} - m_{\nt_1} < 5$~GeV), the final state charged leptons 
resulting from the charged slepton decay into a charged lepton and a
neutralino may not be detected due to the 
small phase space available. 
This region is excluded for the charged sleptons decaying 
indirectly via a \lb\ coupling. 
For the charged sleptons decaying indirectly via a \lbp\ 
coupling, no exclusion was possible in this region. 
The exclusion region for the direct decays is independent of  
$\Delta m$. 

The production cross-section for left-handed charged sleptons is always 
larger than that for right-handed charged sleptons; therefore, 
whenever applicable for the superpotential form given 
in Equation~(\ref{lagrangian}), 
results are conservatively quoted for right-handed charged sleptons.

\subsection{Selectron limits}

\begin{figure}[htbp]
\centering
\begin{tabular}{c}
\epsfig{file=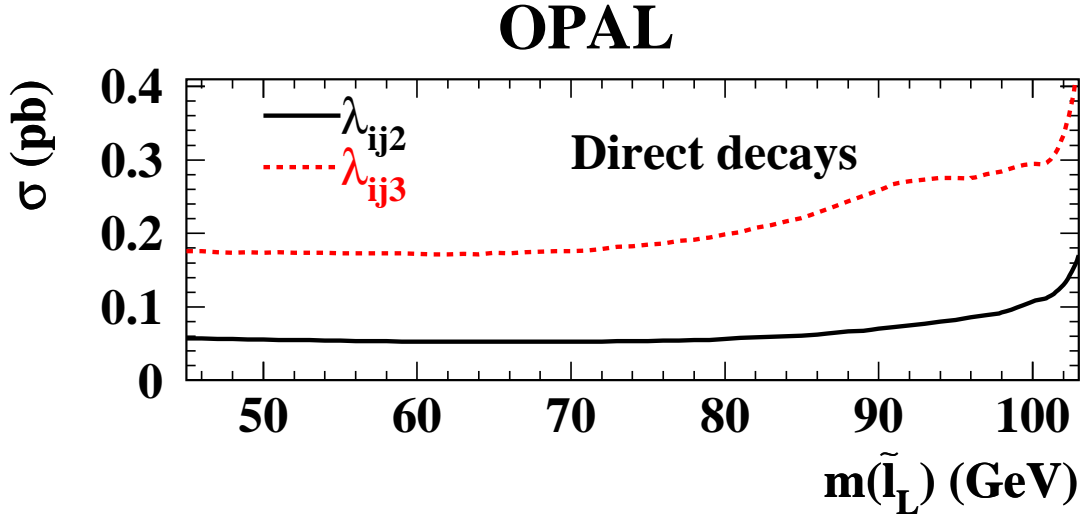       ,width=15.0cm} \\
\end{tabular}
\caption[]{\it
Charged slepton direct decays via a \lb\ coupling: 
upper limits at the 95\% C.L.
on the pair-production cross-sections 
of left-handed sleptons decaying directly.  The dashed line shows 
the lowest upper limit ($\tau \nu \tau \nu$ final states) while the 
solid line shows the highest one ($\mu \nu \mu \nu$ final states). 
}
\label{fig:cross_selectron_dir_lb}
\end{figure}

Upper limits on the cross-sections of pair-produced selectrons followed
by the direct decay via a \lb\ coupling are shown 
in Figure~\ref{fig:cross_selectron_dir_lb}.  
Due to the structure of the $LL{\overline E}$ term and 
the presence of a neutrino in the decay products, only
left-handed sleptons can decay via this term, the other decays being 
suppressed. 
Since the selection efficiency for this analysis only depends on 
the final states and not on the parent slepton type, this limit also 
holds for the pair production of smuons and staus decaying directly 
via a \lb\ coupling.

\begin{figure}[htbp]
\centering
\begin{tabular}{c}
\epsfig{file=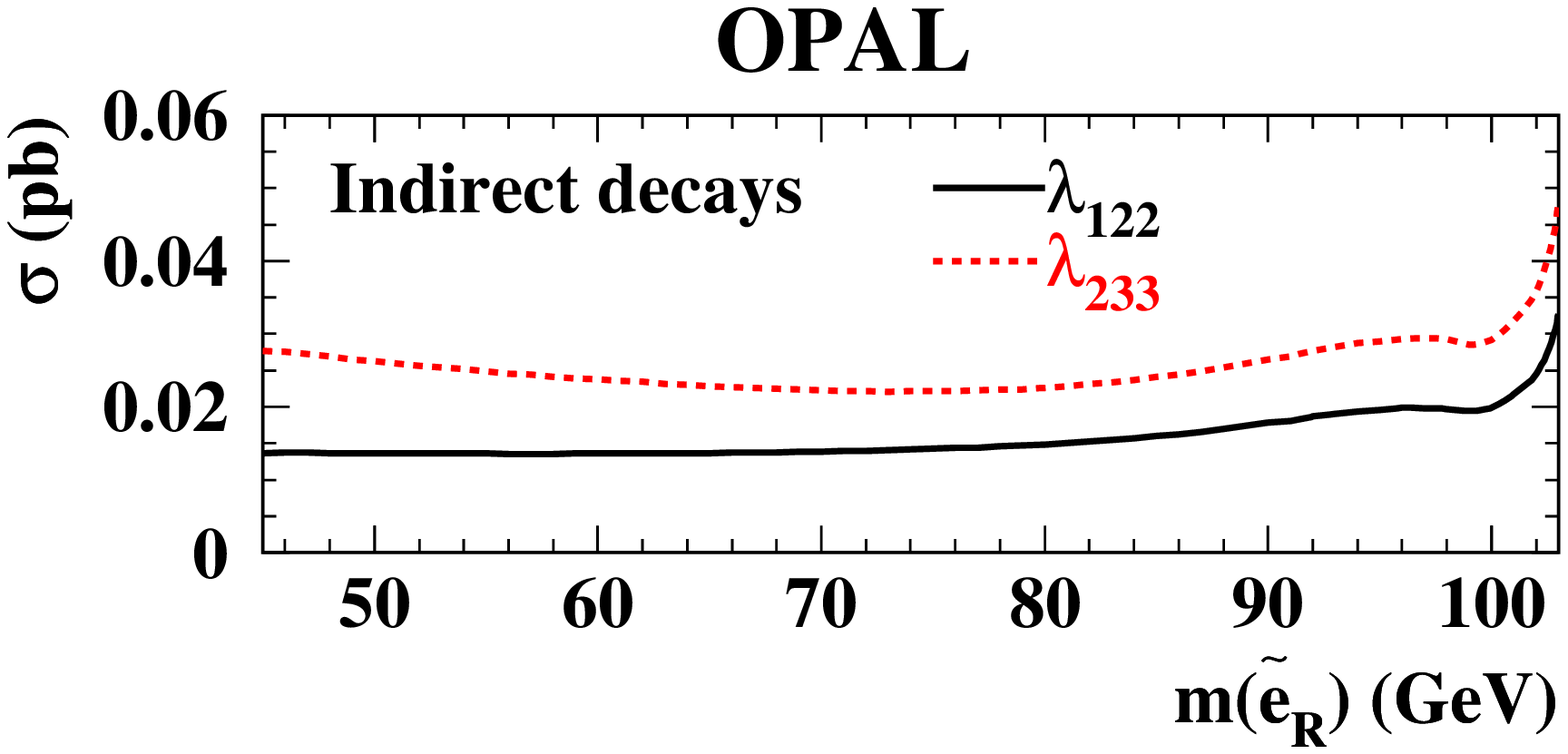       ,width=15.0cm} \\
\end{tabular}
\caption[]{\it
Selectron indirect decays via a \lb\ coupling: 
upper limits at the 95\% C.L.
on the pair-production 
cross-sections 
of a right-handed $\sele_R$ decaying indirectly.
The dashed line shows 
the lowest upper limit ($e \tau \tau \nu e\tau \tau \nu$ final state) 
while the solid line shows the highest one ($e \mu \mu \nu e \mu \mu \nu$
final state). 
}
\label{fig:cross_selectron_ind_lb}
\end{figure}

Upper limits on the cross-sections of pair-produced selectrons decaying 
indirectly via a \lb\ coupling are shown in Figure~\ref{fig:cross_selectron_ind_lb}. 

\begin{figure}[htbp]
\centering
\begin{tabular}{c}
\epsfig{file=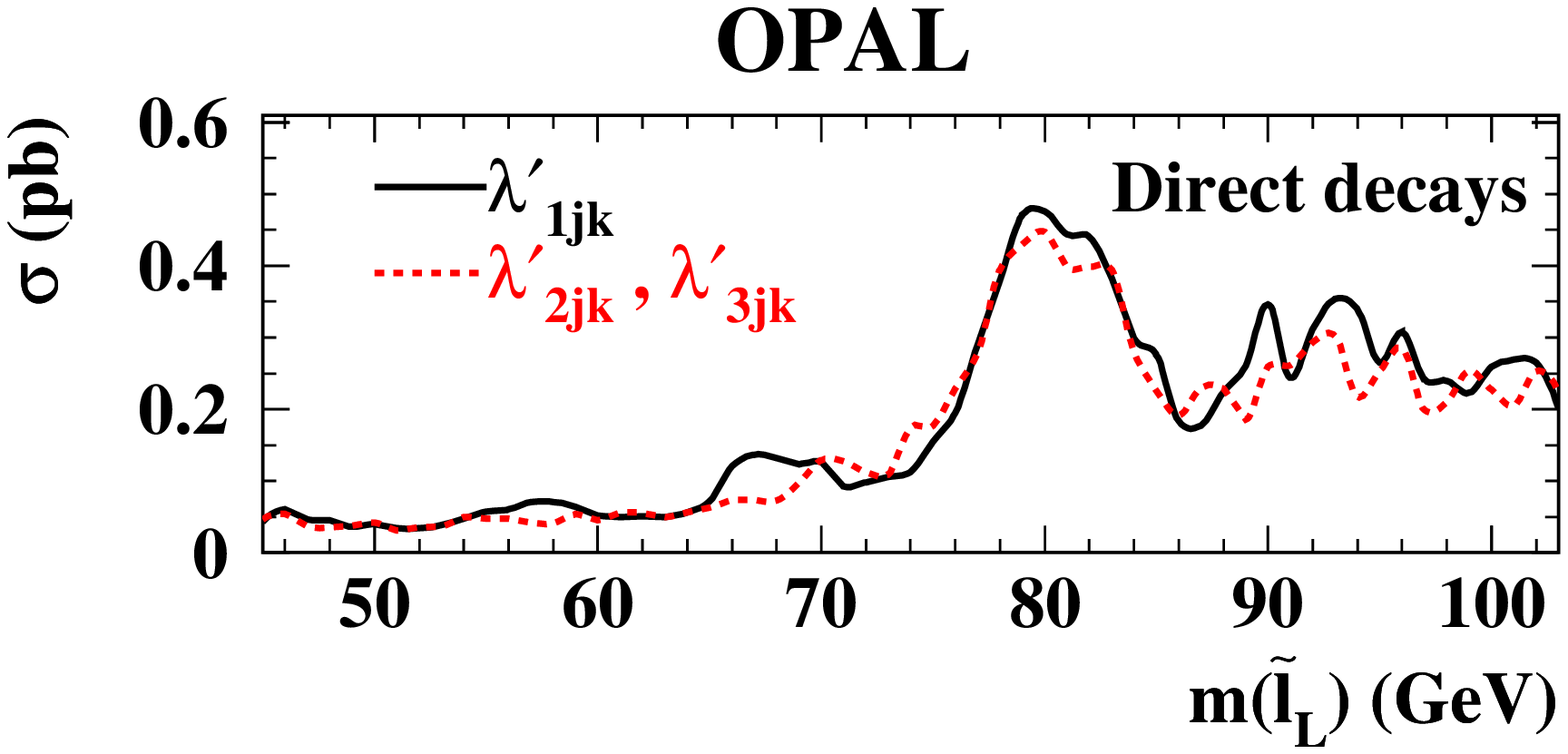,width=15.0cm} \\
\end{tabular}
\caption[]{\it
Charged slepton direct decays via a \lbp\ coupling:
upper limits at the 95\% C.L. 
on the pair-production cross-sections of $\sele$ (solid line)
and $\smu$/$\stau$ (dashed line) decaying directly. 
} 
\label{fig:cross_sel_4jets}
\end{figure}

Upper limits on the cross-sections of pair-produced selectrons 
decaying directly via a \lbp\ coupling to a four-jet final state
are shown in Figure~\ref{fig:cross_sel_4jets}. The peak 
structure visible in the figure at approximately the mass of the 
W-boson comes from irreducible background due to W pair-production.

\begin{figure}[htbp]
\centering
\begin{tabular}{c}
\epsfig{file= 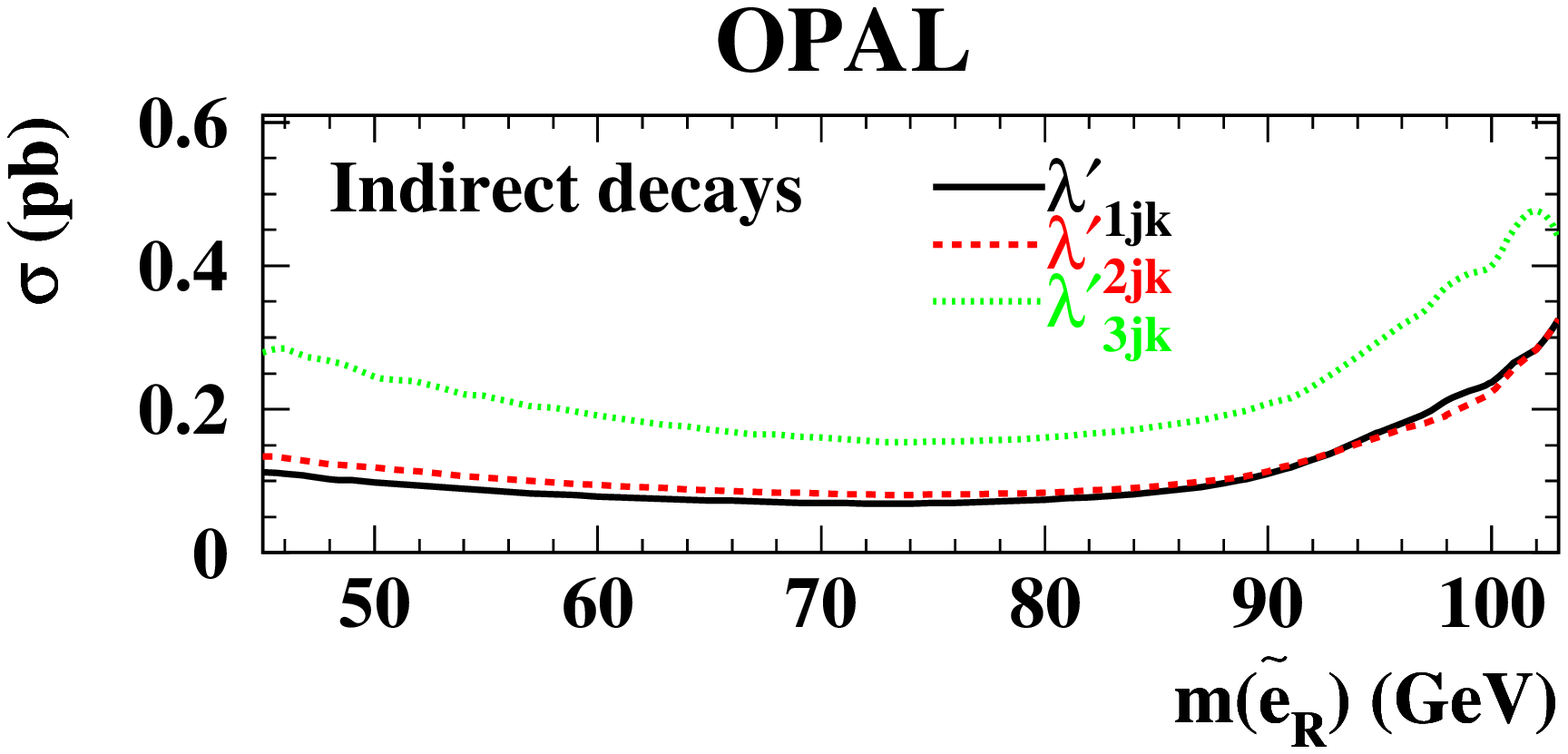    ,width=15.0cm} \\
\end{tabular}
\caption[]{\it
Selectron indirect decays via a \lbp\ coupling:
upper limits at the 95\% C.L. on the pair-production cross-sections
the of a $\sele_R$ decaying 
indirectly in the electron channel (solid line), 
in the muon channel (dashed line) and
in the tau channel (dotted line). 
} 
\label{fig:cross_selectron_lbp}
\end{figure}

Upper limits on the cross-sections of pair-produced selectrons decaying 
indirectly via a \lbp\ coupling are shown in Figure~\ref{fig:cross_selectron_lbp}
for 
the indirect decay of a $\sele_R$ in the electron channel, 
in the muon channel and
in the tau channel. 

\begin{figure}[htbp]
\centering
\begin{tabular}{cc}
\epsfig{file=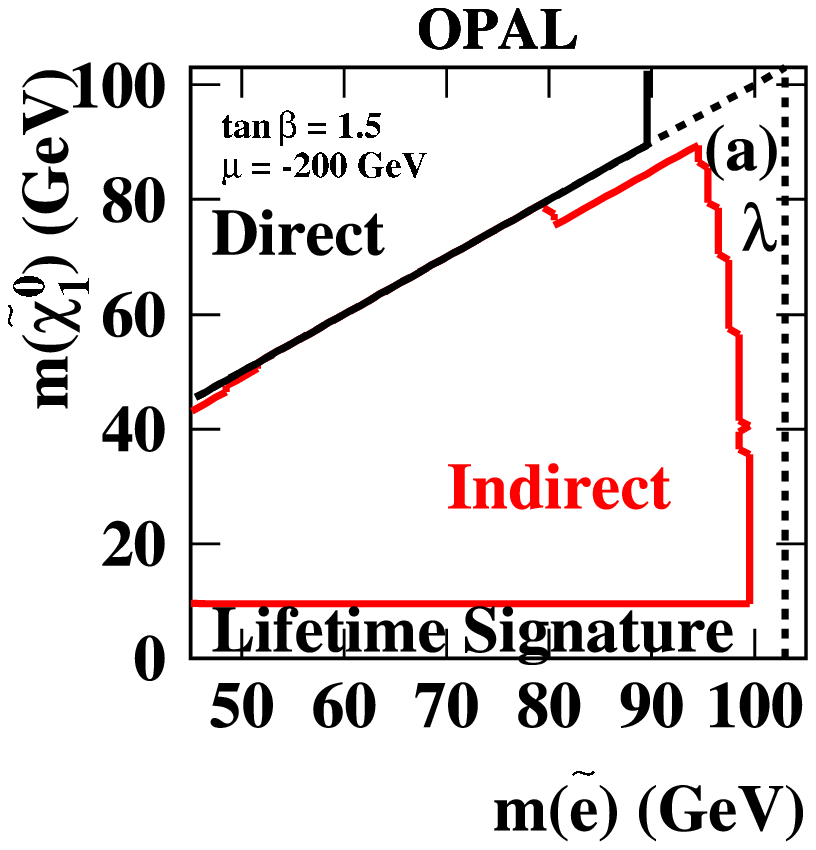  ,width=8.0cm} &
\epsfig{file=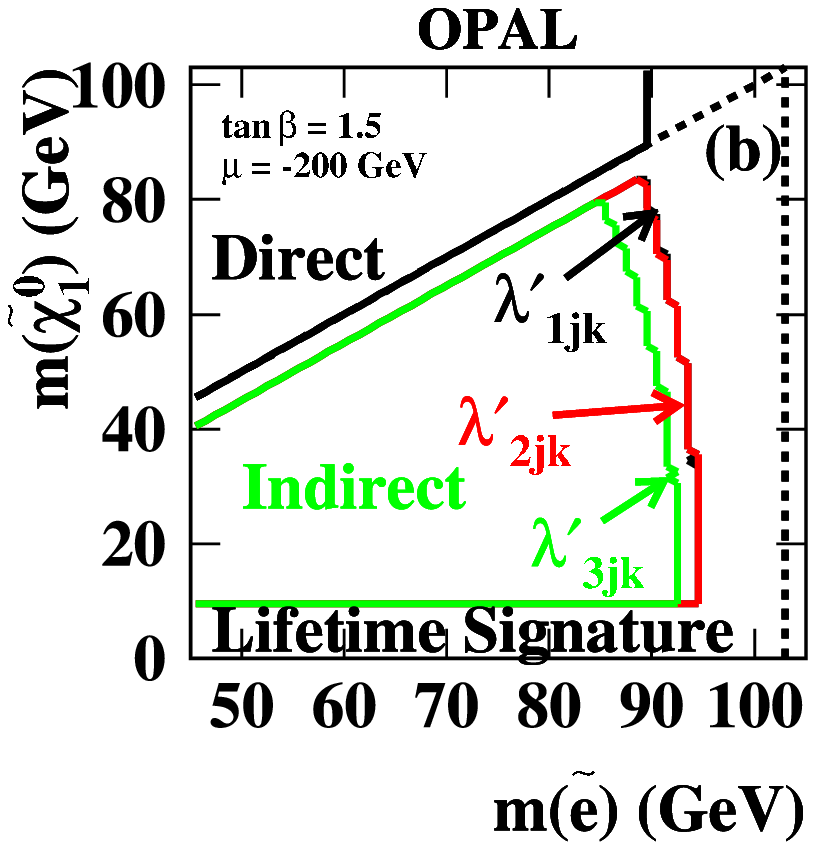, width=8.0cm} \\
\end{tabular}
\caption[]{\it
CMSSM exclusion region for $\sele^+ \sele^-$ 
production in 
the $(m_{\sele}, m_{\nt_1})$ plane at 95\% C.L.  
for  (a) a \lb\ coupling and (b) a \lbp\ coupling. The limits for \lbp\ couplings are presented separately according to the first index.
For the direct decays, the exclusion is shown for the only 
possible case of $\sele_L \sele_L$. 
The kinematic limit is shown by the dashed lines. 
}
\label{fig:mssm_selectron}
\end{figure}

In the CMSSM, the selectron pair-production cross-section
is enhanced by the presence of a $t$-channel diagram.
The exclusion limits for selectrons
decaying via a \lb\ coupling are shown in Figure~\ref{fig:mssm_selectron}(a). 
In the region where the neutralino is heavier than
the selectron, only direct decays are possible. 
When the neutralino is lighter than the selectron, the
indirect decays are expected to be dominant.
The exclusion refers to right-handed 
selectrons for the indirect decays and to left-handed selectrons for 
direct decays.  

The exclusion limits for selectrons
decaying via a \lbp\ coupling are shown in Figure~\ref{fig:mssm_selectron}(b). 
The lower mass limits for selectrons decaying directly or indirectly
via a \lb\ or a \lbp\ coupling are included in 
Table~\ref{tab:slepmasslimits}.
The mass limits for the indirect decays are quoted for a 
low-mass $\nt_1$ (10~GeV).

\begin{table}[htbp]
\begin{center}
\begin{tabular}{|c||c|c||c|c|}
\hline
\multicolumn{5}{|c|}{Charged Slepton Lower Mass Limits(GeV)} \\
\hline
 Species &   \multicolumn{2}{c|}{$\lambda$}    
                 &   \multicolumn{2}{c|}{$\lambda^{'}$}   \\
\hline
                 &    Direct & Indirect   &    Direct & Indirect \\
\cline{2-3}
\cline{4-5}
$\sele$          &    89     &   99       &      89   &   92     \\
$\smu$           &    74     &   94       &      75   &   87     \\
$\stau$          &    74     &   92       &      75   &   -     \\
\hline
\end{tabular}
\end{center}
\caption{\it
Lower mass limits for charged sleptons decaying directly or indirectly
via a \lb\ or a \lbp\ coupling. 
The mass limits for indirect decays are quoted for a 
low-mass $\nt_1$ (10~GeV). 
The limits refer to right-handed 
charged sleptons for the indirect decays and to left-handed charged sleptons 
for 
direct decays.  
}
\label{tab:slepmasslimits}
\end{table}

\subsection{Smuon limits}

Upper limits on the cross-sections of pair-produced smuons 
decaying directly via a \lb\ coupling are shown 
in Figure~\ref{fig:cross_selectron_dir_lb}.  

\begin{figure}[htbp]
\centering
\begin{tabular}{c}
\epsfig{file=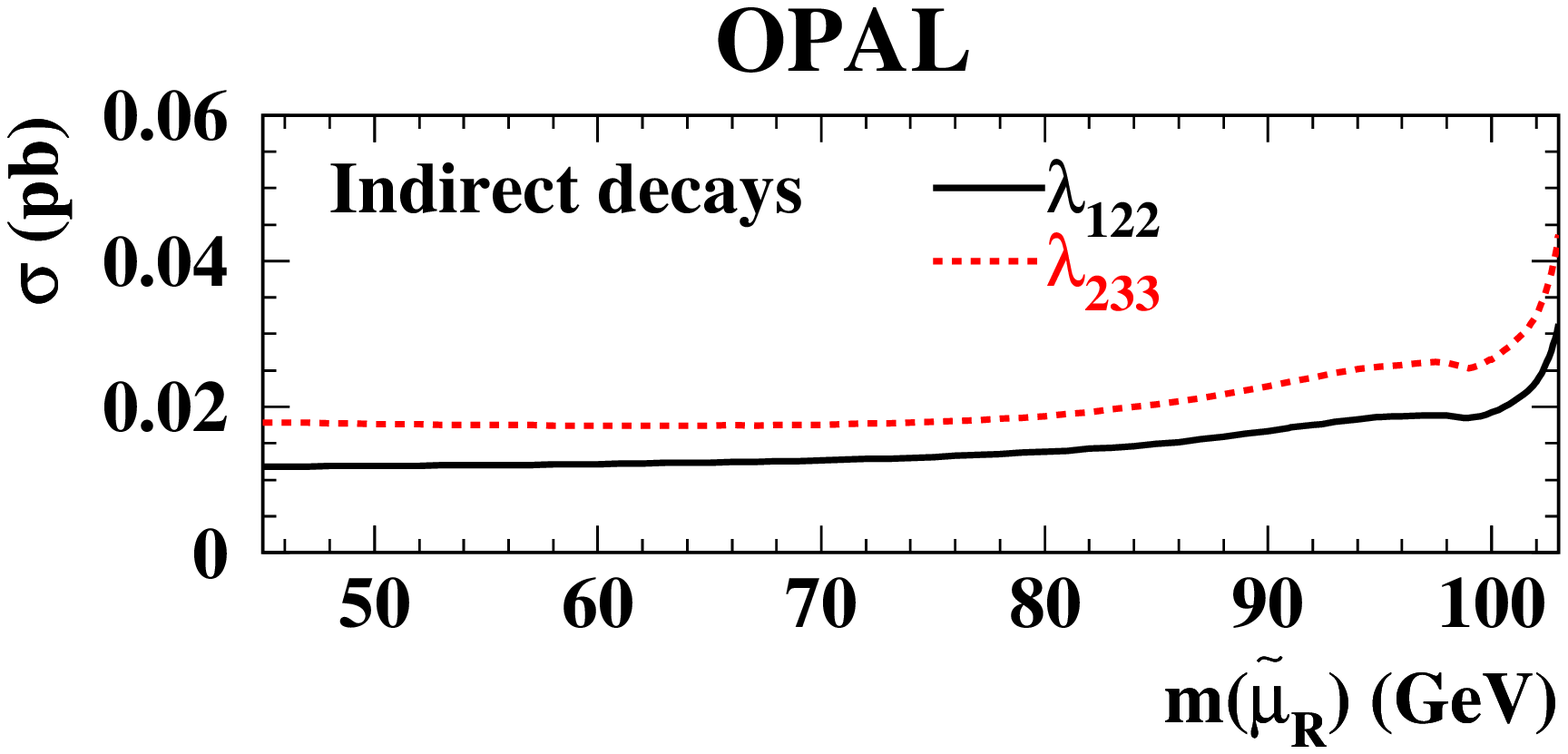       ,width=15.0cm} \\
\end{tabular}
\caption[]{\it
Smuon indirect decays via a \lb\ coupling: 
upper limits at the 95\% C.L.
on the pair-production 
cross-sections 
of right-handed $\smu_R$ decaying indirectly.
The dashed line shows 
the lowest upper limit ($\mu \tau \tau \nu \mu \tau \tau \nu$ final state) 
while the solid line shows the highest one ($\mu  \mu \mu \nu \mu \mu \mu \nu$
final state). 
}
\label{fig:cross_smuon_ind_lb}
\end{figure}

Figure~\ref{fig:cross_smuon_ind_lb} shows 
upper limits on the cross-sections of pair-produced smuons followed
by an indirect decay via a \lb\ coupling.

\begin{figure}[htbp]
\centering
\begin{tabular}{c}
\epsfig{file=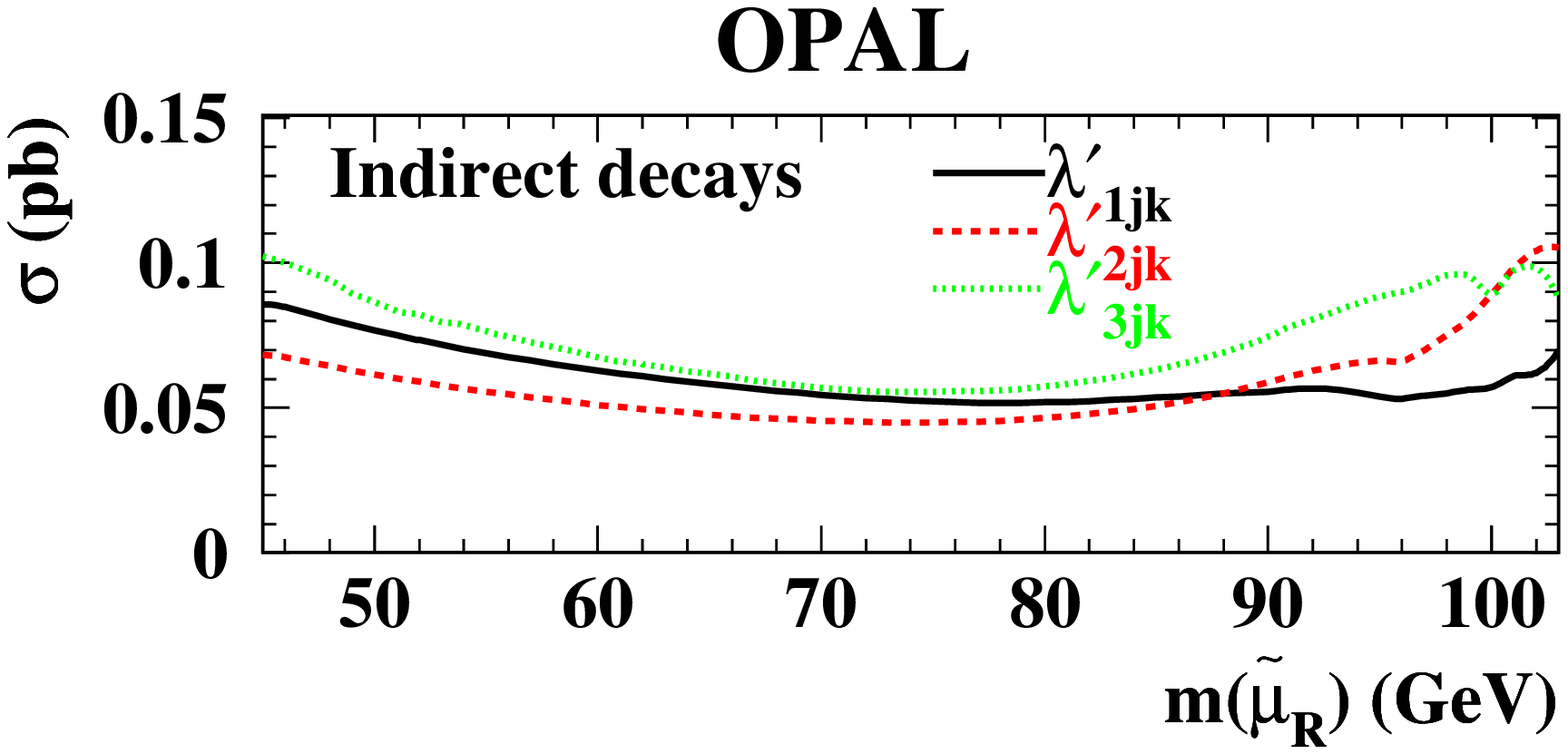     ,width=15.0cm} \\
\end{tabular}
\caption[]{\it
Smuon indirect decays via a \lbp\ coupling:
upper limits at the 95\% C.L. on the pair-production cross-sections of 
$\smu_R$ decaying indirectly in the electron channel (solid line), 
in the muon channel (dashed line) and
in the tau channel (dotted line). 
} 
\label{fig:cross_smuon_lbp}
\end{figure}

Upper limits on the cross-sections for pair-produced smuons  
decaying indirectly via a \lbp\ coupling are shown in Figure~\ref{fig:cross_smuon_lbp}.

Upper limits on the cross-sections of pair-produced smuons 
decaying directly via a \lbp\ coupling to a four-jet final state are shown 
in Figure~\ref{fig:cross_sel_4jets}.

\begin{figure}[htbp]
\centering
\begin{tabular}{cc}
\epsfig{file=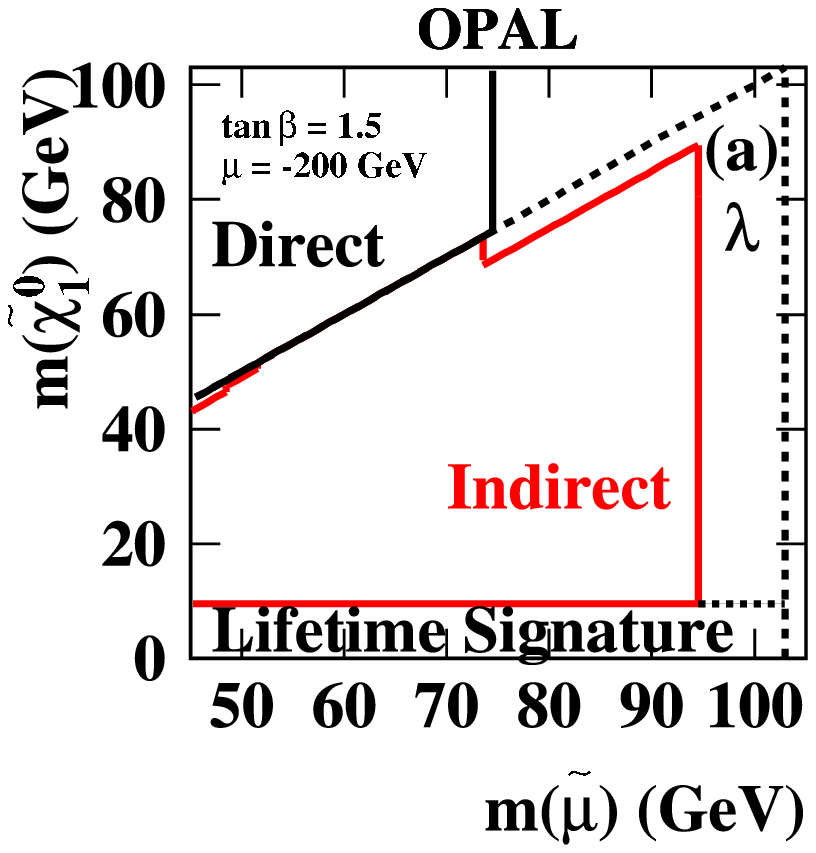, width=8.0cm} 
\epsfig{file=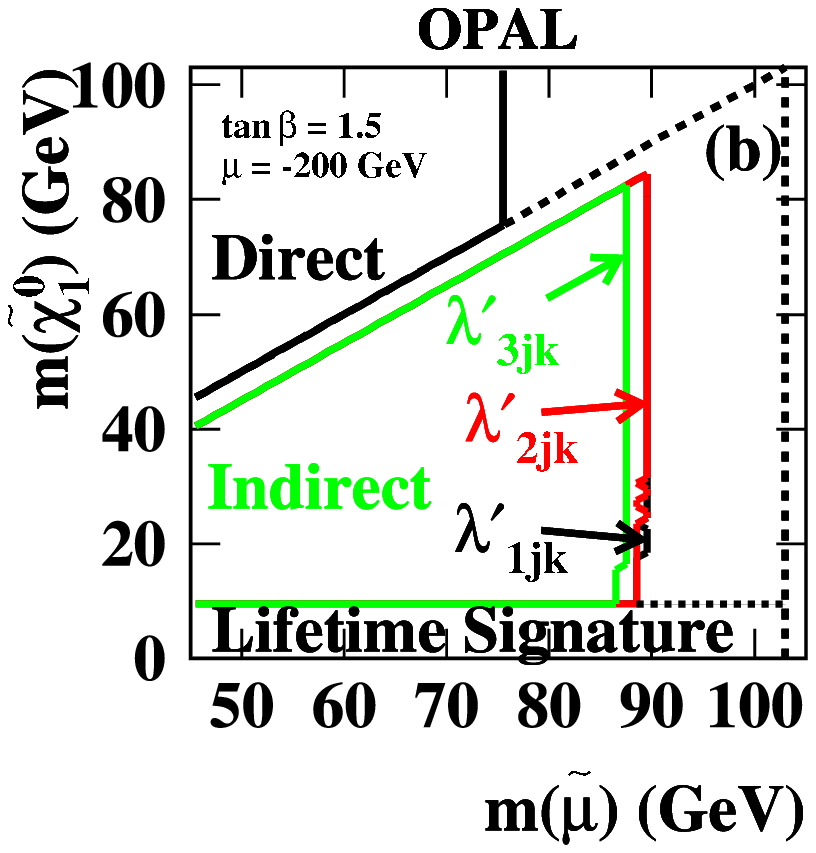, width=8.0cm} 
\end{tabular}
\caption[]{\it
CMSSM exclusion regions for $\smu^+ \smu^-$ 
production in 
the $(m_{\smu}, m_{\nt_1})$ plane at 95\% C.L.  
for  (a) a \lb\ coupling and (b) a \lbp\ coupling.
The limits for \lbp\ couplings are presented separately according 
to the flavour of the lepton in the decay.
For the direct decays, the exclusion region is shown for 
the case $\smu_L \smu_L$. 
The kinematic limit is shown by the dashed lines.}
\label{fig:mssm_smuon}
\end{figure}

The exclusion limits for smuons
decaying via a \lb\ coupling are shown in Figure~\ref{fig:mssm_smuon}(a). 
The exclusion limits for smuons
decaying via a \lbp\ coupling are shown in Figure~\ref{fig:mssm_smuon}(b). 
The lower mass limits for smuons decaying directly or indirectly
via a \lb\ or a \lbp\ coupling are included in 
Table~\ref{tab:slepmasslimits}.


\subsection{Stau limits}

Upper limits on the cross-sections of pair-produced staus 
decaying directly via a \lb\ coupling are shown 
in Figure~\ref{fig:cross_selectron_dir_lb}.  

\begin{figure}[htbp]
\centering
\begin{tabular}{c}
\epsfig{file=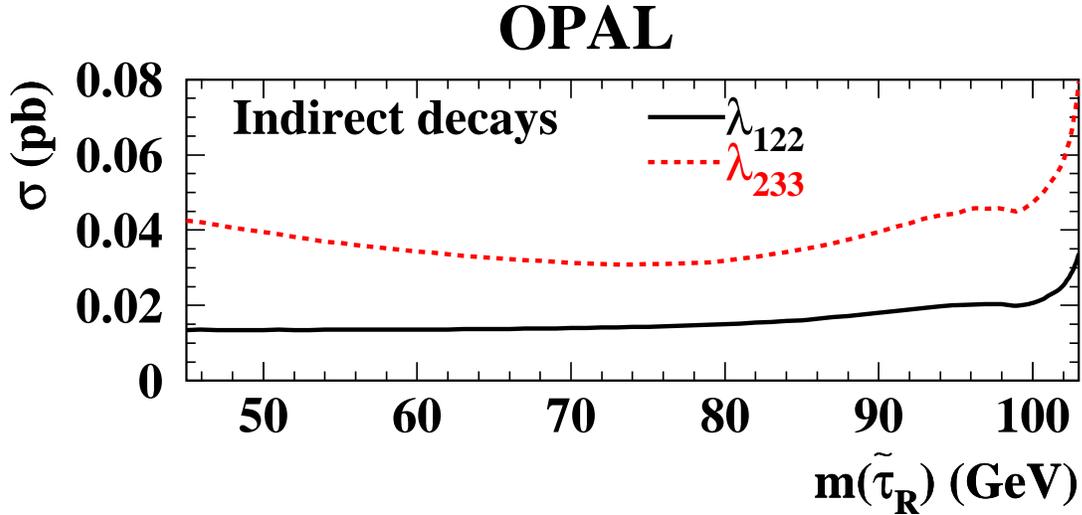       ,width=15.0cm} \\
\end{tabular}
\caption[]{\it
Stau indirect decays via a \lb\ coupling: 
upper limits at the 95\% C.L.
on the pair-production 
cross-sections 
of right-handed $\stau_R$ decaying indirectly.
The dashed line shows 
the lowest upper limit ($\tau \tau \tau \nu \tau \tau \tau \nu$ final state),
while the solid line shows the highest 
one ($\tau  \mu \mu \nu \tau \mu \mu \nu$
final state). 
}
\label{fig:cross_stau_ind_lb}
\end{figure}

Upper limits on the cross-sections of pair-produced staus 
decaying indirectly via a \lb\ coupling are shown in 
Figure~\ref{fig:cross_stau_ind_lb}. 

\begin{figure}[htbp]
\centering
\begin{tabular}{c}
\epsfig{file=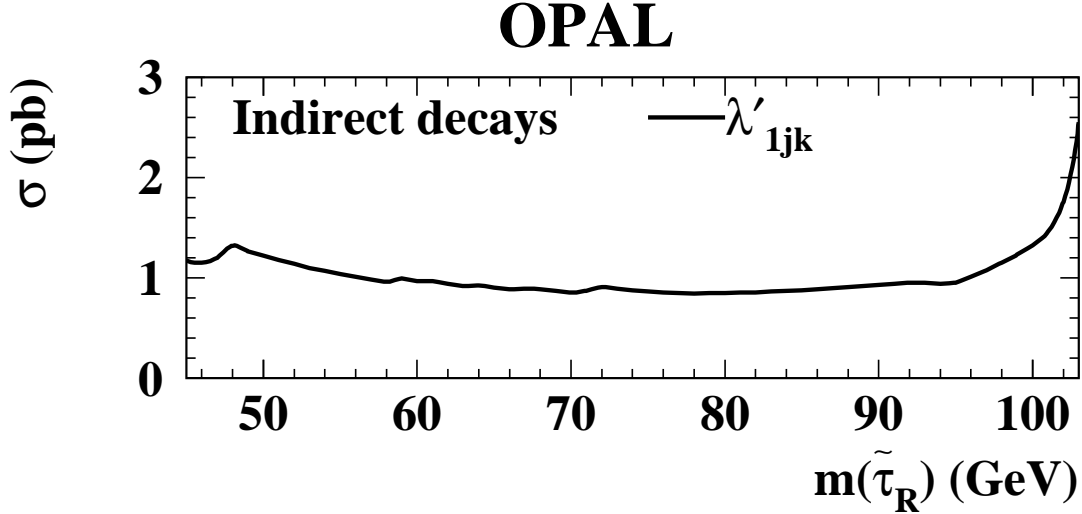     ,width=15.0cm} \\
\end{tabular}
\caption[]{\it
Stau indirect decays via a \lbp\ coupling:
upper limit on the pair-production cross-sections
for the indirect decay of a $\stau_R$ in the electron
channel. 
The indirect decay of a $\stau_R$ in the muon channel and
the indirect decay of a $\stau_R$ in the tau channel yield identical 
results.
}
\label{fig:cross_stau_lbp}
\end{figure}

Figure~\ref{fig:cross_stau_lbp} shows 
upper limits on the cross-sections for pair-produced staus 
decaying indirectly via a \lbp\ coupling in the electron channel. 
Since the neutralino can decay into final states with either a
charged lepton or a neutrino, limits are set by
varying this branching ratio for the neutrino
final state between 0 and 1. For the stau decay in the electron,
muon and tau channels, the lowest limits arise in all cases for
the neutrino branching ratio of 1. This means that the limits
for the muon and tau channels are identical to the electron channel
limit shown.

Upper limits on the cross-sections of pair-produced staus 
decaying directly via a \lbp\ coupling to a four-jet final state
are shown in Figure~\ref{fig:cross_sel_4jets}.

\begin{figure}[htbp]
\centering
\begin{tabular}{cc}
\epsfig{file=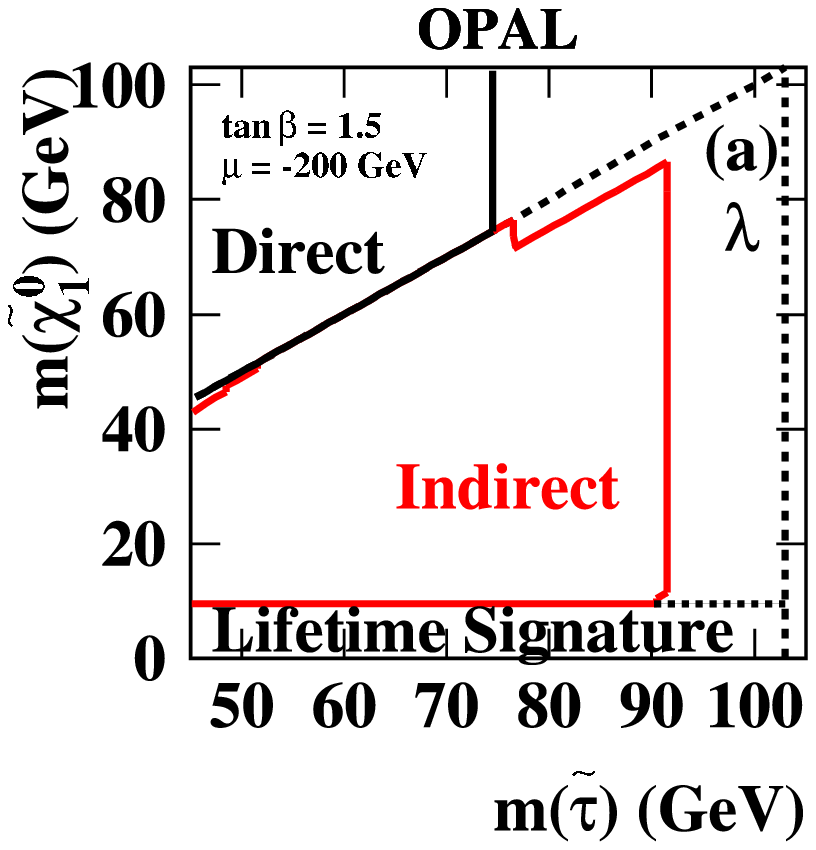, width=8.0cm} &  
\epsfig{file=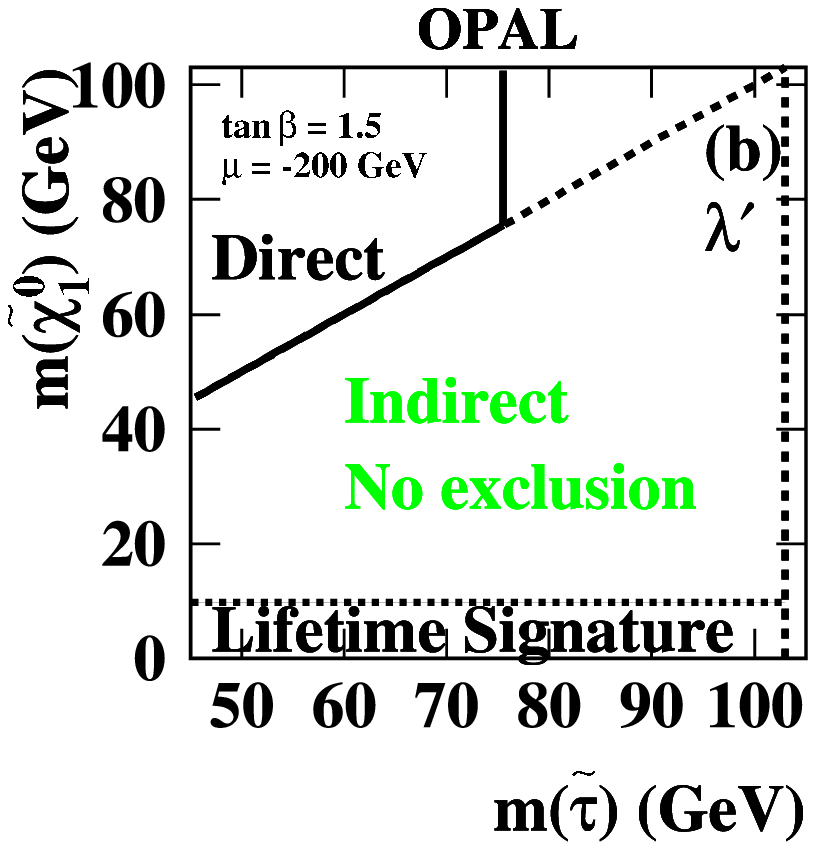, width=8.0cm} \\ 
\end{tabular}
\caption[]{\it
CMSSM exclusion region for $\stau^+ \stau^-$ 
production in 
the $(m_{\stau}, m_{\nt_1})$ plane at 95\% C.L.  
for  (a) a \lb\ coupling
and (b) a \lbp\ coupling.
For indirect decays via a \lbp\ coupling, 
no exclusion was possible since the excluded experimental cross-section is 
always larger than the theoretical cross-sections. 
For direct decays, the exclusion region for $\stau_L 
\stau_L$ is shown. 
The kinematic limit is shown by the dashed lines.  }
\label{fig:mssm_stau}
\end{figure}

The exclusion limits 
in the  $(m_{\stau}, m_{\nt_1})$ plane 
for staus
decaying via a \lb\ coupling are shown in Figure~\ref{fig:mssm_stau}(a). 
The exclusion limits for staus
decaying via a \lbp\ coupling are shown in Figure~\ref{fig:mssm_stau}(b). 
The lower mass limits for staus decaying directly or indirectly
via a \lb\ or a \lbp\ coupling are included in 
Table~\ref{tab:slepmasslimits}.
For indirect decays via a \lbp\ coupling, 
no exclusion was possible since the theoretical cross-sections were
smaller than the experimental limits in all cases. 

\subsection{Sneutrino limits}

\begin{figure}[htbp]
\centering
\begin{tabular}{c}
\epsfig{file=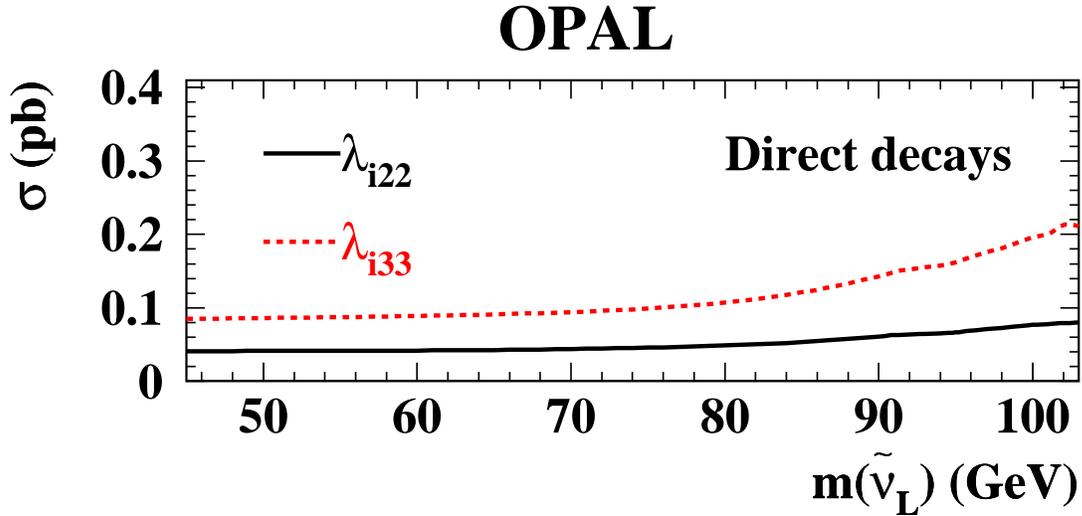       ,width=15.0cm} \\
\end{tabular}
\caption[]{\it
Sneutrino direct decays via a \lb\ coupling: upper limits at the 95\% C.L.
on the pair-production cross-sections 
of sneutrinos decaying directly.  The dashed line shows 
the lowest upper limit (final states with four taus), while the 
solid line shows the highest one (final states with four muons). 
}
\label{fig:cross_sneutrino_dir_lb}
\end{figure}

Upper limits on the cross-sections of pair-produced sneutrinos 
decaying directly via a \lb\ coupling are shown in 
Figure~\ref{fig:cross_sneutrino_dir_lb}.  

\begin{figure}[htbp]
\centering
\begin{tabular}{c}
\epsfig{file=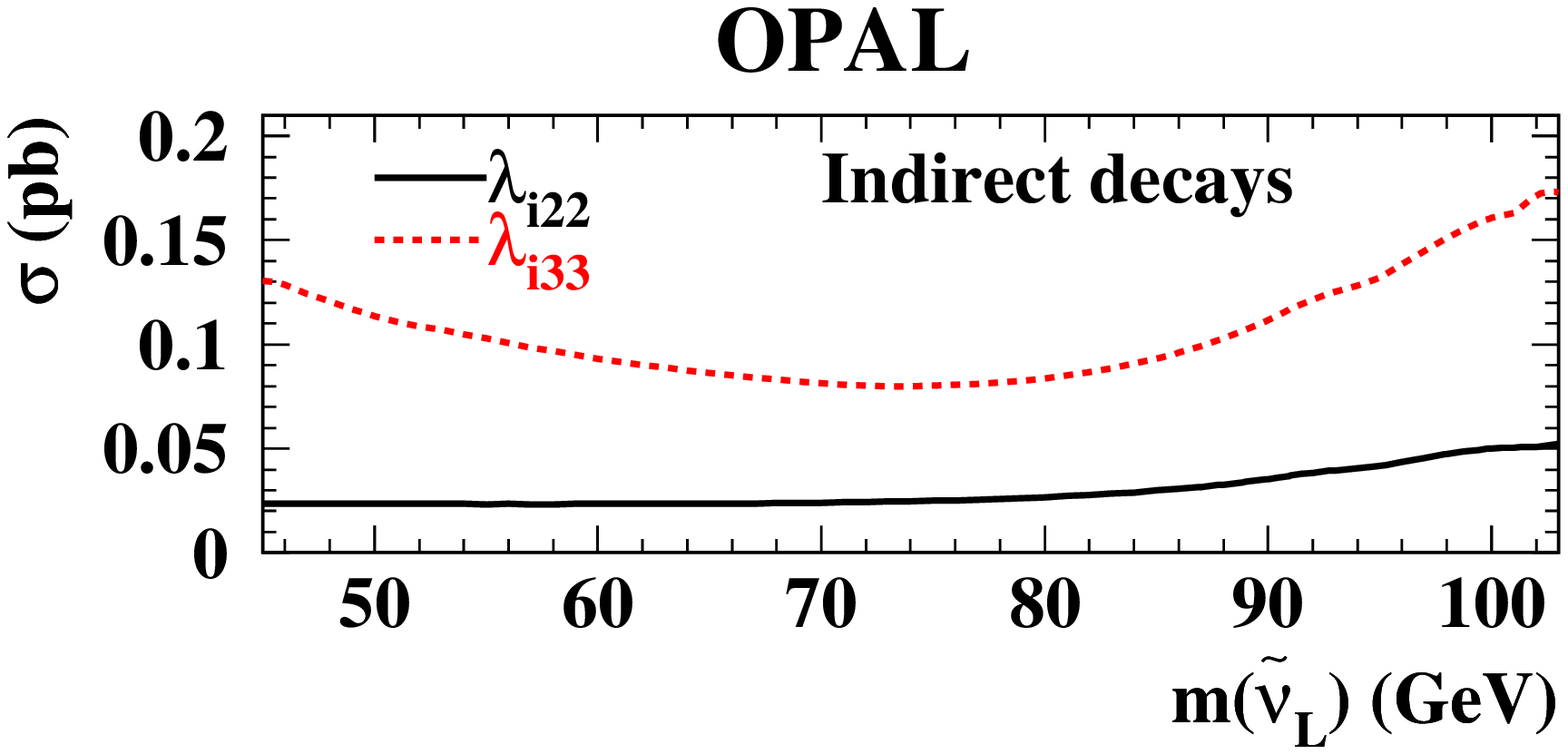       ,width=15.0cm} \\
\end{tabular}
\caption[]{\it
Sneutrino indirect decays via a \lb\ coupling: upper limits at the 95\% C.L.
on the pair-production 
cross-sections 
of sneutrinos decaying indirectly, 
for $\Delta m \leq m_{\snu}/2$.  The dashed line shows 
the lowest upper limit (final states with 4 $\tau$'s and missing energy) 
while the solid line shows the highest one (final states with 4 $\mu$'s 
and missing energy). 
}
\label{fig:cross_sneutrino_ind_lb}
\end{figure}

Upper limits on the cross-sections of pair-produced sneutrinos 
decaying indirectly via a \lb\ coupling are shown 
for $\Delta m \leq m_{\snu}/2$ 
in Figure~\ref{fig:cross_sneutrino_ind_lb}. These limits degrade for
$\Delta m > m_{\snu}/2$ due to the larger fraction of missing energy
and lower energy final state charged leptons.  

\begin{figure}[htbp]
\centering
\begin{tabular}{c}
\epsfig{file=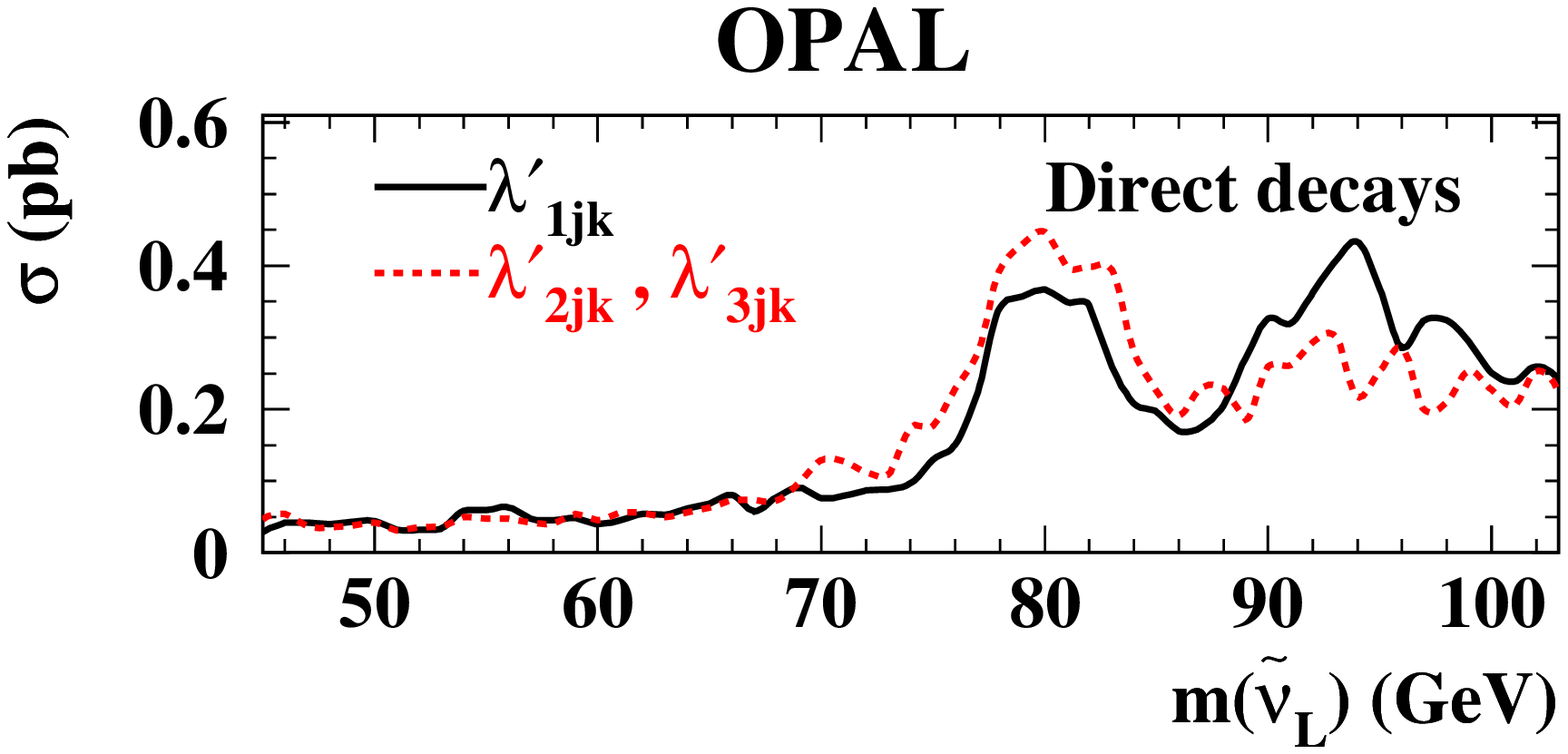     ,width=15.0cm} \\
\end{tabular}
\caption[]{\it
Sneutrino direct decays via a \lbp\ coupling:
upper limits at the 95\% C.L. on the 
pair-production cross-sections of $\snu_{\mathrm e}$ (solid line)
and $\snu_{\mu}$/$\snu_{\tau}$ (dashed line) decaying directly. 
} 
\label{fig:cross_sneu_4jets}
\end{figure}

Upper limits on the cross-sections of pair-produced  
sneutrinos decaying directly via a \lbp\ coupling to a four-jet final state
are shown in Figure~\ref{fig:cross_sneu_4jets}.

\begin{figure}[htbp]
\centering
\begin{tabular}{c}
\epsfig{file=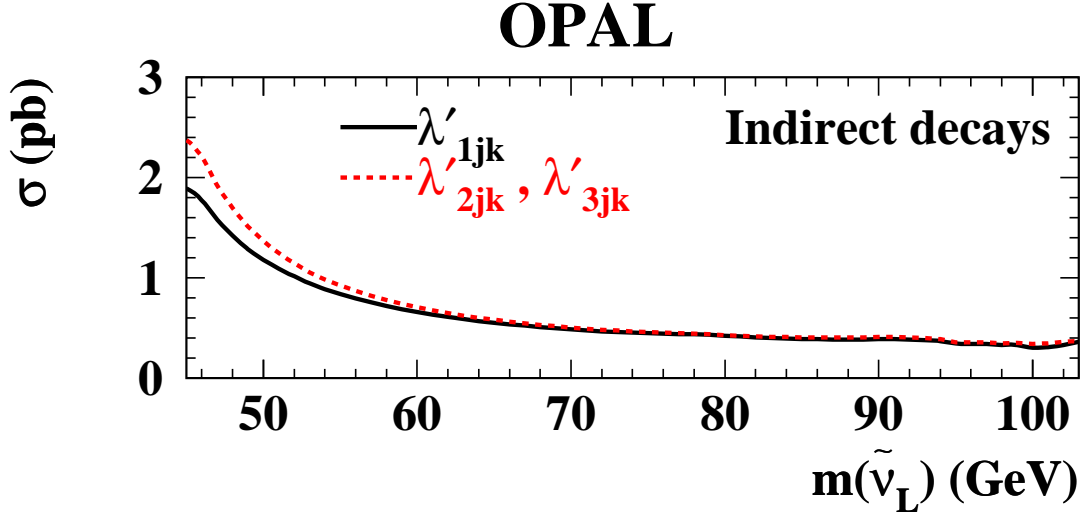 ,width=15.0cm} \\
\end{tabular}
\caption[]{\it
Sneutrino indirect decays via a \lbp\ coupling:
upper limits at the 95\% C.L. on the 
pair-production cross-sections of $\snu_{\mathrm e}$ (solid line)
and $\snu_{\mu}$/$\snu_{\tau}$ (dashed line),
for $\Delta m \leq m_{\snu}/2$.
} 
\label{fig:cross_sneu_4jets_emiss}
\end{figure}

Upper limits on the cross-sections of pair-produced  
sneutrinos decaying indirectly via a \lbp\ coupling to a four-jet final state
with missing energy are shown 
for $\Delta m \leq m_{\snu}/2$
in Figure~\ref{fig:cross_sneu_4jets_emiss}.
These limits degrade for
$\Delta m > m_{\snu}/2$ due to the larger fraction of missing energy
and lower energy final state jets.  

\begin{figure}[htbp]
\centering
\begin{tabular}{cc}
\epsfig{file=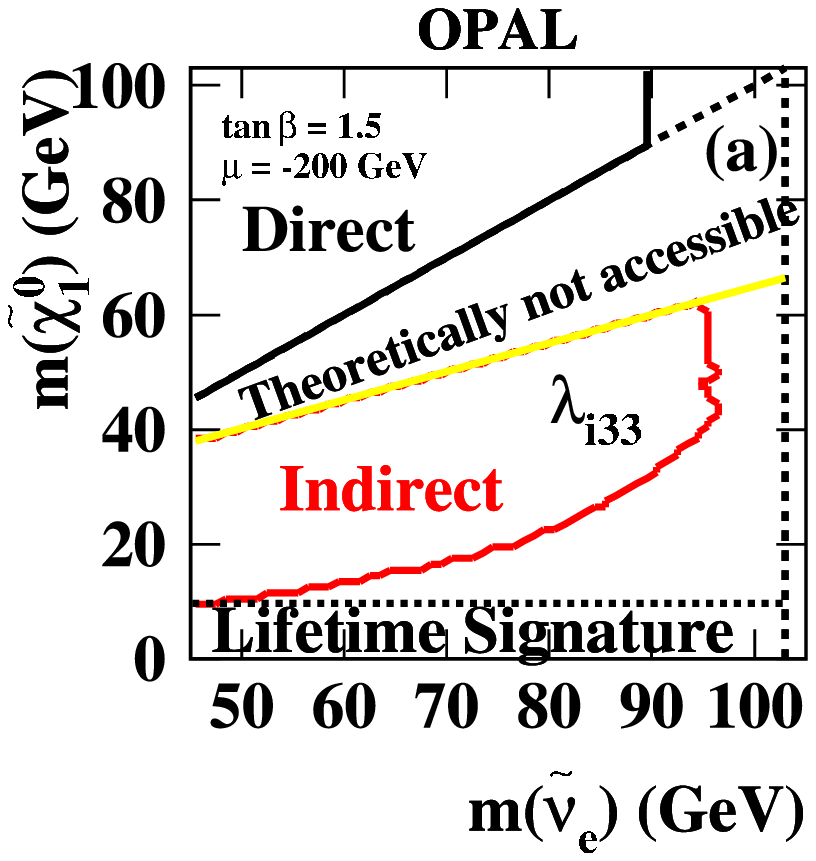,  width=8.0cm} &
\epsfig{file=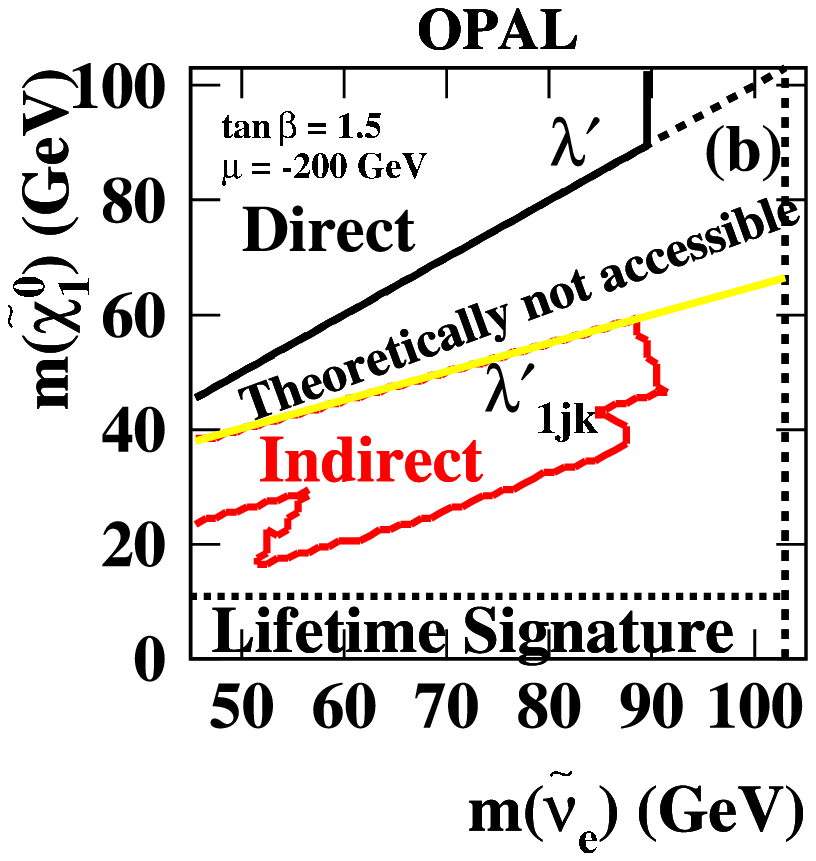, width=8.0cm} \\
\end{tabular}
\caption[]{\it
CMSSM exclusion region for $\snu_e \snu_e$ 
production in 
the $(m_{\snu_e}, m_{\nt_1})$ plane at 95\% C.L.  
for  (a) a \lb\ coupling and (b) a \lbp\ coupling, assuming 
the CMSSM predicted branching ratio for the 
decay  $\sneutrino_e \rightarrow \nu_e \nt_1$. 
The kinematic limit is shown by the dashed lines.  }
\label{fig:mssm_sneutrino_e}
\end{figure}

In the CMSSM, the $\snu_e$ pair-production cross-section  
is enhanced by the presence of the $t$-channel diagram.
The exclusion limits for $\snu_e$
decaying via a \lb\ coupling are 
shown in Figure~\ref{fig:mssm_sneutrino_e}(a).  
For a low-mass $\nt_1$, this excluded region degrades
due to a substantially smaller branching ratio for the 
decay $\sneutrino \rightarrow \nu \nt_1$ 
but also due to a smaller detection efficiency for 
$\Delta m > m_{\snu}/2$. 
The exclusion limits for $\snu_e$
decaying via a \lbp\ coupling are shown 
in Figure~\ref{fig:mssm_sneutrino_e}~(b). 
The lower mass limits for electron sneutrinos decaying directly or indirectly
via a \lb\ or a \lbp\ coupling are summarised in 
Table~\ref{tab:snumasslimits}.

\begin{table}[htbp]
\begin{center}
\begin{tabular}{|c||c|c||c|c|}
\hline
\multicolumn{5}{|c|}{Sneutrino Lower Mass Limits(GeV)} \\
\hline
 Species &   \multicolumn{2}{c|}{$\lambda$}    
                 &   \multicolumn{2}{c|}{$\lambda^{'}$}   \\
\hline
                 &    Direct & Indirect   &    Direct & Indirect \\
\cline{2-3}
\cline{4-5}
$\snu_e$                &    89     &   95          &    89   &   88     \\

$\snu_{\mu}/\snu_{\tau}$&    79     &   81          &      74   &   -     \\

\hline
\end{tabular}
\end{center}
\caption{\it
Lower mass limits for sneutrinos decaying directly or indirectly
via a \lb\ or a \lbp\ coupling. 
For the indirect decays, lower mass limits are quoted for
a large-mass $\nt_1$ (60~GeV).
}
\label{tab:snumasslimits}
\end{table}

Some combinations of neutralino and sneutrino masses do not correspond
to points in the CMSSM parameter space for the specific choice of 
$\mu = -200$~GeV and $\tan \beta$ = 1.5.
Regions where this is the case are
labelled ``Theoretically not accessible'' in the figures. 

\begin{figure}[htbp]
\centering
\begin{tabular}{cc}
\epsfig{file=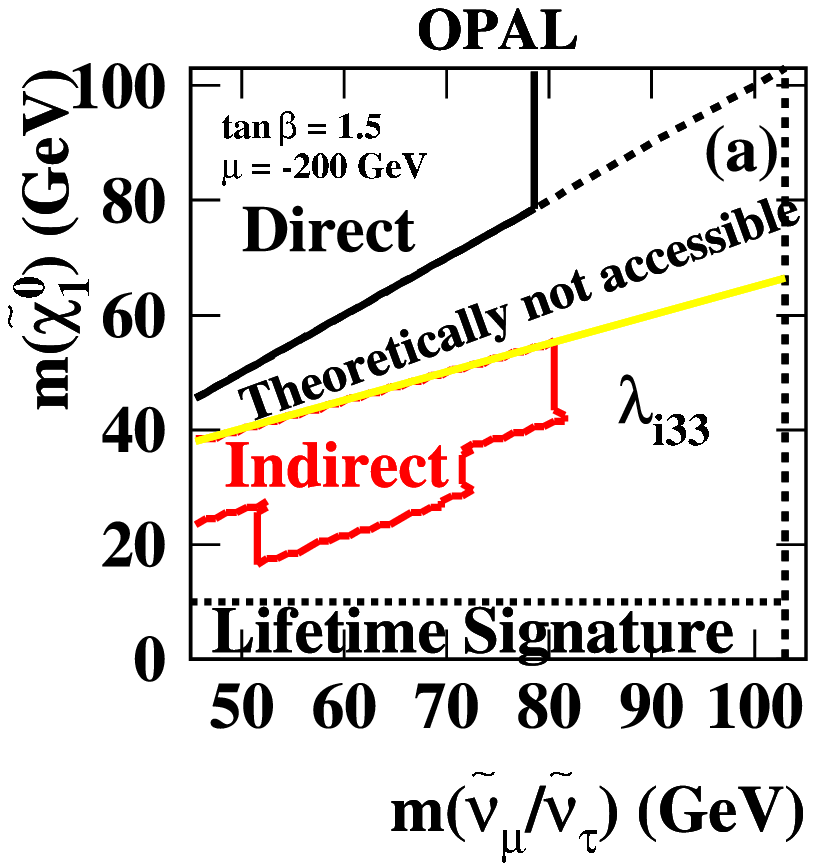,  width=8.0cm} &
\epsfig{file=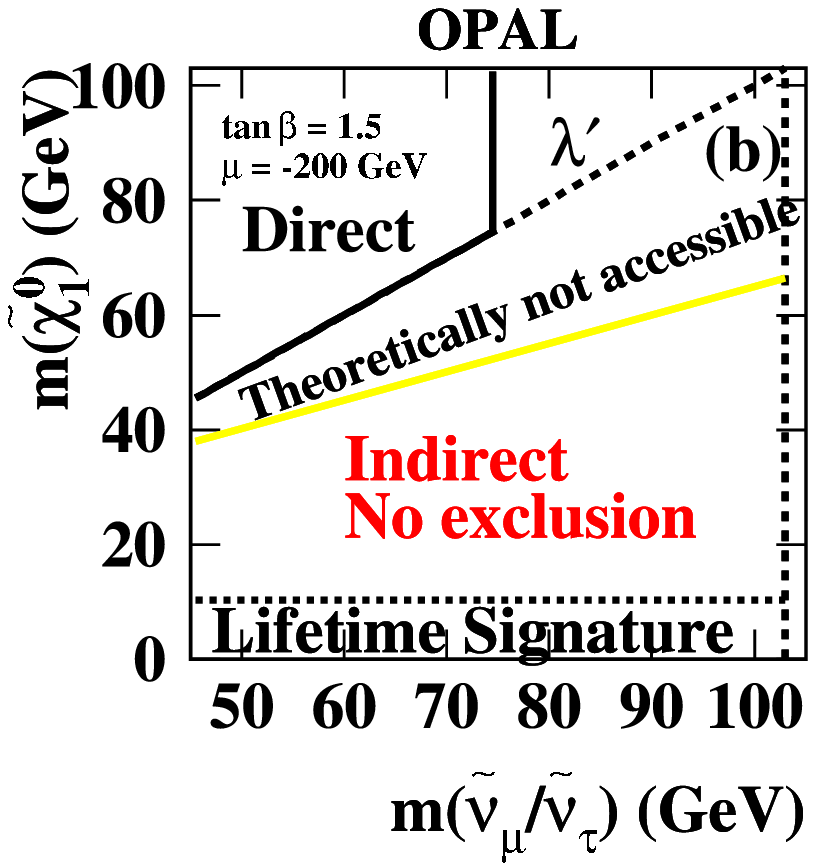, width=8.0cm} \\
\end{tabular}
\caption[]{\it
CMSSM exclusion region for $\snu_{\mu} \snu_{\mu}$, $\snu_{\tau} \snu_{\tau}$
production in 
the $(m_{\snu}, m_{\nt_1})$ plane at 95\% C.L.  
for  (a) a \lb\ coupling and (b) a \lbp\ coupling, assuming 
the CMSSM predicted branching ratio for the 
decay  $\sneutrino \rightarrow \nu \nt_1$. 
The kinematic limit is shown as the dashed line.  }
\label{fig:mssm_sneutrino_mu}
\end{figure}

The theoretical cross-sections for $\snu_{\mu}$ and 
$\snu_{\tau}$ pair-production are smaller than for $\snu_e$. 
The exclusion limits for $\snu_{\mu}$/$\snu_{\tau}$
decaying via a \lb\ coupling are shown in Figure~\ref{fig:mssm_sneutrino_mu}(a).  

The exclusion limits for $\snu_{\mu}$/$\snu_{\tau}$
decaying via a \lbp\ coupling are shown 
in Figure~\ref{fig:mssm_sneutrino_mu}~(b). 
No exclusion is possible for the indirect decays. 
The lower mass limits for muon/tau sneutrinos decaying directly via a 
\lb\ or a \lbp\ are summarised in Table~\ref{tab:snumasslimits}.

\subsection{Stop limits}

\begin{figure}
\centering
\begin{tabular}{c}
\epsfig{file=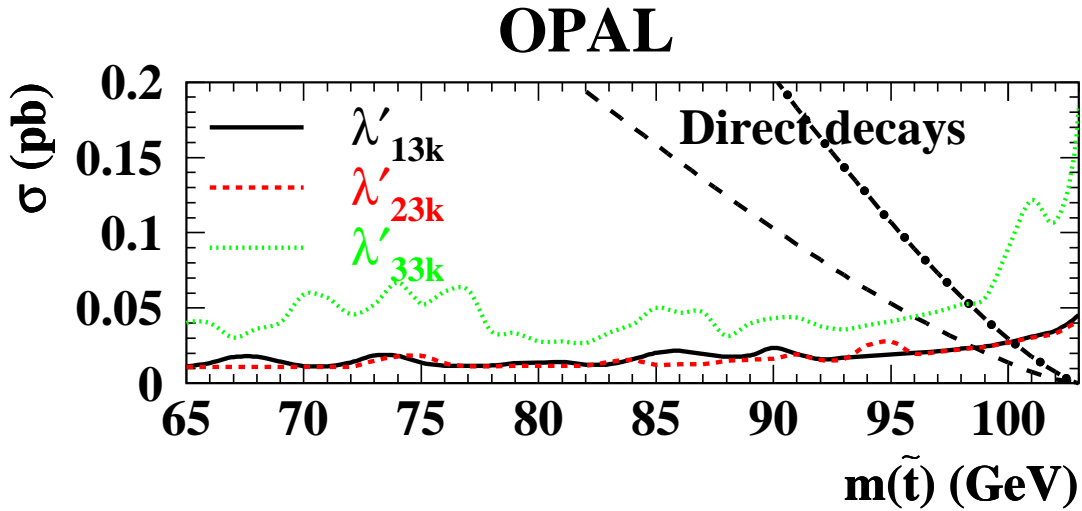,width=15cm} \\
\end{tabular}
\caption[]{\it
Stop direct decays via a \lbp\ coupling: 
upper limits at the 95\% C.L. on the production cross-section of $\stopx$
in the electron channel (solid line), the muon channel (short-dashed line) and 
in the tau channel (dotted line). 
Also shown are the maximum (dashed-dotted line) 
and minimum (long-dashed line)
cross-sections predicted by the CMSSM,
corresponding to mixing angles of 0 rad and 0.98 rad (decoupling 
limit). 
}
\label{fig:limit_stop}
\end{figure}

A cross-section limit of 0.03~pb was derived for the pair-production
of stops decaying directly via \lbp$_{13k}$ or \lbp$_{23k}$ (electron 
and muon channels),
in the mass region 45~GeV $< m_{\stopx} <$100~GeV. The
excluded cross-sections are shown in 
Figure~\ref{fig:limit_stop} as a function of the stop mass
for the electron, muon and tau channels.

If one assumes a stop production cross-section as predicted by the CMSSM, 
masses lower than 98~GeV can be excluded for any mixing 
angle $\theta_{\stopx}$.
A cross-section limit of 0.07~pb was derived for the pair-production
of the stops decaying directly via \lbp$_{33k}$ (tau channel),
in the mass region 45~GeV $< m_{\stopx} <$ 100~GeV. 
In the tau channel, 
masses lower than 96~GeV can be excluded for any mixing 
angle $\theta_{\stopx}$. 
More detailed exclusion limits 
are given in Table~\ref{tab:limit_stop}. 
The value of the mixing angle $\theta_{\stopx} = 0.98$ rad 
corresponds to the decoupling of the stop from the Z$^0$ and 
gives the lowest stop pair-production cross-sections.

\begin{table}[htbp]
\begin{center}
\renewcommand{\arraystretch}{1.3}
\begin{tabular}{|c||c|c|}
\hline
\multicolumn{3}{|c|}{Stop Lower Mass Limits (GeV)} \\
\hline
 Channels &   \multicolumn{2}{c|}{$\theta_{\stopx}$} \\
\hline
                 &    0 rad      &    0.98 rad \\
\cline{2-2}
\cline{3-3}
$ \stopm \rightarrow {\rm e} +$q & 100  & 98  \\
$ \stopm \rightarrow \mu +$q     & 100  & 98  \\
$ \stopm \rightarrow \tau +$q    &  98  & 96  \\
\hline

$ \stopm \rightarrow $ qq        &  88  & 77  \\
\hline
\end{tabular}
\end{center}
\caption{\it
Stop lower mass limits for the two extreme values of the mixing
angle in the electron, muon and tau channels as well as in the 4-jet 
channel.
}
\label{tab:limit_stop}
\end{table}

\begin{figure}
\centering
\begin{tabular}{c}
\epsfig{file=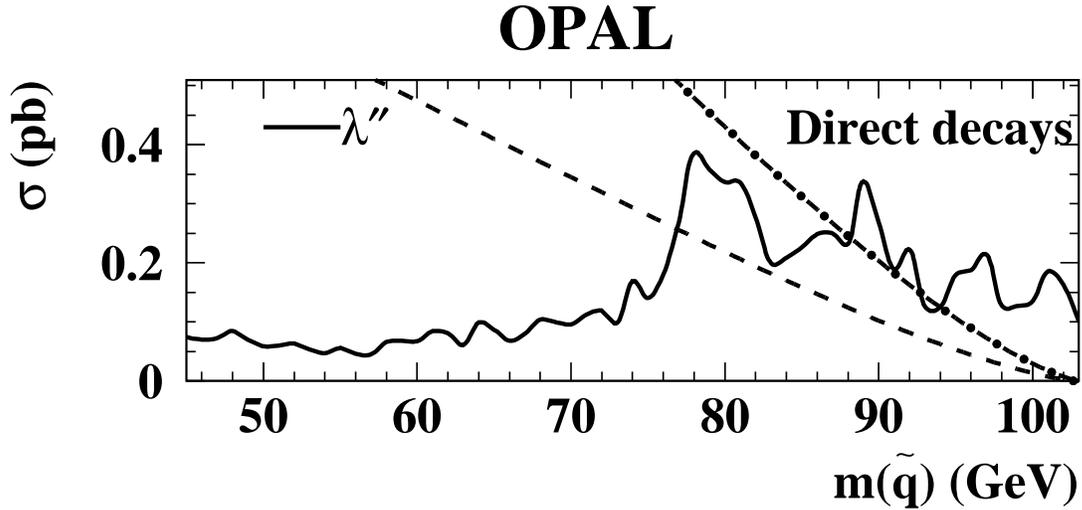,width=15cm} \\
\end{tabular}
\caption[]{\it
Squark direct decays via a \lbpp\ coupling: 
upper limits at the 95\% C.L. on the production cross-section 
of squarks decaying directly.
Also shown are the maximum (dashed-dotted line) and minimum (dashed line)
stop production cross-sections predicted by the CMSSM,
corresponding to mixing angles of 0 rad and 0.98 rad. }
\label{fig:limit_stop_4jets}
\end{figure}

For the light squark decays via \lbpp couplings, 
a cross-section limit of approximately 0.1~pb was 
derived for a squark mass up to $\approx$~70 GeV. 
This limit degrades to 0.39~pb 
in the range of the W mass as shown in Figure~\ref{fig:limit_stop_4jets}.
Assuming the same detection efficiencies for the stop quarks as for the
simulated light squarks, a limit on the stop mass can be established. 
If the stop production cross-section predicted by the CMSSM is assumed, 
masses lower than 77~GeV can be excluded for any mixing 
angle $\theta_{\stopx}$.

\section{Conclusions}
\label{sec:conclusions}

A search was performed for pair-produced 
sfermions with \Rparity\ violating 
decays using the data collected by the OPAL detector at 
centre-of-mass energies of 189$-$209~GeV, 
corresponding to a total luminosity of 
approximately 610~pb$^{-1}$.
Direct and indirect \Rparity\ violating decay modes of $\sell$ and $\snu$ 
via the Yukawa
\lb\ and \lbp\ couplings were considered, as well as 
direct \Rparity\ violating decay modes of
$\stopx$ and $\sq$ via \lbp\ and 
\lbpp.

No significant excess of signal-like events was observed in the data.
Upper limits on the pair-production cross-sections for 
sfermions  
have been computed assuming that only \Rparity\ violating decays occur. 
These cross-section limits
depend only on the masses of the sfermions and 
not on other SUSY parameters.
Mass limits were derived in the framework of the 
Constrained Minimal Supersymmetric Standard Model 
whenever the predicted cross-sections were sufficiently large.

\bigskip\bigskip\bigskip
\appendix
\bigskip\bigskip\bigskip
\appendix
\par
{\Large\bf Acknowledgements}
\par
We particularly wish to thank the SL Division for the efficient operation
of the LEP accelerator at all energies
 and for their close cooperation with
our experimental group.  In addition to the support staff at our own
institutions we are pleased to acknowledge the  \\
Department of Energy, USA, \\
National Science Foundation, USA, \\
Particle Physics and Astronomy Research Council, UK, \\
Natural Sciences and Engineering Research Council, Canada, \\
Israel Science Foundation, administered by the Israel
Academy of Science and Humanities, \\
Benoziyo Center for High Energy Physics,\\
Japanese Ministry of Education, Culture, Sports, Science and
Technology (MEXT) and a grant under the MEXT International
Science Research Program,\\
Japanese Society for the Promotion of Science (JSPS),\\
German Israeli Bi-national Science Foundation (GIF), \\
Bundesministerium f\"ur Bildung und Forschung, Germany, \\
National Research Council of Canada, \\
Hungarian Foundation for Scientific Research, OTKA T-038240, 
and T-042864,\\
The NWO/NATO Fund for Scientific Research, the Netherlands.\\


\end{document}